%% file: main.tex
\documentclass[11pt]{article}
\pdfoutput=1

\usepackage{epsfig}
\usepackage{latexsym}
\usepackage{enumerate}
\usepackage{url}
\usepackage{float}
\usepackage[hypertexnames=false,hyperfootnotes=false]{hyperref}
\usepackage{texnansi}
\usepackage{color}
\usepackage{tikz}
\usepackage[margin=10pt,font=small,labelfont=bf]{caption}
\usepackage[subrefformat=parens,labelformat=parens]{subcaption}
\usepackage{afterpage}
\usepackage{enumitem}
\usepackage[boxed]{algorithm}
\usepackage{algpseudocode}
\usepackage[normalem]{ulem}
\usepackage{lmodern}
\usepackage{booktabs}
\usepackage{graphicx}
\usepackage{sectsty}
\usepackage{ifthen}
\usepackage{amsmath}
\usepackage{amsthm}
\usepackage{amssymb}
\usepackage{amsfonts}
\usepackage{thmtools}
\usepackage{thm-restate}
\usepackage{mathtools}
\usepackage{xspace}
\usepackage{titling}
\usepackage{natbib}
\usepackage{xfrac}
\usepackage{multirow}
\usepackage{array}
\usepackage{bigdelim}
\usepackage{bm}
\usepackage{arydshln}
\usepackage{makecell}
\usepackage{placeins}
\usepackage{cleveref}
\usepackage[textsize=tiny]{todonotes}
\newcommand{\E}{\ensuremath{\mathsf{E}}} 

\DeclareMathOperator{\Var}{Var}

\declaretheoremstyle[headfont=\sffamily\bfseries,bodyfont=\itshape]{thm-sf}

\crefname{assumption}{assumption}{assumptions}

\renewcommand{\thmcontinues}[1]{\hyperref[#1]{continued}}
\newcommand{\paraheader}[1]{\smallskip\noindent{\sffamily\bfseries #1}}

\usetikzlibrary{arrows,patterns,plotmarks,pgfplots.groupplots}
\tikzstyle{every picture} += [>=stealth]
\tikzset{axis/.style={semithick, line join=miter}}
\allsectionsfont{\sffamily}
\makeatletter
\def\@seccntformat#1{\csname the#1\endcsname.\quad}
\makeatother

\floatname{algorithm}{\normalfont\sffamily\bfseries Algorithm}
\floatstyle{ruled}
\newcommand{\emailhref}[1]{\href{mailto:#1}{\tt #1}} 
\graphicspath{ {./figures/} }
\makeatletter
\let\template@orig@includegraphics\includegraphics
\renewcommand{\includegraphics}[2][]{%
  \IfFileExists{#2}{\template@orig@includegraphics[#1]{#2}}{%
    \IfFileExists{figures/#2}{\template@orig@includegraphics[#1]{figures/#2}}{%
      \fbox{\parbox[c][0.35\textheight][c]{0.85\linewidth}{\centering Missing figure: \texttt{\detokenize{#2}}}}%
    }%
  }%
}
\makeatother
\provideboolean{fastcompile}
\newcommand{\hidefastcompile}[1]{\ifthenelse{\boolean{fastcompile}}{}{#1}}
\usepackage{pgfplots}
\usepackage{pgfplotstable}
\usetikzlibrary{calc}
\usepackage{mathtools}
\definecolor{orange}{rgb}{0.85,0.33,0.13} 
\definecolor{green}{rgb}{0.13,0.85,0.33}
\definecolor{purple}{rgb}{0.33,0.13,0.85}
\definecolor{lime}{rgb}{0.65,0.85,0.13}
\definecolor{blue}{rgb}{0.13,0.65,0.85}
\pgfplotscreateplotcyclelist{tricolor}{%
  orange,every mark/.append style={fill=orange!80!black},mark=*\\%
  green,every mark/.append style={fill=green!80!black},mark=square*\\%
  purple,every mark/.append style={fill=purple!80!black},mark=otimes*\\%
  black,mark=star\\%
  orange,every mark/.append style={fill=orange!80!black},mark=diamond*\\%
  green,densely dashed,every mark/.append style={solid,fill=green!80!black},mark=*\\%
  purple,densely dashed,every mark/.append style={solid,fill=purple!80!black},mark=square*\\%
  black,densely dashed,every mark/.append style={solid,fill=gray},mark=otimes*\\%
  orange,densely dashed,mark=star,every mark/.append style=solid\\%
  green,densely dashed,every mark/.append style={solid,fill=green!80!black},mark=diamond*\\%
}
\pgfplotsset{colormap={tricolormap}{color=(orange) color=(green) color=(purple)},
  colormap={quadcolormap}{color=(orange) color=(lime) color=(blue) color=(purple)}}
\pgfplotstableset{%
  font=\small,
  every head row/.style={before row=\toprule[1pt], after row=\midrule},
  every last row/.style={after row=\bottomrule[1pt]}}
\pgfplotsset{compat=1.15}

\usepackage{setspace}
\usepackage[margin=1in]{geometry}

\provideboolean{submissionversion}
\setboolean{submissionversion}{true}

\ifthenelse{\boolean{submissionversion}}{
  \renewcommand{\todo}[2][1]{}
}{
}

\title{Volatility in Prediction Markets:
  A Structural Approach\thanks{CCM is supported by the Briger Family Digital
    Finance Lab at Columbia University. CCM is a research advisor for Paradigm and Uniswap Labs. }
}
\author{Weiye Xi\thanks{Columbia University. \emailhref{wx2305@columbia.edu}.}
\and
Ciamac C.~Moallemi\thanks{Columbia University, Paradigm, and Uniswap Labs. \emailhref{ciamac@gsb.columbia.edu}.}
\and
Mallesh Pai\thanks{Tempo Labs. \emailhref{mallesh.pai@gmail.com}.}
\and
Shouqiao Wang\thanks{Columbia University. \emailhref{shwang27@gsb.columbia.edu}.}}
\date{}
\begin{document}
\maketitle

\input{0_abstract}

\input{1_introduction}
\input{2_model}
\input{3_empirical}

\input{4_primacy}
\input{5_universality}
\input{6_conclusion}

\newpage
\bibliography{references}

\newpage
\appendix
\input{7_appendix}

\end{document}

%% file: 0_abstract.tex
\begin{abstract}
Forward-looking volatility forecasts are central inputs to derivatives pricing, market making, risk management, and volatility-linked trading strategies, with ARCH and GARCH models serving as the canonical workhorses. Such models are natural in standard asset markets, where prices are positive-valued stochastic processes and volatility is typically inferred from return dynamics. Prediction markets have a different structure: prices are bounded probabilities, payoffs are binary, and contracts resolve at known deadlines.

We develop and estimate a volatility model tailored to binary prediction markets. The model combines two economic mechanisms: a Wright-Fisher deadline-resolution component, capturing how remaining binary uncertainty is forced to resolve over time, and a Glosten-Milgrom order-flow component, capturing volatility from informed trading as reflected in spreads and volume. Using a large panel of Kalshi contracts, we show that these structural variables carry substantial forecasting power. Plain ARCH/GARCH benchmarks are dominated by structural specifications; scaling the structural variance by a GARCH multiplier gives the best overall forecasts.The two layers play distinct roles: the structural term sets the level of conditional variance, while the GARCH recursion corrects the timing of interval violations.The model also provides an interpretable measurement framework: volatility is highest near fifty-fifty prices, rises near resolution, and varies across categories with the timing and discreteness of information arrival. Economics contracts are closer to smooth deadline-resolution dynamics, while sports contracts exhibit more event-concentrated, jump-like behavior. Across major categories, re-estimating the model within each category does not improve out-of-sample performance, suggesting that the pooled structural specification transfers beyond the headline result.

\end{abstract}

\medskip


%% file: 1_introduction.tex
\expandafter\ifx\csname ifconf\endcsname\relax\expandafter\newif\csname ifconf\endcsname\fi
\section{Introduction}
\label{sec:intro}

Prediction-market prices are increasingly used as real-time probability forecasts for economic, political, weather, sports, and entertainment events. Exchange-traded platforms such as Kalshi and Polymarket have grown substantially in the past several years, with recent financial press coverage documenting rapid increases in trading volume \citep{wsj2026kalshiprofits}. Prediction-market prices are now routinely referenced as market-implied probabilities. However, a price is only a level forecast: it says what the market currently believes, not how likely that belief is to change. A contract trading at a given price may represent a stable consensus of genuine uncertainty, or it may sit immediately before an information event that is likely to move the price sharply. Distinguishing these two cases requires a forward-looking model of conditional volatility, not a backward-looking realized-volatility summary.

Volatility forecasts matter because decisions depend on how much prices may move, not just where they are now. Market makers, who offer to buy and sell contracts, use these forecasts to choose their quoted prices and how much they are willing to trade: larger anticipated price movements increase the risk of holding contracts between trades. Traders face a related problem when deciding how large a position to take. The same position can produce small or large fluctuations in profit and loss depending on price volatility, so a limit on acceptable risk translates into different position sizes as predicted volatility changes. These considerations matter even for traders who agree on an event's probability, because they may need to sell before the outcome is known. Beyond trading, users who rely on prediction-market probabilities to make decisions need to assess how much those probabilities might change over their decision horizon. Forecasting volatility therefore supports both the operation of prediction markets and the use of their prices as forecasts.

The prediction-market setting, however, is different from the standard asset-market setting. ARCH and GARCH models are natural workhorses for volatility forecasting in ordinary financial assets, where prices are positive-valued stochastic processes and volatility is typically inferred from return dynamics \citep{engle1982autoregressive,bollerslev1986generalized}. A prediction-market contract instead has a price that is a bounded probability, a payoff that is binary, and a resolution date that is known in advance. These features impose structure that the former models do not encode. Order-book variables add further information, since spreads and trading activity reflect adverse selection and informed trading.

In this paper, we develop a structural volatility model tailored to this setting. Suppressing the hourly index, the baseline specification writes conditional variance as the sum of two components:
        \[
          h_t^2 \;=\;
            \underbrace{\frac{p_t(1-p_t)}{\tau_t}}_{\text{deadline-resolution channel}}
            \;+\;
            K \cdot \underbrace{\nu(V_t) \cdot \frac{s_t^2}{4}}_{\text{adverse-selection channel}}
        \]
The first term is a Wright-Fisher deadline-resolution (DR) component: it captures how remaining binary uncertainty, $p_t(1-p_t)$, is released over the time left to resolution, $\tau_t$. Here $p_t$ is the prediction-market price. The second term is a Glosten-Milgrom adverse-selection (AS) component: it links bid-ask spreads ($s_t$) to trade-level adverse-selection variance and scales this variance by a monotone transformation of trading activity ($\nu(V_t)$). The parameter $K$ governs the strength of the adverse-selection channel. We refer to this model as DR-AS.

Empirically, we construct an hourly panel of Kalshi binary contracts from August 2021 through April 2026. For each contract-hour, we measure the market-implied probability, bid-ask spread, trading volume, and time to resolution before the next-hour price is realized. Our headline exercise forecasts the one-hour-ahead volatility of the probability, that is, the conditional standard deviation of the next-hour price change. Each month, we fit the models on all prior observations and test them on that month's contract-hours, yielding roughly 6.8 million out-of-sample forecasts. Because conditional variance is not directly observed, we evaluate each volatility forecast through the prediction interval it implies for the next-hour price. We score these intervals using the Winkler interval score \citep{winkler1972decision}, which penalizes intervals that are too wide as well as intervals that miss the realized price, with interval endpoints calibrated from each model's own out-of-sample standardized innovations rather than a Gaussian multiplier. As a complementary loss on the variance forecast itself, we also report the QLIKE loss, both against squared price changes and against an intraday realized-variance target.

We compare DR-AS with a broad set of structural, dynamic, and hybrid specifications. Plain ARCH/GARCH benchmarks lag the structural specifications: the pure deadline-resolution component already improves substantially on those generic benchmarks, DR-AS with concave volume scaling is the strongest structural specification, and GARCH \(\times\) DR-AS, which scales DR-AS by a GARCH(1,1) multiplier, is the best overall model. In headline numbers, the one-parameter structural specification outperforms a plain GARCH(1,1) by \(15\%\) on the volume-weighted interval score. Under the QLIKE loss \citep{patton2011volatility}, a likelihood-based loss that is robust to noise in the volatility proxy, the structural forecast assigns roughly twice the per-hour likelihood of plain GARCH to the realized price changes. Relative to the Archak--Ipeirotis benchmark, the closest existing model, DR-AS improves the interval score by \(8\%\) and assigns 14 times the per-hour likelihood to the realized price changes, largely because the probit-Brownian shape lets the variance collapse near the price boundaries; with GARCH dynamics on both sides, GARCH \(\times\) DR-AS improves the interval score by \(9\%\) and the likelihood ratio is 1.9.

Another benefit of our model is that it decomposes conditional variance into interpretable components. Volatility is highest near fifty-fifty prices, where binary outcome uncertainty $p(1-p)$ peaks, and intensifies as resolution approaches, consistent with the $1/\tau$ deadline-resolution scaling; both patterns are visible in model-free univariate binning (Section~\ref{sec:stylized}). Spreads and volume are weaker drivers in isolation, but jointly, through the order-flow term, they capture microstructure volatility outside the deadline channel (Section~\ref{sec:horserace}).

Finally, we show that the globally fitted GARCH \(\times\) DR-AS specification is the best or statistically indistinguishable from the best volatility predictor in most major Kalshi categories, including Sports, Politics, Economics, Elections, and Entertainment, which span the bulk of the panel. Although categories differ in absolute volatility level and modeling difficulty, re-estimating parameters within each category makes out-of-sample performance worse on most of the traded volume.

The rest of the paper is organized as follows. Section~\ref{sec:model} derives the structural model. Section~\ref{sec:empirical_design} describes the Kalshi panel, the specifications, and the interval-score evaluation. Section~\ref{sec:horserace} reports the main forecast comparison\ifconf, and Section~\ref{sec:universality} examines category heterogeneity and portability.\else, Section~\ref{sec:universality} examines category heterogeneity and portability, and Section~\ref{sec:conclusion} concludes.\fi

\subsection{Related Literature}
\label{sec:literature}


\paraheader{Volatility forecasting.} The conditional-volatility literature begins with ARCH and GARCH \citep{engle1982autoregressive,bollerslev1986generalized} and extends through realized-volatility approaches such as \citet{andersen2003modeling} and the heterogeneous autoregressive model of \citet{corsi2009simple}, which we include as a realized-volatility benchmark. These models are designed for financial return series and capture persistence in conditional second moments. Our setting differs because the traded object is a bounded probability that resolves to a binary payoff at a known deadline.

\paraheader{Market microstructure and information flow.} Our order-flow component draws on the adverse-selection view of bid-ask spreads in \citet{glosten1985bid}. Related work studies the information content of trades \citep{easley1992time,hasbrouck1991measuring}. The use of volume as an information-arrival proxy is related to the mixture-of-distributions tradition \citep{tauchen1983price,andersen1996return}. In our setting, these ideas enter a binary-contract volatility model through the joint dependence of volatility on spreads and trading activity.

\paraheader{Prediction markets and binary prices.} A large prediction-market literature studies when market prices can be interpreted as probabilities or forecasts \citep{wolfers2004prediction,manski2006interpreting}. Hanson-style market scoring rules provide one mechanism-design foundation for information markets \citep{hanson2003combinatorial}. Closest to our paper, \citet{archak2010modeling} develop a structural volatility model for binary prediction markets based on latent belief dynamics. We use their implied probit-Brownian variance shape as a benchmark, but differ by using a Wright-Fisher/Bernoulli boundary shape, adding a Glosten-Milgrom order-flow channel, allowing multiplicative ARCH/GARCH dynamics, and evaluating the specifications on a large Kalshi panel. More recently, \citet{dalen2025toward} proposes a Black-Scholes-style kernel for prediction-market derivatives based on belief volatility and jump-diffusion dynamics.


%% file: 2_model.tex
\section{A Structural Model for Prediction-Market Volatility}
\label{sec:model}

\ifconf
The YES claim pays \(\Theta\in\{0,1\}\) at resolution time \(T\), and the prediction-market price is the posterior probability of the YES state, \(p_t=\mathbb P(\Theta=1\mid\mathcal F_t)\), so \((p_t)_{0\leq t\leq T}\) is a bounded martingale and terminal settlement gives \(p_T=\Theta\). We decompose market information into two observation channels, \(\mathcal F_t=\mathcal F_t^{P}\vee\mathcal F_t^{Q}\). The \(P\)-channel is the Wright-Fisher deadline-resolution channel, motivated below from weak public signals. The \(Q\)-channel is the Glosten-Milgrom order-flow channel, derived below from informed trading and bid-ask spreads. The next two subsections develop their per-step conditional variances separately; the final subsection combines them.

\else
This section derives a structural volatility model for a binary prediction-market claim. The market price is modeled as a posterior probability, and its one-step conditional variance is decomposed into two information channels: a Wright-Fisher deadline-resolution channel and a Glosten-Milgrom order-flow channel.

Let \((\Omega,\mathcal F,\mathbb P)\) be a probability space with filtration \((\mathcal F_t)_{0\leq t\leq T}\). The YES claim pays
\(
    \Theta\in\{0,1\}
\)
at resolution time \(T\), where \(\Theta=1\) means that the YES event occurs. Interpreting
\(\mathbb P\) as the risk-neutral distribution, the prediction-market price is the posterior
probability of the YES state:
\begin{equation*}
    p_t
    =
    \mathbb P(\Theta=1\mid \mathcal F_t)
    =
    \mathbb E[\Theta\mid \mathcal F_t].
\end{equation*}
By the tower property, \((p_t)_{0\leq t\leq T}\) is a bounded \(\mathcal F_t\)-martingale,
and terminal settlement reveals the true state, so \(p_T=\Theta\in\{0,1\}\).

We decompose market information into two observation channels:
\(
\mathcal F_t
    =
    \mathcal F_t^{P}\vee\mathcal F_t^{Q}
\).
The \(P\)-channel is the Wright-Fisher deadline-resolution channel, motivated below from weak public signals. The \(Q\)-channel is the Glosten-Milgrom order-flow channel, derived below from informed trading and bid-ask spreads. Both channels are signals about the same terminal state \(\Theta\). The next two subsections develop their per-step conditional variances separately; the final subsection combines them.
\fi

\subsection{Wright-Fisher Deadline-Resolution Channel}
\label{subsec:public_channel}
\ifconf
Let \(p_t^{P}=\mathbb P(\Theta=1\mid\mathcal F_t^{P})\) denote the market's posterior probability of YES given public information up to time \(t\). At resolution the claim pays a Bernoulli outcome with success probability \(p_t^{P}\), so the variance that remains to be resolved between \(t\) and \(T\) is \(p_t^{P}(1-p_t^{P})\). Suppose this remaining variance is released at a constant rate over the time to resolution \(T-t\). On the hourly observation grid \(t_i<t_{i+1}\) with step \(\Delta_i=t_{i+1}-t_i\) and time to resolution \(\tau_i=T-t_i\), the public channel then contributes the one-step conditional variance
\begin{equation}
    \Var\!\left(\Delta p_i^{P}\mid\mathcal F_{t_i}^{P}\right)
    =
    \frac{p_{t_i}(1-p_{t_i})}{\tau_i}\,\Delta_i,
    \label{eq:wf_discrete_variance}
\end{equation}
where \(\Delta p_i^{P}=p_{t_{i+1}}^{P}-p_{t_i}^{P}\). This is the deadline-resolution variance component used throughout the rest of the paper, evaluated in the empirical implementation at the observed full-market price \(p_{t_i}\).

In addition, this constant release rate can be micro-founded by the structural learning model in Appendix~\ref{app:wf_structural}, in which the market observes a stream of weak binary public signals about \(\Theta\). As the signal strength vanishes, the posterior converges to the neutral Wright-Fisher diffusion, a martingale on \([0,1]\) that is absorbed at \(0\) and \(1\), matching binary settlement. A deterministic information clock that accelerates as the deadline approaches forces settlement by \(T\), and under this clock the remaining variance \(p_t^{P}(1-p_t^{P})\) decays linearly in calendar time, which is the constant release rate behind \eqref{eq:wf_discrete_variance}.

\else

Let \(\widetilde p_u=\mathbb P(\Theta=1\mid\mathcal F_u^P)\) denote the market's posterior
probability of YES after observing the public-information channel up to information time \(u\). In
Appendix~\ref{app:wf_structural} we present a structural learning model in which the market
observes a stream of weak binary public signals about the eventual settlement state
\(\Theta\). Standard generator-convergence arguments yield the diffusion limit
\begin{equation}
  d\widetilde p_u
  =
  \sqrt{\widetilde p_u(1-\widetilde p_u)}\,d\widetilde B_u^{P},
  \label{eq:public_wf_diffusion}
\end{equation}
which is the neutral Wright-Fisher diffusion in information time \(u\). Here,
$(\widetilde B_u^{P})$ is a Brownian motion.

Two features make \eqref{eq:public_wf_diffusion} a natural object for a prediction-market
posterior. First, \(\widetilde p_u\) is a martingale on \([0,1]\), consistent with the requirement
that the price be the conditional expectation of a bounded payoff. Second, the diffusion
coefficient \(\sigma(p)=\sqrt{p(1-p)}\) vanishes at \(0\) and \(1\), so the boundary points are
absorbing; the only possible limit points are \(0\) and \(1\), matching binary
settlement. Moreover, it is possible to establish via standard arguments that
$\widetilde p_u \rightarrow \widetilde p_\infty \in \{0,1\}$, i.e., the process is absorbed
eventually almost surely.

\paraheader{Deadline clock.}
The SDE \eqref{eq:public_wf_diffusion} absorbs in \(\{0,1\}\) eventually, but not necessarily by any particular calendar time. \input{fc_moves/2_1_deadline_clock}
\paraheader{Related models of prediction-market beliefs.}
Wright-Fisher and related belief-martingale diffusions on \([0,1]\) have been used previously to model prediction-market prices and binary forecasts; see, among others, \cite{archak2010modeling,restocchi2018stylised,dalen2025toward}.

\paraheader{Discretization.}
The rest of the paper works in discrete time on an hourly observation grid \(t_i<t_{i+1}\) with step \(\Delta_i=t_{i+1}-t_i\) and time to resolution \(\tau_i=T-t_i\). Over such a step the variance budget \eqref{eq:wf_fundamental_variance} integrates in closed form and gives the one-step conditional variance
\begin{equation}
    \Var\!\left(\Delta p_i^{P}\mid\mathcal F_{t_i}^{P}\right)
    =
    \frac{p_{t_i}(1-p_{t_i})}{\tau_i}\,\Delta_i,
    \label{eq:wf_discrete_variance}
\end{equation}
where \(\Delta p_i^{P}=p_{t_{i+1}}^{P}-p_{t_i}^{P}\). The identity holds for every \(\Delta_i\leq\tau_i\). Appendix~\ref{app:variance_budget} gives the derivation. This is the deadline-resolution variance component used throughout the rest of the paper. In the empirical implementation, this public-channel expression is evaluated at the observed full-market price \(p_{t_i}\); this treats the current information from both channels as summarized by the common market posterior while keeping the two variance contributions separate.
\fi

\subsection{Glosten-Milgrom Order-Flow Channel}
\label{sec:model_flow}
\ifconf
The second channel captures private information revealed through order flow. Fix time \(t\) and write \(p=\mathbb P(\Theta=1\mid\mathcal F^{Q}_t)\) for the order-flow-channel posterior. An information-sensitive order arrives, either a buy or a sell of the YES claim. With probability \(\alpha\) the trader is informed, observes \(\Theta\), and trades in its direction; with probability \(1-\alpha\) the trader is uninformed and buys or sells with probability \(1/2\) each. A buy order is therefore positive evidence for the YES state, and Bayes' rule gives the posterior \(p^{+}\) after a buy and \(p^{-}\) after a sell. A competitive market maker quotes ask \(a=p^{+}\) and bid \(b=p^{-}\), so the spread \(s=a-b\) is the distance between the two possible posteriors. Because the updated posterior is a conditional martingale, its one-event conditional variance is
\begin{equation}
    v^{Q}(p,\alpha)
    =\operatorname{Var}(p_{\mathrm{after}}-p\mid\mathcal F^{Q}_t)
    =\frac{s^2}{4}\left[1-\alpha^2(2p-1)^2\right].
    \label{eq:gm_exact_variance}
\end{equation}
Appendix~\ref{app:gm_derivation} gives the derivation. The squared half-spread \(s^2/4\) is read off the book. The correction \(1-\alpha^2(2p-1)^2\) lies in \((0,1]\) and depends on the informed-trader share \(\alpha\), which is not observed.

To aggregate per-event variance to a per-step variance over the hourly grid of \eqref{eq:wf_discrete_variance}, let \(N_t^{Q}\) be the counting process of information-sensitive order-flow events and assume \(\mathbb E[dN_t^{Q}\mid\mathcal F^{Q}_t]=\kappa\,\nu(V_t)\,dt\), where \(V_t\) is observed volume, \(\nu(V_t)\) maps measured volume into an event-arrival proxy, and \(\kappa\ge0\) captures the effective intensity of information-sensitive order flow after controlling for \(V_t\). Multiplying the arrival rate by the one-event variance \eqref{eq:gm_exact_variance} gives the variance rate of the channel, \(\kappa\,\nu(V_t)\,(s_t^2/4)\,[1-\alpha_t^2(2p_t-1)^2]\). Only the spread in this expression is observed. The intensity \(\kappa\) and the correction factor enter it only through their product, so we estimate that product, \(\kappa\,[1-\alpha^2(2p-1)^2]\), as a single scale parameter \(K\). Treating the per-step state variables as locally constant on \([t_i,t_{i+1})\), the per-step conditional variance is
\begin{equation}
    \Var\!\left(\Delta p_i^{Q}\mid\mathcal F_{t_i}^{Q}\right)
    \approx
    K\,\nu(V_{t_i})\,\frac{s_{t_i}^2}{4}\,\Delta_i.
    \label{eq:gm_discrete_variance}
\end{equation}
This reduced-form expression treats \(s_{t_i}^2/4\), the squared half-spread, as the adverse-selection price-impact scale and lets \(K\) absorb the GM correction factor, trade aggregation, and the conversion from volume to information-event intensity. \(K\) is therefore not a structural estimate of the informed-trader share.

\else

The second channel captures private information revealed through order flow. Fix time \(t\) and write
\[
    p=\mathbb P(\Theta=1\mid\mathcal F^{Q}_t)
\]
for the order-flow-channel posterior.
\input{fc_moves/2_2_gm_derivation}Because \(p_{\mathrm{after}}=\mathbb P(\Theta=1\mid\mathcal F^{Q}_t,B/S)\) is a posterior, it is a conditional martingale, and its one-event conditional variance is
\begin{align}
    v^{Q}(p,\alpha)
    &=\operatorname{Var}(p_{\mathrm{after}}-p\mid\mathcal F^{Q}_t) \nonumber \\
    &=\mathbb P(B\mid\mathcal F^{Q}_t)(p^{+}-p)^2
      +\mathbb P(S\mid\mathcal F^{Q}_t)(p^{-}-p)^2 \nonumber \\
    &=\frac{4\alpha^2p^2(1-p)^2}{1-\alpha^2(2p-1)^2} \nonumber \\
    &=\frac{s^2}{4}\left[1-\alpha^2(2p-1)^2\right].
    \label{eq:gm_exact_variance}
\end{align}
Equation \eqref{eq:gm_exact_variance} is the exact binary Glosten-Milgrom one-event posterior variance. It splits that variance into two factors. The squared half-spread \(s^2/4\) is read off the book. The correction \(1-\alpha^2(2p-1)^2\) carries the informed-trader share \(\alpha\), which is not observed; it lies in \((0,1]\) and equals one at \(p=1/2\).

To aggregate per-event variance to a per-step variance over the same hourly grid \([t_i,t_{i+1})\) used in \eqref{eq:wf_discrete_variance}, let \(N_t^{Q}\) be the counting process of information-sensitive order-flow events, with arrival rate \(\lambda_t\) defined by \(\mathbb E[dN_t^{Q}\mid\mathcal F^{Q}_t]=\lambda_t\,dt\). By \eqref{eq:gm_exact_variance}, the variance rate of the channel is
\begin{equation*}
    \lambda_t\,v^{Q}(p_t,\alpha_t)
    =
    \frac{s_t^2}{4}\,\lambda_t\!\left[1-\alpha_t^2(2p_t-1)^2\right],
\end{equation*}
in which only the spread is observed. The arrival rate and the correction enter the variance rate only through their product, so neither has to be identified separately. We model that product, an effective arrival rate of information-sensitive order flow, as
\begin{equation*}
    \lambda_t\!\left[1-\alpha_t^2(2p_t-1)^2\right]
    =
    K\nu(V_t),
\end{equation*}
where \(V_t\) is observed volume, \(\nu(V_t)\) maps measured volume into an event-arrival proxy, and \(K\ge0\) is a scale parameter. Because the correction depends on \(p\) and \(\alpha\), \(K\) should be interpreted as an average price-impact multiplier rather than as a structural estimate of the informed-trader share. Treating the per-step state variables as locally constant on \([t_i,t_{i+1})\), the per-step conditional variance is
\begin{equation}
    \Var\!\left(\Delta p_i^{Q}\mid\mathcal F_{t_i}^{Q}\right)
    \approx
    K\,\nu(V_{t_i})\,\frac{s_{t_i}^2}{4}\,\Delta_i.
    \label{eq:gm_discrete_variance}
\end{equation}
This reduced-form expression treats \(s_{t_i}^2/4\), the squared half-spread, as the adverse-selection price-impact scale and lets \(K\) absorb the GM correction factor, trade aggregation, and the conversion from volume to information-event intensity.
\fi

\subsection{Combining the Two Channels}
\label{subsec:discrete_variance}
\ifconf
Both channels update the same posterior, so on the hourly grid the one-step price innovation \(\varepsilon_i=p_{t_{i+1}}-p_{t_i}\) is the sum of the two channel innovations, \(\Delta p_i^{P}+\Delta p_i^{Q}\). The two innovations are uncorrelated, so their variances \eqref{eq:wf_discrete_variance} and \eqref{eq:gm_discrete_variance} add:
\begin{equation}
    \Var\!\left(\varepsilon_i\mid\mathcal F_{t_i}\right)
    \approx
    h_i^2(K)
    =
    \left[
        \frac{p_{t_i}(1-p_{t_i})}{\tau_i}
        +
        K\,\nu(V_{t_i})\frac{s_{t_i}^2}{4}
    \right]\Delta_i .
    \label{eq:main-model}
\end{equation}
Equation~\eqref{eq:main-model} is an approximation to the conditional variance. The deadline-resolution term is exact, and the approximation lies in the order-flow term, which evaluates the spread and the trading activity at the forecast origin; the exact expression is derived in Appendix~\ref{app:variance_decomposition}. The covariance vanishes because the two channels never move the price at the same instant. Public information moves the price continuously, in many small steps, while order flow moves it only at the arrival of an informative order, by the jump \(p^{+}-p\) or \(p^{-}-p\). At those arrival instants the public channel does not move the price, and between them order flow does not, so the two contributions have no movement in common. Appendix~\ref{app:variance_decomposition} writes the price as a jump diffusion and gives the proof. This predictor is the structural object taken to the data in the rest of the paper. The choice of activity proxy \(\nu(\cdot)\), the estimation of \(K\), and the design of the forecast evaluation are discussed in Section~\ref{sec:empirical_design}; the out-of-sample results are in Section~\ref{sec:horserace}. In standardized form the price innovation is
\begin{equation}
    \varepsilon_i=h_i(K)\,z_i,
    \qquad
    \mathbb E[z_i\mid\mathcal F_{t_i}]=0,
    \qquad
    \operatorname{Var}(z_i\mid\mathcal F_{t_i})=1,
    \label{eq:innovation_equation}
\end{equation}
with \(z_i\) a unit-variance standardized shock on which no parametric distribution is imposed. Throughout, we refer to \(h_i=\sqrt{h_i^2(K)}\) as the \emph{volatility} of the one-step price innovation.

\else

Both channels update the same posterior. Collect the order-flow jumps into the pure-jump martingale
\begin{equation*}
    J_t^{Q}=\sum_{k=1}^{N_t^{Q}}Y_k,
\end{equation*}
where \(Y_k\) is the posterior jump at the \(k\)-th event, equal to \(p^{+}-p\) or \(p^{-}-p\) evaluated just before it. Evaluating both channels at the full-market price, the price follows the jump diffusion
\begin{equation}
    dp_t
    =
    \sqrt{\frac{p_{t-}(1-p_{t-})}{T-t}}\,dW_t^{P}
    +
    dJ_t^{Q},
    \label{eq:main-sde}
\end{equation}
with \(W^{P}\) the Brownian motion of \eqref{eq:wf_calendar_sde}. On the hourly grid, write the one-step price innovation as the sum of the two channel innovations:
\begin{equation*}
\varepsilon_{i} = p_{t_{i+1}} - p_{t_i}
    =
    \Delta p_i^{P}+\Delta p_i^{Q},
\end{equation*}
where \(\Delta p_i^{P}\) and \(\Delta p_i^{Q}\) are the conditional martingale increments generated by the Wright-Fisher deadline-resolution channel and the Glosten-Milgrom order-flow channel, respectively.

\paraheader{Variance decomposition.}
The per-step variances of the two channels in \eqref{eq:wf_discrete_variance} and \eqref{eq:gm_discrete_variance} add:
\begin{equation}
    \Var\!\left(\varepsilon_i\mid\mathcal F_{t_i}\right)
    \approx
    h_i^2(K)
    =
    \left[
        \frac{p_{t_i}(1-p_{t_i})}{\tau_i}
        +
        K\,\nu(V_{t_i})\frac{s_{t_i}^2}{4}
    \right]\Delta_i .
    \label{eq:main-model}
\end{equation}
Appendix~\ref{app:variance_decomposition} derives this decomposition. The two channels add because they never move the price at the same instant. The deadline channel moves it continuously, while order flow moves it only when an informative order arrives.

The approximation in \eqref{eq:main-model} evaluates spreads and activity at the forecast origin, since their path over the coming hour is not observed. This predictor is the structural object taken to the data in the rest of the paper. It combines a single deadline-resolution term, a single spread-scaled order-flow term, and one nonnegative scale parameter \(K\). The choice of activity proxy \(\nu(\cdot)\), the estimation of \(K\), and the design of the forecast evaluation are discussed in Section~\ref{sec:empirical_design}; the out-of-sample results are in Section~\ref{sec:horserace}.

Under the structural conditional-variance specification, the price innovation can be written in standardized form as
\begin{equation}
    \varepsilon_i=h_i(K)\,z_i,
    \qquad
    \mathbb E[z_i\mid\mathcal F_{t_i}]=0,
    \qquad
    \operatorname{Var}(z_i\mid\mathcal F_{t_i})=1,
    \label{eq:innovation_equation}
\end{equation}
so that \(h_i(K)\) is the conditional scale of the price innovation and \(z_i\) is a unit-variance standardized shock. This decomposition does not impose a Gaussian distribution, or any other parametric distribution, on \(z_i\); under \eqref{eq:main-sde} the hourly innovation is a Gaussian increment plus a random number of order-flow jumps. Throughout, we refer to the conditional standard deviation \(h_i=\sqrt{h_i^2(K)}\), the square root of the structural conditional variance, as the \emph{volatility} of the one-step price innovation.
\fi


%% file: fc_moves/2_1_deadline_clock.tex
To force terminal resolution at the deadline \(T\), introduce the deterministic information clock
\begin{equation*}
    A(t)
    =\int_0^t\frac{dr}{T-r}
    =-\log\!\left(\frac{T-t}{T}\right),
    \qquad
    A'(t)=\frac{1}{T-t},
\end{equation*}
which accelerates as resolution approaches and satisfies \(A(t)\to\infty\) as \(t\uparrow
T\). Define the calendar-time public posterior \(p_t^{P}=\widetilde p_{A(t)}\). After the time
change, \(p_t^{P}\) satisfies
\begin{equation}
  dp_t^{P}
  =
  \sqrt{\frac{p_t^{P}(1-p_t^{P})}{T-t}}\,dW_t^{P},
  \label{eq:wf_calendar_sde}
\end{equation}
where $(W^{P}_t)$ is a calendar-time Brownian motion.
\ifconf The next paragraphs establish\else Appendix~\ref{app:variance_budget} establishes\fi{} the calendar-time
variance-budget identity
\begin{equation}
  \Var[p_\tau^{P} \mid \mathcal{F}^P_t]
  =
  \mathbb{E}\!\left[
    \int_t^\tau
    \frac{p_s^{P}(1-p_s^{P})}{T-s}\,
    ds
    \,\middle|\,\mathcal{F}^P_t
  \right],
  \qquad t \leq \tau \leq T.
  \label{eq:wf_fundamental_variance}
\end{equation}
The integrand \(p_s^{P}(1-p_s^{P})\) is remaining binary uncertainty and \(1/(T-s)\) is the deadline
information-release rate. As \(\tau\uparrow T\), the right-hand side converges to \(p_t^{P}(1-p_t^{P})\),
recovering \(\Var[p_T^{P} \mid \mathcal{F}^P_t] = \Var[\Theta \mid \mathcal{F}^P_t] = p_t^{P}(1-p_t^{P})\); this
formalizes the statement that the deadline clock spends exactly the remaining binary uncertainty
by time~\(T\).

%% file: fc_moves/2_2_gm_derivation.tex
An information-sensitive order-flow event arrives. The order is either a buy order for the YES claim, denoted by \(B\), or a sell order, denoted by \(S\). With probability \(\alpha\), the trader is informed and observes \(\Theta\); with probability \(1-\alpha\), the trader is uninformed. The informed trader buys YES when \(\Theta=1\) and sells YES when \(\Theta=0\). The uninformed trader buys or sells with probability \(1/2\) each. Therefore
\begin{gather*}
    \mathbb P(B\mid\Theta=1)=\frac{1+\alpha}{2},
    \quad
    \mathbb P(B\mid\Theta=0)=\frac{1-\alpha}{2},
    \\
    \mathbb P(S\mid\Theta=1)=\frac{1-\alpha}{2},
    \quad
    \mathbb P(S\mid\Theta=0)=\frac{1+\alpha}{2}.
\end{gather*}

The likelihood ratio of a buy order is \((1+\alpha)/(1-\alpha)>1\), so a buy order is positive evidence for the YES state. The unconditional buy probability is
\begin{equation*}
    \mathbb P(B\mid\mathcal F^{Q}_t)
    =p\frac{1+\alpha}{2}+(1-p)\frac{1-\alpha}{2} =\frac{1+\alpha(2p-1)}{2}.
\end{equation*}
Bayes' rule gives the posterior after a buy order:
\begin{equation*}
    p^{+}
    =\mathbb P(\Theta=1\mid B,\mathcal F^{Q}_t)
    =\frac{p(1+\alpha)}{1+\alpha(2p-1)},
\end{equation*}
and after a sell order,
\begin{equation*}
    p^{-}
    =\mathbb P(\Theta=1\mid S,\mathcal F^{Q}_t)
    =\frac{p(1-\alpha)}{1-\alpha(2p-1)}.
\end{equation*}

A competitive risk-neutral market maker breaks even conditional on order direction: quoting ask
price \(a=\mathbb E[\Theta\mid B,\mathcal F^{Q}_t]=p^{+}\) and bid price
\(b=\mathbb E[\Theta\mid S,\mathcal F^{Q}_t]=p^{-}\). The spread is
\begin{equation*}
    s
    =a-b=p^{+}-p^{-}
    =\frac{4\alpha p(1-p)}{1-\alpha^2(2p-1)^2}.
\end{equation*}
The two possible posterior jumps are
\begin{equation*}
    p^{+}-p
    =\frac{2\alpha p(1-p)}{1+\alpha(2p-1)},
    \qquad
    p^{-}-p
    =-\frac{2\alpha p(1-p)}{1-\alpha(2p-1)}.
\end{equation*}

%% file: 3_empirical.tex
\expandafter\ifx\csname ifconf\endcsname\relax\expandafter\newif\csname ifconf\endcsname\fi
\section{Data and Forecast Evaluation}
\label{sec:empirical_design}

This section describes the empirical object and the forecast evaluation. We first
construct an hourly Kalshi panel and define the one-hour volatility target. We
then summarize the model families in the forecast comparison, describe the
expanding-window estimation and the interval-score and likelihood-based
evaluation, and close with
model-free stylized facts that corroborate the structural variables of
Section~\ref{sec:model}.

\subsection{Hourly forecasting panel}
\label{sec:dataset}

The data are contract-hour observations from Kalshi between August 2021 and
April 2026. Each observation is one listed yes/no claim during one calendar
hour. Markets with several mutually exclusive outcomes appear on Kalshi as
separate binary claims, one per candidate outcome, so the unit of observation is
the claim-level contract rather than the broader event. For each contract-hour
we observe price information, trading volume, and the best bid and ask at the close of the hour.

\medskip

\ifconf
At the close \(t_i\) of each hour we observe the mid-quote
\(p_{t_i}=\tfrac12(b_{t_i}+a_{t_i})\), the quoted spread
\(s_{t_i}=a_{t_i}-b_{t_i}\), the volume \(V_{t_i}\) traded during the hour, and
the time to resolution \(\tau_i=T-t_i\), and we forecast the scale of the
one-hour innovation \(\varepsilon_i=p_{t_{i+1}}-p_{t_i}\).
Appendix~\ref{app:model-classes} gives the full definitions of the
forecast-origin state and the forecast target.
\else
\input{fc_moves/3_1_state_target}\fi

\paraheader{Sample.}
We retain contracts with at least 48 hourly observations with a valid closing
quote, roughly two days of quoted hours, which removes very short-lived or
sparsely quoted contracts. We also exclude Kalshi's multi-leg (KXMVE) markets,
whose quoted bid and ask are frequently placeholders at 0 and 1 rather than
executable quotes; they make up the entire Exotics category and \(0.8\%\) of
contract-hours. After the monthly split, the pooled out-of-sample panel
contains \(6{,}753{,}373\) contract-hours from \(41{,}348\) contracts for
structural specifications and \(6{,}664{,}822\) for specifications with lagged
ARCH/GARCH recursions, which drop the first hour of each contract to
initialize the recursion.

\subsection{Model classes}
\label{sec:specifications}

\ifconf
We compare three classes of models (Table~\ref{tab:spec-families}). Structural predictors use only prediction-market state variables: deadline resolution alone, the binary-market benchmark of \cite{archak2010modeling}, and DR-AS. Generic time-series benchmarks, plain ARCH/GARCH and the HAR-RV model of \citet{corsi2009simple}, are the standard tools for any price series and condition only on the contract's own price history, so the gap between them and the structural predictors measures what the state variables \((p,\tau,s,V)\) add. The third class scales a structural baseline by a unit-mean ARCH/GARCH component in the multiplicative form of \citet{engle2008spline} and \citet{engle2012forecasting}. The structural term sets the level of conditional variance, and the dynamic component captures clustering around it.

Within DR-AS we compare four activity proxies, \(\nu(V)\in\{1,\log(1+V),\sqrt V,V\}\). The activity-scaled deadline family instead rescales the deadline clock by \(a(V)=\nu(V)/\bar V\). Appendix~\ref{app:model-classes} describes the model classes in full, and Appendix~\ref{app:estimation_details} gives the estimation details and recursions.
\else
\input{fc_moves/3_2_classes_intro}\fi

\begin{table}[t]
\centering
\caption{Specification families in the forecast comparison. Models are grouped
by their economic role.}
\label{tab:spec-families}
\scriptsize
\setlength{\tabcolsep}{4pt}
\ifconf\def\fcTA{0.24}\def\fcTB{0.24}\def\fcTC{0.45}\def\fcRowSep{0.15em}\else\def\fcTA{0.23}\def\fcTB{0.34}\def\fcTC{0.35}\def\fcRowSep{0.4em}\fi
\begin{tabular}{>{\raggedright\arraybackslash}p{\fcTA\linewidth}
                >{\raggedright\arraybackslash}p{\fcTB\linewidth}
                >{\raggedright\arraybackslash}p{\fcTC\linewidth}}
\toprule
Family & Representative specification & Purpose \\
\midrule
Deadline resolution
&
\(p(1-p)/\tau\)
&
Binary uncertainty and fixed-deadline structure only
\\[\fcRowSep]

Archak--Ipeirotis benchmark \cite{archak2010modeling}
&
\(\varphi^2(\Phi^{-1}(p))/\tau\)
&
Latent-belief binary-market volatility benchmark
\\[\fcRowSep]

DR-AS
&
\(p(1-p)/\tau + K\nu(V)s^2/4\)
&
Adds spread-scaled order-flow variance
\\[\fcRowSep]

Activity-scaled deadline resolution
&
\(a(V)\,p(1-p)/\tau\)
&
Tests whether activity rescales the deadline clock
\\[\fcRowSep]

Plain ARCH/GARCH
&
ARCH(1), GARCH(1,1)
&
Generic volatility persistence without prediction-market state variables
\\[\fcRowSep]

Realized-volatility benchmark
&
HAR-RV \cite{corsi2009simple}
&
Persistence in lagged intraday realized variance, no prediction-market state variables
\\[\fcRowSep]

Multiplicative component dynamics
&
ARCH/GARCH \(\times\) structural baseline
&
Residual clustering after conditioning on binary-market structure
\\
\bottomrule
\end{tabular}
\end{table}

\ifconf
\else
\input{fc_moves/3_2_classes_details}\fi

\subsection{Estimation and forecast evaluation}
\label{sec:estimation}

Each specification produces, for every retained contract-hour \(i\), a
volatility forecast \(h_i\), equivalently a conditional variance forecast
\(h_i^2\). We estimate all specifications in a monthly expanding-window design:
each calendar month from September 2021 through April 2026 serves once as the
test window, with all earlier observations used for training. This yields 56
out-of-sample test months. A contract may appear in both a training window and a
later test month, but no observation from a test month is used to fit that
month's parameters.

\ifconf
Parameters are estimated separately for each test month by volume-weighted
Gaussian quasi-maximum likelihood, and the same volume weights are used in
evaluation. Appendix~\ref{app:model-classes} gives the estimation details.
\else
\input{fc_moves/3_3_estimation}\fi

\ifconf
We evaluate each forecast in two ways. The interval score VW-IS-EQ
\citep{winkler1972decision,gneiting2007strictly} scores the 95\% prediction
interval for the next-hour price implied by \(h_i\), using each model's own
empirical quantiles of the standardized innovation, so it rewards narrow
intervals and penalizes misses; VW-Cov-EQ reports the coverage of these
intervals. VW-QLIKE is the quasi-likelihood loss \citep{patton2011volatility}
of the variance forecast against the squared innovation, and VW-QLIKE-TSRV
replaces the squared innovation by the two-scale realized variance of the
mid-quote \citep{zhang2005tale}. All scores are volume-weighted, and lower is
better except for coverage. Appendix~\ref{app:model-classes} gives the
formulas.
\else
\input{fc_moves/3_3_evaluation}\fi

\subsection{Stylized facts}
\label{sec:stylized}

\ifconf
Model-free binning shows the two primitive inputs of the deadline-resolution
term directly in the data. We sort observations by each candidate driver and
compute the volume-weighted root-mean-square next-hour innovation within each
bin. Time to resolution produces the largest contrast across bins, a factor of
\(22.6\). Volatility rises sharply in the final hours, as the \(1/\tau\)
deadline clock implies, and is roughly flat more than a day from resolution,
where the order-flow term dominates the deadline term.
Along the price axis, volatility peaks near 50 cents and falls toward the
boundaries, matching the inverted-U of \(p(1-p)\).
Appendix~\ref{app:stylized} reports the binning figures and the ranking of all
candidate drivers, and Appendix~\ref{app:binning_figures} gives the full
diagnostics.
\else
\input{fc_moves/3_4_stylized}\fi

%% file: fc_moves/3_1_state_target.tex
\paraheader{Forecast-origin state.}
Let \(t_i\) denote the close of hour \(i\). The main price variable is the
end-of-hour mid-quote
\[
    p_{t_i}=\tfrac12(b_{t_i}+a_{t_i}),
\]
where \(b_{t_i}\) and \(a_{t_i}\) are the best YES bid and ask at the close of
hour \(t_i\), normalized to dollars in \([0,1]\). The quoted spread is
\(s_{t_i}=a_{t_i}-b_{t_i}\), \(V_{t_i}\) is volume traded during the hour ending
at \(t_i\), and \(\tau_i=T-t_i\) is time to resolution. These are all
forecast-origin variables, observed before the next-hour horizon
\([t_i,t_{i+1})\). We use the mid-quote rather than the last trade because it
reflects the closing order book and is less exposed to bid--ask bounce or stale
prints.

\medskip
\paraheader{Forecast target.}
At the close of hour \(i\), after observing the order book at \(t_i\), we
forecast the signed one-hour innovation
\begin{equation}
    \varepsilon_i
    =
    p_{t_{i+1}}-p_{t_i},
    \qquad
    t_{i+1}=t_i+1 .
    \label{eq:eps_empirical}
\end{equation}
The conditional scale of this innovation is the empirical object modeled by the
structural variance \(h_i^2(K)\) in \eqref{eq:main-model}, with \(\Delta_i=1\)
on the hourly grid. The conditional variance itself is latent, so evaluation requires a proxy. The
primary proxy is the squared innovation \(\varepsilon_i^2\); as a second proxy,
Section~\ref{sec:estimation} also uses the intraday realized variance of the
mid-quote over the same hour \((t_i,t_{i+1}]\), built from one-minute quotes.

%% file: fc_moves/3_2_classes_intro.tex
We compare three broad classes of models, aiming to isolate what prediction-market structure adds. The first class contains structural predictors, including deadline resolution alone, the binary-market benchmark of \cite{archak2010modeling}, and DR-AS. The second contains plain ARCH/GARCH models and a heterogeneous autoregressive model of realized variance (HAR-RV), which serve as generic time-series benchmarks with no prediction-market state variables. These two benchmarks are the standard tools a practitioner would apply to any price series without knowledge of the contract. ARCH/GARCH is the canonical parametric model of conditional variance driven by past squared innovations; HAR-RV is the standard forecasting benchmark once intraday data are available, using realized variance over the previous hour, day, and week, and is typically hard to beat in equity and foreign-exchange markets. Both condition only on the contract's own price history, so the gap between them and the structural specifications measures what the prediction-market state variables \((p,\tau,s,V)\) add over the best generic time-series approach. The third combines a structural baseline with ARCH/GARCH dynamics in the multiplicative component form of \citet{engle2008spline} and \citet{engle2012forecasting}, in which the structural term sets the level of conditional variance and a unit-mean ARCH/GARCH component captures clustering around it, asking whether volatility clustering remains after conditioning on price, time to resolution, spread, and volume. Table~\ref{tab:spec-families} groups the main specification families by the economic question they address.

%% file: fc_moves/3_2_classes_details.tex
For the DR-AS family we compare four activity proxies, \(\nu(V)\in\{1,\log(1+V),\sqrt V,V\}\). The constant proxy isolates the spread channel, the concave proxies allow the marginal effect of activity to diminish with volume, and the linear proxy corresponds to volume being proportional to the arrival rate of information-sensitive trades.

For the activity-scaled deadline resolution family, we use $ a(V_i)=\nu(V_i)/\bar V  (\bar V>0)$ with \(\nu(V)\in\{V,\sqrt V,\log(1+V)\}\). This family asks whether trading activity primarily changes the speed of the deadline clock rather than adding an adverse-selection component. A constant-activity version would amount only to a fitted rescaling of the parameter-free deadline benchmark, so it is not treated as a separate activity-scaled deadline specification.

The plain dynamic benchmarks use lagged squared innovations but no prediction-market state variables,
\[
    h_i^2=\omega+\alpha\varepsilon_{i-1}^2
    \qquad\text{and}\qquad
    h_i^2=\omega+\alpha\varepsilon_{i-1}^2+\beta h_{i-1}^2 ,
\]
and the HAR-RV benchmark \citep{corsi2009simple} regresses the variance on realized variance over the previous hour, day, and week,
\[
    h_i^2 = b_0 + b_h\,\mathrm{RV}^{(h)}_i + b_d\,\mathrm{RV}^{(d)}_i + b_w\,\mathrm{RV}^{(w)}_i ,
\]
where \(\mathrm{RV}^{(h)}_i\) is the five-minute realized variance of the mid-quote over the hour ending at \(t_i\) and \(\mathrm{RV}^{(d)}_i\), \(\mathrm{RV}^{(w)}_i\) are its averages over the preceding 24 and 168 hours.

For the multiplicative component family, a structural baseline \(B_i\), such as the DR-AS variance \(h_i^2(K)\) in \eqref{eq:main-model}, is scaled by a unit-mean ARCH/GARCH component,
\[
    h_i^2 = c\,B_i\,g_i,
    \qquad
    g_i = (1-\alpha-\beta) + \alpha z_i^2 + \beta g_{i-1},
    \qquad
    z_i^2 = \frac{\varepsilon_{i-1}^2}{c\,B_{i-1}},
\]
with \(g_1=1\) at each contract's first observation and \(\beta=0\) for the ARCH(1) version. The standardized shock \(z_i\) is the previous hour's innovation divided by the previous hour's structural scale, so \(g_i\) rises above one after a surprisingly large move and decays back toward one at rate \(\alpha+\beta\). The structural term sets the state-dependent level of conditional variance, and the ARCH/GARCH component captures clustering around that level. Plain ARCH/GARCH benchmarks correspond to a constant baseline \(B_i\).

%% file: fc_moves/3_3_estimation.tex
Parameters are estimated separately for each test month using forecast-origin
volume weights. The same weights are used in evaluation, so the reported scores
emphasize contract-hours in which more risk is traded. DR-AS has a
single free parameter, the nonnegative order-flow scale \(K\), which we estimate
by one-dimensional volume-weighted Gaussian quasi-maximum likelihood for the
standardized innovation in \eqref{eq:innovation_equation}; the
deadline-resolution and Archak--Ipeirotis predictors have no free parameters.

In the multiplicative specifications, the parameters
\((c,\alpha,\beta)\), with \(c\ge0\), \(\alpha,\beta\ge0\), and
\(\alpha+\beta<1\), are estimated by volume-weighted quasi-maximum likelihood in
each test month; the scale \(K\) inside the structural baseline \(B_i\) is
calibrated first from the structural fit and held fixed.
Appendix~\ref{app:estimation_details} provides the quasi-likelihood objective
and the implementation details for the recursions and for the HAR-RV benchmark.

%% file: fc_moves/3_3_evaluation.tex
\paraheader{Prediction intervals with calibrated quantiles.}
The conditional variance is latent, so we evaluate a volatility forecast through
the prediction interval it implies for the next-hour price. From the
forecast-origin scale \(h_i\), we form a nominal 95\% interval around
\(p_{t_i}\) and clip it to the admissible range \([0,1]\):
\[
    L_i=\max\{0,\,p_{t_i}+\hat q_{0.025}\,h_i\},
    \qquad
    U_i=\min\{1,\,p_{t_i}+\hat q_{0.975}\,h_i\}.
\]
The multipliers \(\hat q_{0.025}\) and \(\hat q_{0.975}\) are each model's own
volume-weighted empirical \(2.5\%\) and \(97.5\%\) quantiles of the standardized
innovation \(u_i=\varepsilon_i/h_i\), computed on the out-of-sample forecasts of
all earlier test months. These quantiles minimize
the expected interval score for a given scale forecast
\citep{gneiting2007strictly}, and calibrating them for each model keeps the
comparison free of differences in tail shape.

Writing \(p_i^+=p_{t_{i+1}}=p_{t_i}+\varepsilon_i\) for the realized next-hour
price, we score \([L_i,U_i]\) with the Winkler interval score
\citep{winkler1972decision,gneiting2007strictly},
\begin{equation}
  \mathrm{IS}_i
  =
  (U_i-L_i)
  +
  \frac{2}{\eta}(L_i-p_i^+)\,\mathbb I\{p_i^+<L_i\}
  +
  \frac{2}{\eta}(p_i^+-U_i)\,\mathbb I\{p_i^+>U_i\},
  \qquad \eta=0.05 .
  \label{eq:winkler}
\end{equation}
Lower scores are better. The width term rewards sharpness, while the penalty
terms charge for intervals that miss the realized price move. The score
therefore penalizes both intervals that are too wide and intervals that are too
narrow to cover.

The headline interval metric is the volume-weighted interval score with
calibrated quantiles, reported alongside volume-weighted coverage,
\begin{equation}
    \mathrm{VW\text{-}IS\text{-}EQ}
    =
    \frac{\sum_i V_{t_i}\,\mathrm{IS}_i}{\sum_i V_{t_i}},
    \qquad
    \mathrm{VW\text{-}Cov\text{-}EQ}
    =
    \frac{\sum_i V_{t_i}\,\mathbb I\{L_i\le p_i^+\le U_i\}}{\sum_i V_{t_i}}.
    \label{eq:vwis}
\end{equation}
Coverage is a calibration diagnostic rather than the ranking metric, since a model can obtain high
coverage by making intervals too wide. A useful model posts a low interval score while keeping
coverage near the nominal 95

\paraheader{QLIKE.}
As a complementary loss on the variance forecast itself, we use QLIKE
\citep{patton2011volatility},
\begin{equation}
    \mathrm{QLIKE}_i
    =
    \log \tilde h_i^2 + \frac{\varepsilon_i^2}{\tilde h_i^2},
    \qquad
    \tilde h_i^2=\max\{h_i^2,\,\delta^2\},
    \label{eq:qlike}
\end{equation}
volume-weighted as in \eqref{eq:vwis} to give VW-QLIKE. The floor \(\delta\),
set to half a tick, keeps the loss finite when a forecast variance collapses
toward zero near the price boundaries. QLIKE is the loss implied by the
Gaussian quasi-likelihood. As a second proxy, we replace \(\varepsilon_i^2\) by the two-scale realized variance of the
mid-quote over \((t_i,t_{i+1}]\) \citep{zhang2005tale}, which averages
five-minute realized variances over offset subgrids and corrects them for
microstructure noise at the one-minute scale, giving VW-QLIKE-TSRV.

%% file: fc_moves/3_4_stylized.tex
Before fitting the multivariate specifications, we ask whether the mechanisms in
\eqref{eq:main-model} are visible directly in the data. We sort observations into
bins by each candidate driver and compute the volume-weighted root-mean-square
next-hour innovation (VW-RMS) within each bin. Appendix~\ref{app:binning_figures}
gives the exact binning statistic and additional figures.

\begin{figure}[!htbp]
\centering
\begin{subfigure}{0.48\textwidth}
\includegraphics[width=\linewidth]{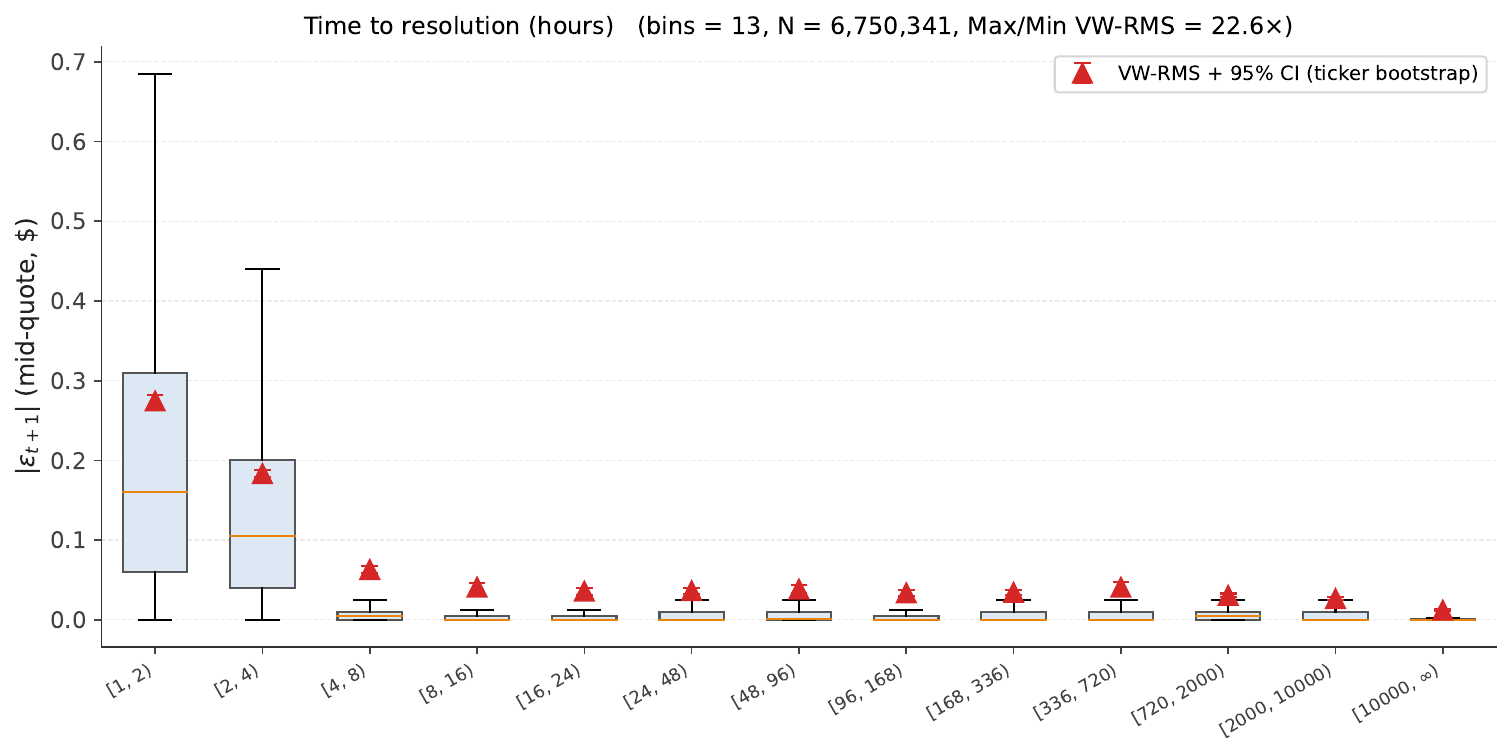}
\caption{Time to resolution in hours. Max/Min \(=22.6\times\).}
\end{subfigure}
\hfill
\begin{subfigure}{0.48\textwidth}
\includegraphics[width=\linewidth]{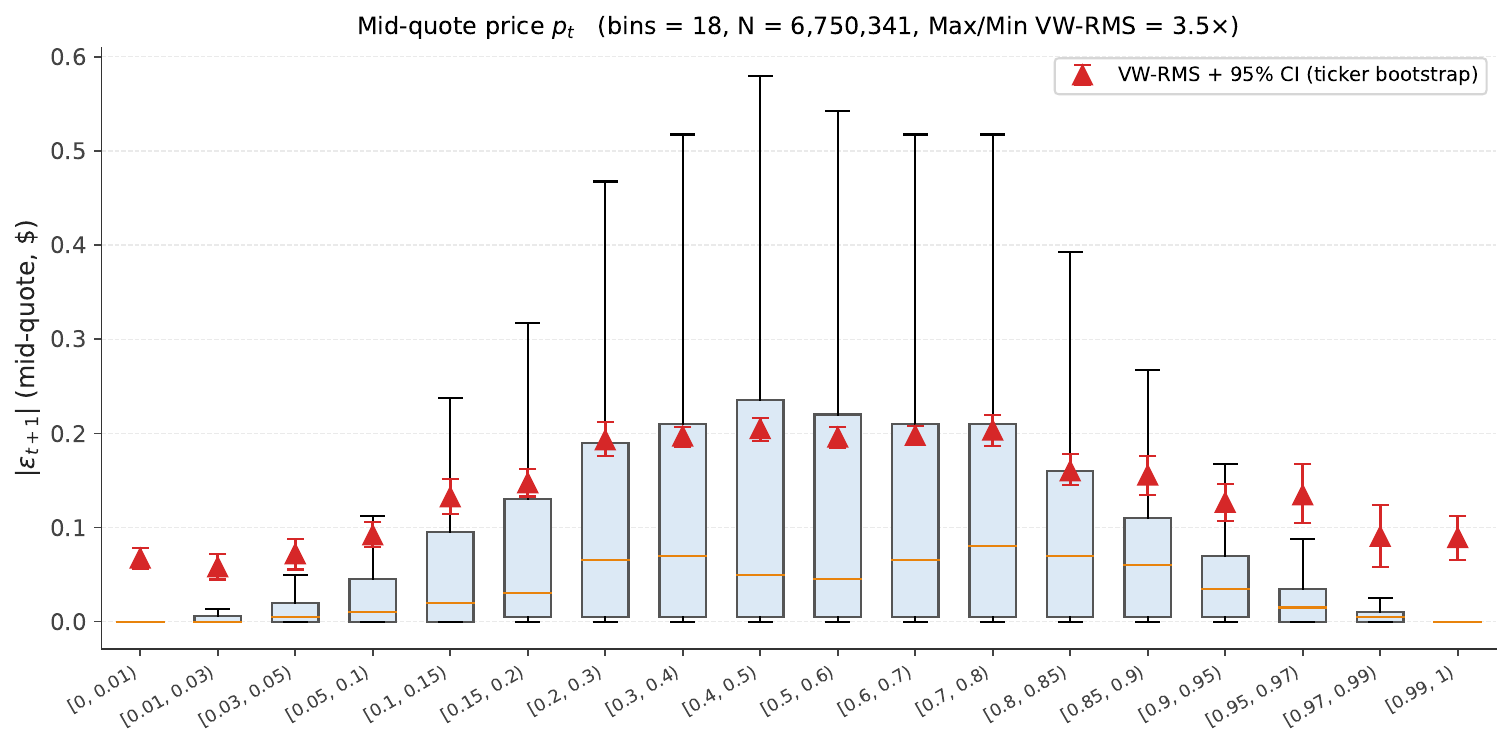}
\caption{Price \(p\). Max/Min \(=3.5\times\).}
\end{subfigure}
\caption{Model-free binning diagnostics for the two primitive inputs of the
deadline-resolution term. Boxes are volume-weighted quartiles of
\(|\varepsilon_i|\) within each bin; red markers are the volume-weighted RMS
next-hour innovation with 95\% contract-cluster bootstrap confidence
intervals.}
\label{fig:primitive-bins}
\end{figure}

\begin{table}[!htbp]
\centering
\caption{Model-free driver-strength ratios. Each ratio is the maximum bin
VW-RMS over the minimum. The table omits \(1/\sqrt{\tau}\), a monotone
re-expression of time to resolution rather than a separate primitive driver; the full
eight-driver table is in Appendix~\ref{app:binning_figures}.}
\label{tab:driver-summary}
\small
\begin{tabular}{clcl}
\toprule
Rank & Driver & Ratio & Interpretation \\
\midrule
1 & Time to resolution \(\tau\) & \(22.6\times\) & Deadline effect \\
2 & Time \% to resolution & \(9.0\times\) & Normalized time-to-resolution proxy \\
3 & Category & \(8.0\times\) & Proxy for duration / market type \\
4 & Volume \(V\) & \(7.5\times\) & Activity / information-flow proxy \\
5 & Open interest & \(3.8\times\) & Participation proxy \\
6 & Price \(p\) & \(3.5\times\) & Boundary shape in \(p(1-p)\) \\
7 & Spread \(s\) & \(2.2\times\) & Adverse-selection / liquidity proxy \\
\bottomrule
\end{tabular}
\end{table}

Table~\ref{tab:driver-summary} ranks the candidate drivers by the ratio of the
largest bin VW-RMS to the smallest. Time to resolution leads by a wide margin.
Volatility rises sharply in the final hours, as the \(1/\tau\) deadline clock
implies, but is roughly flat for contracts more than a day from resolution,
where a pure \(1/\tau\) clock would still predict a steady decline. There the
deadline term is small relative to the order-flow term. Along the price axis, realized volatility peaks near 50 cents and
falls toward the binary boundaries, matching the inverted-U
of \(p(1-p)\), as seen in Figure~\ref{fig:primitive-bins}, although the
boundary bins do not collapse to zero: near the
boundaries the mid-quote is still revised through the thin side of the book
even when little trading occurs. Category
also has a large univariate contrast, but as we argue \ifconf in Appendix~\ref{app:binning_figures}\else below\fi, it is a composite
proxy for market type and duration mix rather than a primitive state variable.

Among the microstructure variables, volume is a strong and monotone driver,
while spread is weak in isolation and not monotone. This does not contradict
the order-flow channel, because DR-AS uses volume and spread jointly through
\(\nu(V)s^2/4\). Section~\ref{sec:horserace} separates the deadline and
order-flow channels in the forecast comparison.

%% file: 4_primacy.tex
\expandafter\ifx\csname ifconf\endcsname\relax\expandafter\newif\csname ifconf\endcsname\fi
\section{Main Results}
\label{sec:horserace}

This section compares the out-of-sample forecasts from the model families in
Section~\ref{sec:specifications}. Table~\ref{tab:main-horserace} reports
VW-IS-EQ, VW-Cov-EQ, VW-QLIKE,\footnote{Differences in VW-QLIKE can be read as
likelihood ratios: QLIKE is twice the Gaussian negative log-likelihood up to a
constant, so a gap \(\Delta\) between two models corresponds to a
volume-weighted average per-hour quasi-likelihood ratio of
\(\exp(\Delta/2)\).} and VW-QLIKE-TSRV for the main specifications on the
mid-quote panel. On VW-IS-EQ the point estimates form three economic tiers:
the generic benchmarks perform worst, structural models perform substantially
better, and multiplicative structural-dynamic models perform best. On the two
QLIKE scores the tiers are less clean. DR-AS with \(\sqrt V\), \(\log(1+V)\),
or constant activity, and GARCH \(\times\) DR-AS, beat every generic
benchmark, but deadline resolution alone, the Archak--Ipeirotis benchmark, and
several of their multiplicative versions score no better than plain ARCH(1). The lowest
VW-IS-EQ, VW-QLIKE, and VW-QLIKE-TSRV are all attained by
GARCH \(\times\) DR-AS\,(\(\sqrt V\)). Because several specifications are close
in level, we use paired bootstrap comparisons\ifconf\else\ below\fi{} to assess whether the
ranking is statistically separated rather than a point-estimate artifact.

\ifconf\begin{table}[t]\else\begin{table}[htbp]\fi
\centering
\caption{Out-of-sample forecast comparison on the mid-quote panel, ranked by
VW-IS-EQ within each block. Results are pooled across 56 expanding-window test
months. Lower is better for VW-IS-EQ, VW-QLIKE, and VW-QLIKE-TSRV; VW-Cov-EQ is
the volume-weighted coverage of the calibrated 95\% interval. Pairwise
bootstrap comparisons for the ranking claims are reported in
Table~\ref{tab:pairwise-ranking}\ifconf{} in Appendix~\ref{app:pairwise}\fi.\ifconf{} DR denotes deadline resolution.\fi}
\label{tab:main-horserace}
\ifconf\footnotesize\else\small\fi
\setlength{\tabcolsep}{4pt}
\begin{tabular}{lcccc}
\toprule
\ifconf Specification & \makecell{VW-IS-\\EQ \(\downarrow\)} & \makecell{VW-Cov-\\EQ} & \makecell{VW-\\QLIKE \(\downarrow\)} & \makecell{VW-QLIKE-\\TSRV \(\downarrow\)} \\
\else Specification & VW-IS-EQ \(\downarrow\) & VW-Cov-EQ & VW-QLIKE \(\downarrow\) & VW-QLIKE-TSRV \(\downarrow\) \\
\fi
\midrule
\multicolumn{5}{l}{\textit{Multiplicative component dynamics}} \\
\textbf{GARCH \(\times\) DR-AS\,(\(\sqrt V\))}   & \textbf{0.5913} & 0.951 & \textbf{\(-4.232\)} & \textbf{\(-3.486\)} \\
\ifconf GARCH \(\times\) activity-scaled DR\,(\(\sqrt V\))\else GARCH \(\times\) activity-scaled deadline\,(\(\sqrt V\))\fi & 0.5955 & 0.956 & \(-3.493\) & \(-2.787\) \\
\ifconf GARCH \(\times\) DR\else GARCH \(\times\) deadline resolution\fi             & 0.6073 & 0.955 & \(-4.029\) & \(-3.057\) \\
ARCH \(\times\) DR-AS\,(\(\sqrt V\))             & 0.6172 & 0.939 & \(-4.131\) & \(-3.264\) \\
\ifconf ARCH \(\times\) DR\else ARCH \(\times\) deadline resolution\fi              & 0.6360 & 0.975 & \(-2.452\) & \(-1.686\) \\
GARCH \(\times\) Archak--Ipeirotis               & 0.6467 & 0.970 & \(-2.953\) & \(-1.695\) \\
ARCH \(\times\) Archak--Ipeirotis                & 0.6487 & 0.970 & \(-2.589\) & \(-1.191\) \\
\midrule
\multicolumn{5}{l}{\textit{Structural predictors}} \\
\textbf{DR-AS\,(\(\sqrt V\))}                    & \textbf{0.6227} & 0.951 & \textbf{\(-3.873\)} & \(-3.207\) \\
DR-AS\,(\(\log(1+V)\))                           & 0.6246 & 0.956 & \(-3.817\) & \textbf{\(-3.337\)} \\
DR-AS\,(constant activity)                       & 0.6259 & 0.958 & \(-3.763\) & \(-3.317\) \\
DR-AS\,(linear \(V\))                            & 0.6267 & 0.946 & \(-3.702\) & \(-2.663\) \\
\ifconf Activity-scaled DR\,(\(\log(1+V)\))\else Activity-scaled deadline\,(\(\log(1+V)\))\fi        & 0.6366 & 0.957 & \(-2.671\) & \(-1.372\) \\
\ifconf Activity-scaled DR\,(\(\sqrt V\))\else Activity-scaled deadline\,(\(\sqrt V\))\fi          & 0.6390 & 0.959 & \(-1.312\) & \(\phantom{-}0.188\) \\
\ifconf DR only\else Deadline resolution only\fi                         & 0.6457 & 0.968 & \(-2.416\) & \(\phantom{-}0.901\) \\
\ifconf Archak--Ipeirotis\else Archak--Ipeirotis benchmark\fi                      & 0.6795 & 0.969 & \(\phantom{-}1.464\) & \(\phantom{-}6.338\) \\
\midrule
\multicolumn{5}{l}{\textit{Generic benchmarks}} \\
GARCH(1,1) plain                                 & 0.7358 & 0.952 & \(-2.407\) & \(-2.356\) \\
ARCH(1) plain                                    & 0.7570 & 0.945 & \(-2.638\) & \(-2.561\) \\
HAR-RV                                           & 0.6872 & 0.939 & \(-3.079\) & \(-2.993\) \\
\bottomrule
\end{tabular}
\end{table}

\ifconf
Appendix~\ref{app:pairwise} tests these ranking claims with paired
contract-cluster bootstrap comparisons and a model confidence set
\citep{hansen2011model}. Every difference that this section describes as
significant is significant under these tests.
\else
\input{fc_moves/4_0_pairwise}\fi

\subsection{Where the forecast gains come from}
\label{sec:three-comparisons}

With the main ranking and its paired uncertainty established, we read Table~\ref{tab:main-horserace} through three
comparisons: generic dynamics versus structural state variables, the two
ingredients inside the structural model, and multiplicative dynamics after
conditioning on structure.

\paraheader{Structural variables versus generic dynamics.}
The central comparison is between plain GARCH(1,1) and the structural DR-AS
model. Plain GARCH(1,1) attains VW-IS-EQ \(=0.736\), while DR-AS\,(\(\sqrt V\))
attains \(0.623\), a \(15\%\) improvement. On VW-QLIKE the gap is \(1.5\), which means
DR-AS assigns \(2.1\) times the per-hour likelihood of plain GARCH to the
realized price changes. The gap is \(0.88\) on VW-QLIKE-TSRV, which scores the
forecasts against the intraday realized variance rather than the squared price
change; all three differences are significant, so the ranking does not depend
on the variance proxy. The same holds against HAR-RV, the stronger of the two
generic benchmarks: DR-AS improves on it by \(0.065\) on VW-IS-EQ and assigns
\(1.5\) times its per-hour likelihood to the realized price changes. The generic
benchmarks sit far outside the structural tiers in
Table~\ref{tab:main-horserace}, so this is not a marginal ranking. This gain comes from the forecast-origin state
variables \((p,\tau,s,V)\), not from a richer time-series recursion. The bulk
of the forecasting improvement therefore comes from contemporaneous
prediction-market state variables rather than from return persistence alone.

\paraheader{The two structural ingredients.}
Within the structural family, the progression from the Archak--Ipeirotis
benchmark to deadline resolution to DR-AS separates two design choices along
this ordering. The Archak--Ipeirotis benchmark uses the probit-Brownian shape
\(\varphi^2(\Phi^{-1}(p))\); deadline resolution replaces it with the
Wright--Fisher/Bernoulli shape \(p(1-p)\); and DR-AS adds the order-flow term
\(K\nu(V)s^2/4\).

Replacing the Archak--Ipeirotis price shape with \(p(1-p)\) improves VW-IS-EQ
from \(0.680\) to \(0.646\), a \(5\%\) reduction, and raises the per-hour
likelihood by a factor of \(7\). Both shapes vanish at the binary boundaries,
but \(p(1-p)\) is the variance of the binary payoff itself, whereas the
probit-Brownian shape lets the variance collapse much faster near the
boundaries, where the realized price still moves. Empirically, the simple
binary-outcome shape fits the heterogeneous Kalshi panel better than the
probit-Brownian shape, and most of the likelihood gain comes from the
boundary regions.

Adding the order-flow term further improves VW-IS-EQ from \(0.646\) to
\(0.623\), a \(4\%\) reduction, with a likelihood ratio of \(2.1\). This term
enters through the squared half-spread and trading activity: the spread
measures the adverse-selection price-impact scale, while volume proxies for
the frequency of information-sensitive order flow. Thus the deadline channel
captures smooth calendar-time resolution of binary uncertainty, and the
order-flow channel captures additional variance from information entering
through trading. The price shape is the larger of the two steps on every
score; the order-flow term adds a smaller but significant further gain.
Taken together, DR-AS improves on the Archak--Ipeirotis benchmark by \(8\%\)
on the interval score and assigns \(14\) times the per-hour likelihood of that benchmark to the realized price changes.

\begin{figure}[htbp]
\centering
\begin{subfigure}{0.49\textwidth}
\includegraphics[width=\linewidth]{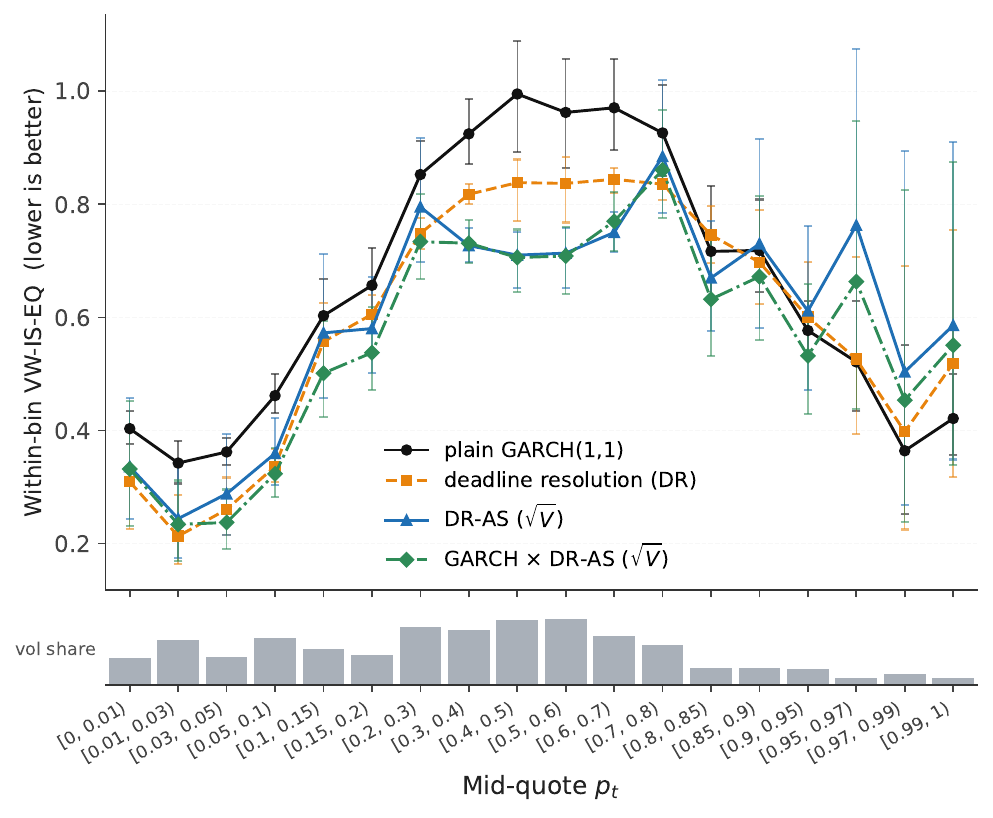}
\caption{By price \(p_t\).}
\end{subfigure}
\hfill
\begin{subfigure}{0.49\textwidth}
\includegraphics[width=\linewidth]{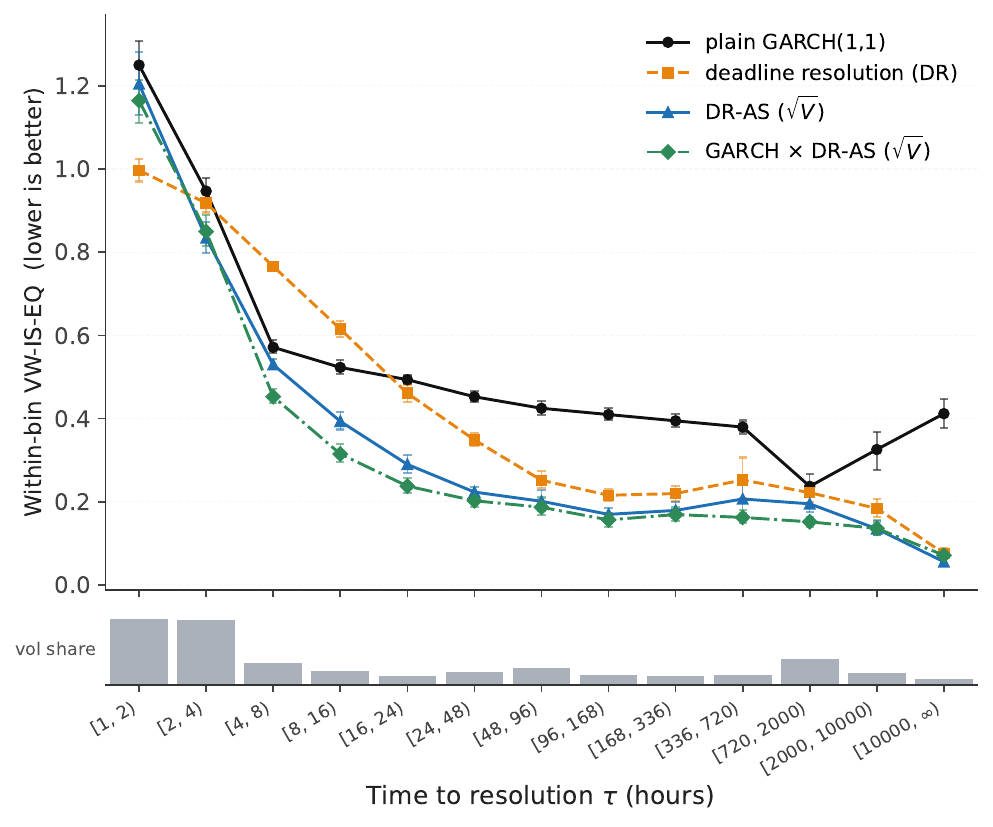}
\caption{By time to resolution.}
\end{subfigure}
\caption{\ifconf Within-bin VW-IS-EQ (lower is better) by price and time to resolution,
with 95\% contract-cluster bootstrap intervals. Bars below each panel show the
volume share of each bin.\else Within-bin out-of-sample VW-IS-EQ (lower is better) for plain GARCH,
deadline resolution alone, DR-AS, and GARCH \(\times\) DR-AS, against price
(left) and time to resolution (right). Error bars are 95\% contract-cluster
bootstrap confidence intervals; the bars below each panel show the volume
share of each bin.\fi}
\label{fig:advantage-bins}
\end{figure}

\paraheader{State-space patterns.}
Figure~\ref{fig:advantage-bins} shows the forecast-score analogue of the
model-free patterns in Section~\ref{sec:stylized}. The gap between plain GARCH
and the structural curves shows where binary-market structure helps; the gap
between deadline resolution and DR-AS shows where the order-flow term adds
information; and the gap between DR-AS and GARCH \(\times\) DR-AS shows where
the dynamic layer adds.

The price panel shows the clearest pattern. The structural advantage over plain
GARCH is largest near \(p=1/2\), where binary uncertainty is highest, and shrinks
toward the boundaries, where little uncertainty remains to be resolved and
plain GARCH becomes competitive. This is consistent with the \(p(1-p)\) shape
of the deadline-resolution variance.

The time-to-resolution panel separates the two structural components. In the
final two hours before resolution, which carry a quarter of all volume,
deadline resolution alone is the best of the four specifications. Adding the
order-flow term makes the forecast worse there, so the term is not negligible
in these hours; it adds variance on top of a deadline term that already
accounts for the resolution risk. Beyond a few
hours the picture reverses. DR-AS improves on deadline resolution at every
horizon from four hours on, and the gap is largest between four hours and four
days, where deadline resolution alone falls behind even plain GARCH. There the
deadline force is weak and the order-flow term accounts for most of the
forecast variance; unlike the deadline term, it does not scale with \(1/\tau\),
so spreads and activity matter throughout the contract's life. The GARCH
multiplier adds a roughly uniform further improvement beyond the final hours.

\paraheader{Multiplicative dynamics.}
Scaling DR-AS by a GARCH(1,1) multiplier lowers VW-IS-EQ from \(0.623\) to
\(0.591\), a \(5\%\) reduction, and raises the per-hour likelihood by a factor
of \(1.18\); both differences are significant. Residual time-series
predictability remains, but it is secondary. Relative to plain GARCH,
GARCH \(\times\) DR-AS improves VW-IS-EQ by \(20\%\) and assigns \(2.5\) times the per-hour
likelihood of plain GARCH to the realized price changes. The structural baseline accounts for most of that gain, with
the GARCH layer providing the remainder. GARCH is therefore complementary to
the structural state variables, not a substitute for them. Every difference
reported in this subsection is also significant on VW-QLIKE-TSRV; the levels
are in Table~\ref{tab:main-horserace}.

\ifconf
Appendix~\ref{app:activity} compares the four activity proxies within DR-AS.
None differs significantly from \(\sqrt V\) on the interval score, concave
proxies dominate linear scaling on the likelihood-based scores, and we keep
\(\sqrt V\) as the headline proxy because it is significantly better on
VW-QLIKE.
\else
\input{fc_moves/4_2_activity}\fi

\subsection{Independence of interval violations}
\label{sec:conditional-coverage}

A good interval forecast must do more than cover the realized price 95\% of
the time. In the conditional-coverage framework of
\citet{christoffersen1998evaluating}, violations, the hours in which the
realized price falls outside the interval, should also be independent over
time. If violations arrive in clusters, the forecast has the average size of
price moves right but fails to raise the interval when volatility picks up and
lower it when volatility subsides. A structural model without a time-series
recursion is exposed to exactly this failure.

\ifconf
We measure independence with \(C_{\mathrm{rate}}\), the excess number of
consecutive violation pairs per transition within each contract relative to a
random placement of that contract's violations. It is positive under
clustering and negative under anti-clustering, and, unlike Christoffersen's
pooled likelihood-ratio test, it separates serial dependence from differences
in violation rates across contracts. Appendix~\ref{app:conditional_coverage}
gives its definition, its variance, and the pooled Christoffersen statistics;
confidence intervals are event-cluster bootstrap intervals, because contracts
on the same underlying event may share shocks.
\else
\input{fc_moves/4_3_crate_statistic}\fi

\ifconf\begin{figure}[t]\else\begin{figure}[htbp]\fi
\centering
\ifconf\vspace{-6pt}\includegraphics[width=0.62\linewidth]{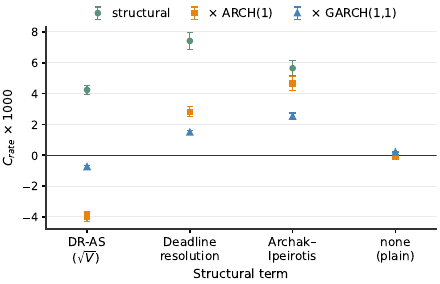}\vspace{-6pt}\else\includegraphics[width=0.75\linewidth]{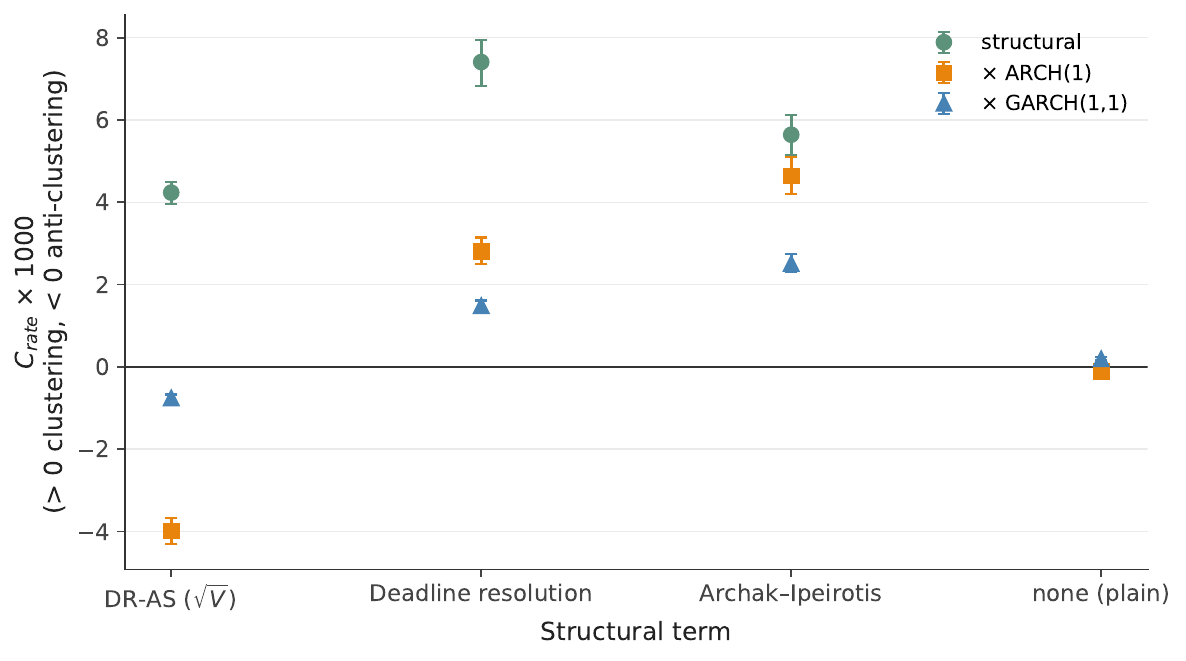}\fi
\caption{\ifconf Clustering of violations of the calibrated 95\% intervals,
\(C_{\mathrm{rate}}\times1000\); positive values indicate clustering. Error bars
are 95\% event-cluster bootstrap intervals.\else Violation clustering on the calibrated 95\% intervals.
\(C_{\mathrm{rate}}\times1000\) for each structural term alone and scaled by
the ARCH(1) and GARCH(1,1) multipliers, and for the plain dynamic benchmarks.
Positive values indicate clustering, negative values anti-clustering. Error
bars are 95\% event-cluster bootstrap intervals.\fi}
\label{fig:christoffersen}
\end{figure}

Figure~\ref{fig:christoffersen} reports \(C_{\mathrm{rate}}\) for each
structural term alone and scaled by the ARCH and GARCH multipliers. The
structural DR-AS forecast leaves clear clustering,
\(C_{\mathrm{rate}}=+0.0042\). The GARCH multiplier reduces it to
\(-0.0008\), about one-fifth of the structural value in magnitude, with
hour-level coverage of 95.5\%. This residual is slightly negative and
statistically different from zero, so the multiplier leaves a small amount of
anti-clustering rather than exact independence. The ARCH multiplier overshoots
into anti-clustering of similar magnitude to the original clustering,
\(-0.0040\). GARCH \(\times\) deadline
resolution (\(+0.0015\)) and GARCH \(\times\) Archak--Ipeirotis (\(+0.0025\))
remain significantly clustered. Plain GARCH has a \(C_{\mathrm{rate}}\) near
zero only because its intervals cover 99\% of hours.

The structural term sets the level of conditional variance, and the GARCH
multiplier helps to fix the timing of departures from it. Only the combination
of DR-AS and the GARCH multiplier comes close to independence. With deadline
resolution or the Archak--Ipeirotis shape as the structural term, the same
multiplier reduces the clustering but leaves it significantly positive.

%% file: fc_moves/4_0_pairwise.tex
\paraheader{Pairwise ranking evidence.}
The entries in Table~\ref{tab:main-horserace} are pooled point estimates. To assess the ranking claims, we compute paired bootstrap confidence intervals for score differences. For two specifications \(A\) and \(B\) and a score \(S\), define
\[
    \Delta_{B-A}
    =
    S(B)-S(A),
    \qquad S\in\{\mathrm{VW\text{-}IS\text{-}EQ},\ \mathrm{VW\text{-}QLIKE}\},
\]
computed on the common evaluation support. Positive values favor specification \(A\). The bootstrap resamples contract-level clusters with replacement and recomputes both scores on the same resampled contracts in each draw. This paired design uses the fact that competing specifications are evaluated on the same contract-hours, so much of the sampling variation is common across specifications.

Table~\ref{tab:pairwise-ranking} reports the comparisons that determine the headline ranking within the multiplicative family (Panel A), the structural interpretation of the structural model (Panel B), and the three-layer structure across families (Panel C). The purpose is not to attach economic meaning to every lower-tier ordering, but to test whether the main ranking claims are separated from sampling variation.

\begin{table}[htbp]
\centering
\caption{Pairwise bootstrap comparisons that determine the main ranking claims
on the mid-quote panel. The reported difference is
\(\Delta_{B-A}=S(B)-S(A)\), so positive values favor specification \(A\).
Confidence intervals are paired contract-cluster bootstrap intervals.}
\label{tab:pairwise-ranking}
\footnotesize
\setlength{\tabcolsep}{3pt}
\begin{tabular}{>{\raggedright\arraybackslash}p{0.27\linewidth}>{\raggedright\arraybackslash}p{0.27\linewidth}cc}
\toprule
Specification \(A\) & Specification \(B\)
& \(\Delta_{B-A}\) VW-IS-EQ [95\% CI] & \(\Delta_{B-A}\) VW-QLIKE [95\% CI] \\
\midrule
\multicolumn{4}{l}{\textit{Panel A. Multiplicative ranking}} \\
GARCH\(\times\)DR\mbox{-}AS\,(\(\sqrt V\)) & ARCH\(\times\)DR\mbox{-}AS\,(\(\sqrt V\))
& \(0.026\) [\(0.020,\;0.032\)] & \(0.10\) [\(0.06,\;0.14\)] \\
GARCH\(\times\)DR\mbox{-}AS\,(\(\sqrt V\)) & GARCH\(\times\)deadline resolution
& \(0.016\) [\(0.006,\;0.026\)] & \(0.20\) [\(0.16,\;0.26\)] \\
GARCH\(\times\)DR\mbox{-}AS\,(\(\sqrt V\)) & GARCH\(\times\)Archak--Ipeirotis
& \(0.055\) [\(0.043,\;0.067\)] & \(1.28\) [\(1.14,\;1.45\)] \\
GARCH\(\times\)DR\mbox{-}AS\,(\(\sqrt V\)) & ARCH\(\times\)Archak--Ipeirotis
& \(0.057\) [\(0.045,\;0.070\)] & \(1.64\) [\(1.52,\;1.78\)] \\
\midrule
\multicolumn{4}{l}{\textit{Panel B. Structural core}} \\
DR\mbox{-}AS\,(\(\sqrt V\)) & Deadline resolution only
& \(0.023\) [\(0.006,\;0.038\)] & \(1.46\) [\(1.23,\;1.73\)] \\
Deadline resolution only & Archak--Ipeirotis benchmark
& \(0.034\) [\(0.030,\;0.037\)] & \(3.88\) [\(3.11,\;4.82\)] \\
DR\mbox{-}AS\,(\(\sqrt V\)) & Archak--Ipeirotis benchmark
& \(0.057\) [\(0.039,\;0.072\)] & \(5.34\) [\(4.49,\;6.35\)] \\
\midrule
\multicolumn{4}{l}{\textit{Panel C. Across families}} \\
DR\mbox{-}AS\,(\(\sqrt V\)) & GARCH(1,1) plain
& \(0.119\) [\(0.097,\;0.140\)] & \(1.49\) [\(1.27,\;1.75\)] \\
DR\mbox{-}AS\,(\(\sqrt V\)) & HAR-RV
& \(0.065\) [\(0.045,\;0.083\)] & \(0.79\) [\(0.68,\;0.92\)] \\
GARCH\(\times\)DR\mbox{-}AS\,(\(\sqrt V\)) & DR\mbox{-}AS\,(\(\sqrt V\))
& \(0.025\) [\(0.013,\;0.040\)] & \(0.33\) [\(0.28,\;0.38\)] \\
ARCH\(\times\)DR\mbox{-}AS\,(\(\sqrt V\)) & DR\mbox{-}AS\,(\(\sqrt V\))
& \(-0.001\) [\(-0.013,\;0.014\)] & \(0.23\) [\(0.19,\;0.27\)] \\
GARCH\(\times\)DR\mbox{-}AS\,(\(\sqrt V\)) & GARCH(1,1) plain
& \(0.145\) [\(0.129,\;0.159\)] & \(1.82\) [\(1.61,\;2.08\)] \\
GARCH\(\times\)DR\mbox{-}AS\,(\(\sqrt V\)) & HAR-RV
& \(0.091\) [\(0.077,\;0.105\)] & \(1.13\) [\(1.01,\;1.25\)] \\
GARCH\(\times\)DR\mbox{-}AS\,(\(\sqrt V\)) & Archak--Ipeirotis benchmark
& \(0.084\) [\(0.071,\;0.096\)] & \(5.70\) [\(4.84,\;6.73\)] \\
\bottomrule
\end{tabular}
\begin{flushleft}
\footnotesize
Notes: Differences are computed on the common evaluation support for each pair.
They need not equal the difference between rounded entries in
Table~\ref{tab:main-horserace}. Panel C pairs a structural specification with a
dynamic one; the structural specification scores the first hour of each
contract, which the dynamic recursions skip, so these pairs are matched by
contract rather than by hour.
\end{flushleft}
\end{table}

Panel A shows that GARCH \(\times\) DR-AS\,(\(\sqrt V\)) is not merely the lowest point estimate. It is statistically separated on both scores from the same DR-AS baseline with ARCH dynamics, from GARCH scaling the deadline-resolution baseline, and from GARCH or ARCH scaling the Archak--Ipeirotis benchmark. A model confidence set \citep{hansen2011model} gives the same reading. GARCH \(\times\) DR-AS\,(\(\sqrt V\)) is the only member of the 95\% set on VW-QLIKE and belongs to the set on VW-IS-EQ, which excludes every Archak-based and every plain dynamic model.

Panel B shows the structural pattern in the structural horse race. First, adding the spread-flow term to deadline resolution is not a point-estimate artifact: DR-AS\,(\(\sqrt V\)) significantly improves on deadline resolution alone on both scores. Second, the deadline-resolution shape improves on the Archak--Ipeirotis benchmark, and the full DR-AS specification is separated even more strongly from Archak--Ipeirotis. We therefore treat DR-AS\,(\(\sqrt V\)) as the stable structural core; the choice of activity proxy within DR-AS is examined in \ifconf Appendix\else Section\fi~\ref{sec:horserace-volume}.

Panel C reports the cross-family comparisons behind the three-layer reading in Section~\ref{sec:three-comparisons}. Each layer is statistically separated from the one below on both scores, with one instructive exception. Scaling DR-AS by an ARCH(1) multiplier produces no statistically detectable difference in the interval score and improves VW-QLIKE modestly but significantly. One interpretation is that the structural core already captures most of what a one-lag reaction to the last innovation can add. It is the GARCH multiplier, with its persistent component, that adds a further significant gain on both scores.

%% file: fc_moves/4_2_activity.tex
\ifconf\else\subsection{Activity scaling within DR-AS}\fi
\label{sec:horserace-volume}

The remaining modeling choice within the DR-AS form is how measured activity
enters the order-flow channel. Table~\ref{tab:main-horserace} reports the levels for the
four activity proxies used in the structural DR-AS specifications, and
Table~\ref{tab:volume-sensitivity} the paired bootstrap differences relative to \(\sqrt V\).

\begin{table}[htbp]
\centering
\caption{Activity scaling within the structural DR-AS family: paired
contract-cluster bootstrap differences relative to \(\sqrt V\),
\(\Delta = S(\nu)-S(\sqrt V)\), so positive values favor \(\sqrt V\). Levels
are reported in Table~\ref{tab:main-horserace}.}
\label{tab:volume-sensitivity}
\footnotesize
\setlength{\tabcolsep}{6pt}
\begin{tabular}{lccc}
\toprule
\(\nu(V)\) proxy & \(\Delta\) VW-IS-EQ [95\% CI] & \(\Delta\) VW-QLIKE [95\% CI] & \(\Delta\) VW-QLIKE-TSRV [95\% CI] \\
\midrule
\(\log(1+V)\)    & \(0.002\) [\(-0.003,\;0.007\)] & \(0.06\) [\(0.02,\;0.11\)] & \(-0.13\) [\(-0.19,\;-0.07\)] \\
constant         & \(0.003\) [\(-0.002,\;0.009\)] & \(0.11\) [\(0.05,\;0.20\)] & \(-0.11\) [\(-0.18,\;-0.03\)] \\
\(V\)            & \(0.004\) [\(-0.000,\;0.008\)] & \(0.17\) [\(0.11,\;0.25\)] & \(\phantom{-}0.55\) [\(0.47,\;0.63\)] \\
\bottomrule
\end{tabular}
\end{table}

On the interval score, none of the three alternatives differs significantly
from \(\sqrt V\); the differences are at most \(0.004\). The scores that
evaluate the variance level separate them. On
VW-QLIKE, \(\sqrt V\) is significantly better than each alternative; on the
realized-variance target, \(\log(1+V)\) and constant activity are
significantly better than \(\sqrt V\), and linear \(V\) is the worst proxy on
both. Concave activity proxies therefore dominate linear scaling, while the
ranking among the concave proxies depends on the variance proxy. We keep
\(\sqrt V\) as the headline proxy because it is significantly better on
VW-QLIKE and no worse on the interval score, at the cost of a significant
deficit on the realized-variance target.

%% file: fc_moves/4_3_crate_statistic.tex
We test independence with a statistic adapted to a panel of many short
contract series. Let \(I_{it}\in\{0,1\}\) indicate a violation of contract
\(i\)'s interval at hour \(t\), let \(\mathcal E_i\) be the set of pairs of
consecutive observed hours of contract \(i\), \(L_i=|\mathcal E_i|\), and
\(n_{11,i}=\sum_{(u,v)\in\mathcal E_i} I_{iu}I_{iv}\) the number of
consecutive violation pairs. Conditional on the \(K_i\) violations of contract
\(i\) over its \(N_i\) observed hours, random placement of those violations
yields \(\mu_i=L_i\,K_i(K_i-1)/[N_i(N_i-1)]\) such pairs, and
\begin{equation}
    C_{\mathrm{rate}}
    =
    \frac{\sum_i \left(n_{11,i}-\mu_i\right)}{\sum_i L_i}
    \label{eq:crate}
\end{equation}
is the excess number of consecutive violation pairs per transition, positive
under clustering and negative under anti-clustering. Pooling the transitions
of all contracts into Christoffersen's likelihood-ratio test would confound
within-contract dependence with differences in violation rates across
contracts. A violation in the current hour mostly identifies a contract that
violates often, which raises the pooled probability of a violation in the
next hour whether or not violations are serially dependent. For plain GARCH,
for example, the pooled probability of a violation is \(0.007\) after a
non-violation but \(0.074\) after a violation, yet \(C_{\mathrm{rate}}\) is
essentially zero. \ifconf The rest of this appendix gives\else Appendix~\ref{app:conditional_coverage} gives\fi{} the
derivation and the exact variance of \eqref{eq:crate}, and also reports the
pooled Christoffersen statistics for the same specifications. Confidence intervals
are event-cluster bootstrap intervals, because contracts on the same
underlying event may share shocks.

%% file: 5_universality.tex
\expandafter\ifx\csname ifconf\endcsname\relax\expandafter\newif\csname ifconf\endcsname\fi
\section{Category Heterogeneity and Model Portability}
\label{sec:universality}
\ifconf
The pooled comparison in Section~\ref{sec:horserace} treats the Kalshi panel as
one forecasting target. This section asks how forecast performance differs
across Kalshi categories. We use the 11 categories with at least 400 contracts,
which together account for \(98\%\) of contract-hours and \(99.6\%\) of volume.
Two results stand out. Forecast difficulty differs sharply across categories,
and Sports is the hardest (Figure~\ref{fig:category-4specs}). Within Sports, the
difficulty comes from single-game markets, where most of the information is
revealed during the game. The model family itself travels across categories.
The promoted specifications win or tie in most major categories, and
re-estimating them within a category does not improve their forecasts.
Appendix~\ref{app:category} reports the sample composition by category, the
per-category paired comparisons, and the per-category refitting.
\else

The pooled comparison in Section~\ref{sec:horserace} treats the Kalshi panel as
one forecasting target. This section asks whether that result travels across
categories, or instead reflects the large, liquid categories that dominate the
volume-weighted criterion. The answer has three parts. The promoted
specifications win or tie in most major categories, re-estimating them within
a category does not improve their forecasts, and categories differ sharply in
difficulty and in the concentration of large price moves.
\fi

\ifconf
\else
\input{fc_moves/5_1_2_composition_portability}\fi

\ifconf
\begin{figure}[t]
\centering
\begin{subfigure}[t]{0.49\textwidth}
\centering
\includegraphics[width=\linewidth]{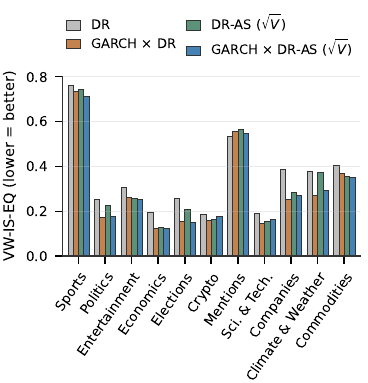}
\caption{VW-IS-EQ of four specifications.}
\label{fig:category-4specs}
\end{subfigure}
\hfill
\begin{subfigure}[t]{0.49\textwidth}
\centering
\includegraphics[width=\linewidth]{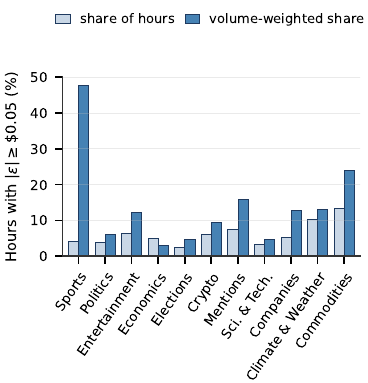}
\caption{Share of hours with \(|\varepsilon_i|\ge \$0.05\).}
\label{fig:category-regimes}
\end{subfigure}
\caption{Forecast difficulty and large price moves across the 11 major
categories, ordered by volume.}
\end{figure}
\else
\begin{figure}[htbp]
\centering
\includegraphics[width=0.75\linewidth]{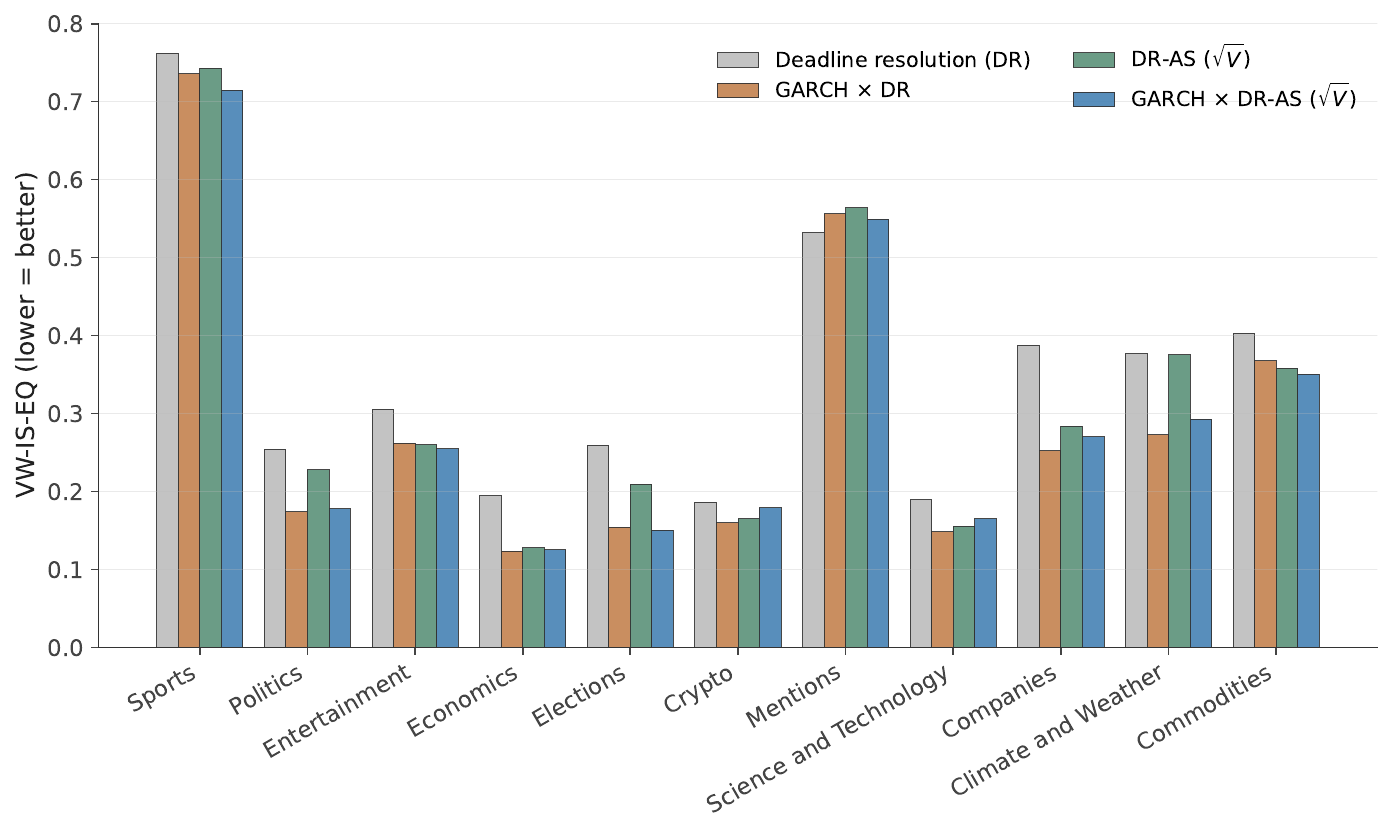}
\caption{Within-category VW-IS-EQ (lower is better) under the global fit for
four representative specifications, deadline resolution alone,
GARCH \(\times\) deadline resolution, DR-AS, and GARCH \(\times\) DR-AS, across
the 11 major categories, ordered by volume.}
\label{fig:category-4specs}
\end{figure}
\fi

\ifconf
\else
\input{fc_moves/5_2_race_text}\fi

\ifconf
\else
\input{fc_moves/5_3_catfit}\fi

\subsection{Smooth resolution and event-concentrated risk}
\label{sec:cat-information}

\ifconf
Forecast difficulty differs sharply across categories. The best global-fit
VW-IS-EQ ranges from \(0.12\) in Economics to \(0.71\) in Sports, a roughly
sixfold difference (Figure~\ref{fig:category-4specs}), and
Figure~\ref{fig:category-regimes} ties this spread to how information arrives.
Economics contracts resolve around scheduled macroeconomic releases, and
Economics is the only major category whose volume-weighted share of
large-move hours, \(3.0\%\), is below its share of hours, \(4.9\%\). In Sports,
only \(4.1\%\) of hours are large-move hours, but they carry \(47.8\%\) of
volume, so nearly half of the volume-weighted evaluation falls on jump-like
hours driven by discrete game events. The smooth conditional-variance
approximation therefore fits smooth-resolution categories best and is least
complete in event-concentrated markets, where an event clock or jump component for public information
would be a natural extension. Appendix~\ref{app:category} gives the full
discussion.
\else
\input{fc_moves/5_4_regimes}\fi

\ifconf
\else
\begin{figure}[!htbp]
\centering
\includegraphics[width=0.72\linewidth]{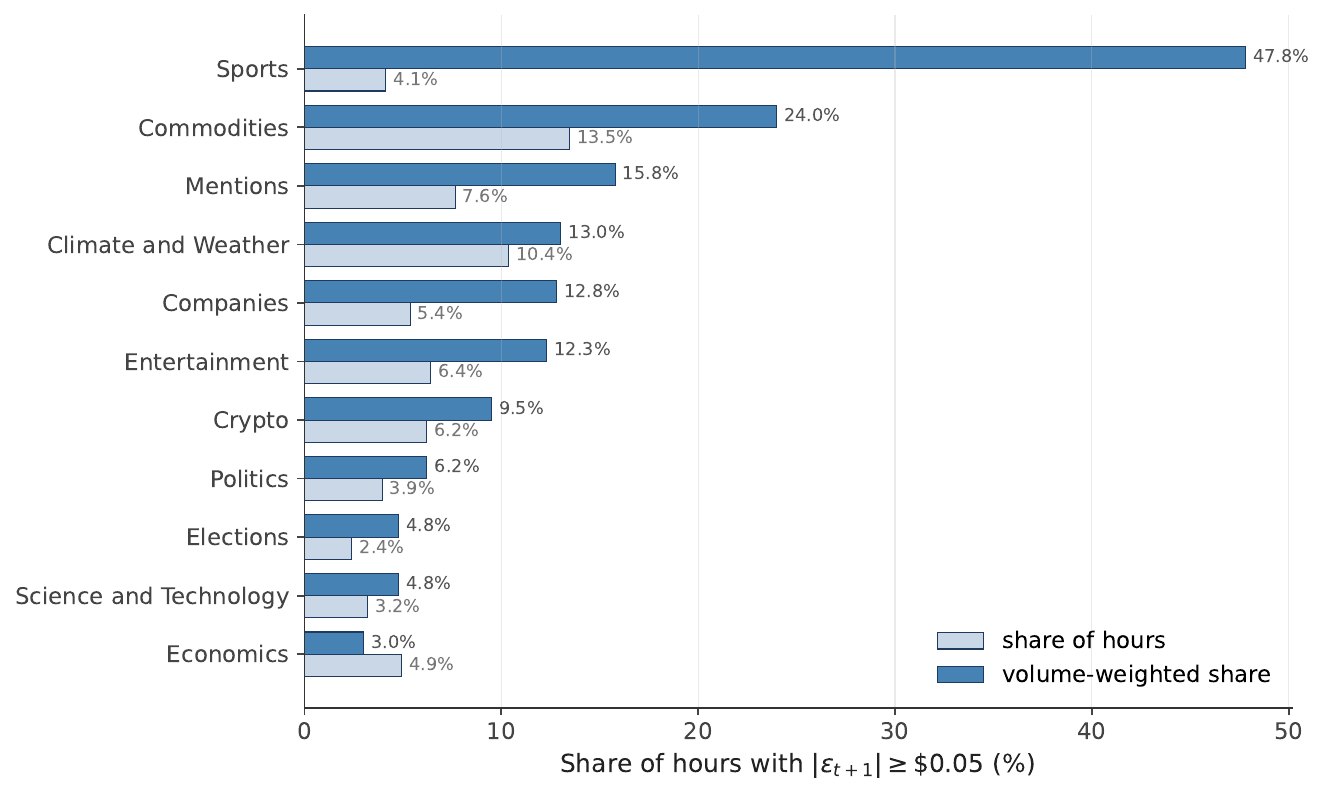}
\caption{Share of hourly observations with \(|\varepsilon_i|\ge \$0.05\) across
the 11 major categories; light bars are shares of hours and dark bars are
volume-weighted shares.}
\label{fig:category-regimes}
\end{figure}
\fi

\subsection{Sports: single-game and other markets}
\label{sec:sports-single-game}

Single-game markets are contracts whose outcomes are determined by one game,
match, fight, or race, typically lasting one to four hours, including player
props, half- or quarter-game markets, and single-game parlays. Other Sports markets cover
single-day sessions or longer events such as championships, series, and season
awards. Single-game prices move in bursts. A typical single-game contract
makes \(54\%\) of its total mid-quote movement in its three largest hourly
moves, compared with \(31\%\) for other Sports contracts.
We label each Sports series as single-game or other by hand, keep the global
fits, and split only the test observations.

\begin{table}[htbp]
\ifconf\vspace{-14pt}\fi
\centering
\caption{Single-game and other Sports markets under the global fit. Panel A
reports sample composition. Panel B reports VW-IS-EQ and VW-QLIKE; the
last two rows report the paired contract-cluster bootstrap difference DR-AS
minus GARCH \(\times\) DR-AS, so positive values favor the GARCH layer, and
its 95\% interval.}
\label{tab:sports-single-game}
\ifconf\renewcommand{\arraystretch}{0.85}\fi
\footnotesize
\setlength{\tabcolsep}{3pt}
\resizebox{\linewidth}{!}{%
\begin{tabular}{lcccccc}
\toprule
\textit{Panel A. Sample} & \multicolumn{2}{c}{All} & \multicolumn{2}{c}{Single-game} & \multicolumn{2}{c}{Other} \\
\midrule
Contracts              & \multicolumn{2}{c}{20{,}901} & \multicolumn{2}{c}{12{,}809} & \multicolumn{2}{c}{8{,}092} \\
Share of hours         & \multicolumn{2}{c}{100\%}    & \multicolumn{2}{c}{42.3\%}   & \multicolumn{2}{c}{57.7\%} \\
Share of volume        & \multicolumn{2}{c}{100\%}    & \multicolumn{2}{c}{76.5\%}   & \multicolumn{2}{c}{23.5\%} \\
\midrule
\textit{Panel B. Forecast accuracy} & IS-EQ & QLIKE & IS-EQ & QLIKE & IS-EQ & QLIKE \\
\midrule
GARCH\(\times\)DR-AS   & 0.714 & \(-3.462\) & 0.838 & \(-2.676\) & 0.322 & \(-5.961\) \\
GARCH\(\times\)DR      & 0.736 & \(-3.184\) & 0.869 & \(-2.434\) & 0.313 & \(-5.570\) \\
\cmidrule(lr){1-7}
DR-AS                  & 0.743 & \(-3.167\) & 0.878 & \(-2.430\) & 0.302 & \(-5.565\) \\
DR                     & 0.762 & \(-2.729\) & 0.884 & \(-2.421\) & 0.366 & \(-3.730\) \\
Archak--Ipeirotis      & 0.807 & \(0.303\)  & 0.936 & \(0.808\)  & 0.389 & \(-1.340\) \\
\midrule
DR-AS \(-\) GARCH\(\times\)DR-AS & \(0.023\) & \(0.273\) & \(0.037\) & \(0.236\) & \(-0.020\) & \(0.394\) \\
95\% Confidence Interval & {\scriptsize[\(0.007,0.042\)]} & {\scriptsize[\(0.219,0.322\)]} & {\scriptsize[\(0.017,0.062\)]} & {\scriptsize[\(0.176,0.292\)]} & {\scriptsize[\(-0.040,-0.004\)]} & {\scriptsize[\(0.281,0.498\)]} \\
\bottomrule
\end{tabular}}
\ifconf\vspace{-16pt}\fi
\end{table}

\ifconf
\else
\input{fc_moves/5_5_sports_figure}\fi

Table~\ref{tab:sports-single-game} shows that DR-AS has an interval score of
\(0.878\) in single-game markets, almost three times its \(0.302\) in other
Sports markets. With \(76.5\%\) of Sports volume, single-game markets
account for about \(90\%\) of the pooled DR-AS score of \(0.743\). These
markets also account for the concentration of large moves documented in
Section~\ref{sec:cat-information}.

The structural results of Section~\ref{sec:three-comparisons} carry over to other
Sports markets. There DR-AS
improves on both deadline resolution and Archak--Ipeirotis on the interval
score, and assigns \(2.5\) and \(8.3\) times their per-hour likelihood to the
realized price changes.

The GARCH layer plays a different role in the two subsets. It raises the
per-hour likelihood of DR-AS in both, by \(13\%\) in single-game markets and
\(22\%\) in other Sports markets, but it improves the intervals only in
single-game markets, lowering the interval score from \(0.878\) to \(0.838\)
(\ifconf Table~\ref{tab:sports-single-game}\else Figure~\ref{fig:sports-single-game}\fi). In other Sports markets, DR-AS alone
already gives the best intervals, and adding GARCH makes them slightly worse.
This matches the division of labor in Section~\ref{sec:conditional-coverage}.
The structural term sets the level of conditional variance, and the GARCH
multiplier corrects the timing of departures from it. Timing matters most when
information arrives in bursts, as it does during a game. When prices move
gradually toward a distant resolution, the structural level alone is enough.

%% file: fc_moves/5_1_2_composition_portability.tex
\subsection{Category composition}
\label{sec:category-composition}

The per-category exercises \ifconf in this appendix and in Section~\ref{sec:universality}\else below\fi{} use the 11 categories with at least 400
contracts, enough for category-level estimation and for contract-cluster
inference. The remaining five named categories are shown in
Table~\ref{tab:category-summary} for sample-composition accounting but are
too small for stable category-specific fits.

\begin{table}[htbp]
\centering
\caption{Sample composition by Kalshi category, ordered by share of volume within each block. The dashed line separates the eleven major categories used in the per-category analysis from the five smaller categories. Volume totals are in millions of contracts. The final column reports the specification in Table~\ref{tab:main-horserace} with the lowest VW-IS-EQ in each category under the global fit; DR is deadline resolution. A further \(19{,}045\) contract-hours carry no category label and are omitted.}
\label{tab:category-summary}
\footnotesize
\setlength{\tabcolsep}{4pt}
\begin{tabular}{lrrrrl}
\toprule
Category & Contract-hours & Share of obs & Volume (M) & Share of vol & Best global spec \\
\midrule
Commodities              &    48{,}781 &  0.7\% &      29 &  0.1\% & GARCH\(\times\)DR-AS \\
Climate and Weather      &   120{,}464 &  1.8\% &      41 &  0.2\% & GARCH\(\times\)DR \\
Companies                &   147{,}990 &  2.2\% &      76 &  0.4\% & GARCH\(\times\)activity-scaled DR \\
Science and Technology   &   175{,}103 &  2.6\% &     109 &  0.5\% & GARCH\(\times\)DR \\
Mentions                 &   399{,}331 &  5.9\% &     193 &  0.9\% & GARCH\(\times\)Archak--Ipeirotis \\
Crypto                   &   274{,}860 &  4.1\% &     264 &  1.2\% & GARCH\(\times\)DR \\
Entertainment            &   862{,}013 & 12.8\% &     616 &  2.9\% & ARCH\(\times\)DR-AS \\
Economics                &   467{,}961 &  6.9\% &     677 &  3.2\% & ARCH\(\times\)DR-AS \\
Elections                &   477{,}069 &  7.1\% &     741 &  3.5\% & GARCH\(\times\)DR-AS \\
Politics                 &   954{,}658 & 14.2\% &  2{,}058 &  9.7\% & GARCH\(\times\)DR \\
Sports                   & 2{,}680{,}489 & 39.8\% & 16{,}226 & 76.8\% & GARCH\(\times\)DR-AS \\
\hdashline
Transportation           &     3{,}441 &  0.1\% &       1 &  0.0\% & GARCH\(\times\)DR \\
Social                   &     4{,}529 &  0.1\% &       1 &  0.0\% & DR-AS \\
World                    &    13{,}091 &  0.2\% &       4 &  0.0\% & ARCH\(\times\)DR-AS \\
Health                   &    22{,}708 &  0.3\% &       6 &  0.0\% & GARCH\(\times\)DR \\
Financials               &    81{,}833 &  1.2\% &      74 &  0.4\% & GARCH\(\times\)DR \\
\bottomrule
\end{tabular}
\end{table}

The category distribution is highly concentrated. Sports alone accounts for
\(40\%\) of contract-hours and \(77\%\) of volume, and Sports and Politics
together for \(54\%\) and \(87\%\). The 11 major categories account for
\(98\%\) of contract-hours and \(99.6\%\) of volume, while the five smaller
categories together have \(125{,}602\) contract-hours and little weight in the
volume-weighted criterion.

\subsection{Model portability across major categories}
\label{sec:category-family-transfer}

The final column of Table~\ref{tab:category-summary} shows that the category
winners do not fragment into unrelated models. Among the 11 major categories,
the best specification retains the deadline-resolution backbone
\(p(1-p)/\tau\) in ten; the exception is Mentions, where
GARCH \(\times\) Archak--Ipeirotis is best. The winners differ only in whether
that backbone is augmented by the order-flow term \(K\nu(V)s^2/4\), which
appears in five of the eleven, by the activity rescaling, which appears in one,
and by GARCH or ARCH dynamics, which appear in all eleven. No major category is
best served by a plain ARCH/GARCH specification. Category heterogeneity
therefore changes the preferred member of the family, not the broad
volatility architecture.

\begin{table}[htbp]
\centering
\caption{Per-category paired comparisons under the global fit. Each entry is
\(\Delta_{B-A}=S(B)-S(A)\) for the pair \(A\) versus \(B\) named in the column
header, so positive values favor the paper's specification \(A\); an asterisk
marks a paired contract-cluster bootstrap 95\% interval that excludes zero. DR is deadline resolution alone and G abbreviates GARCH.}
\label{tab:category-race}
\footnotesize
\setlength{\tabcolsep}{4pt}
\begin{tabular}{lcccccc}
\toprule
& \multicolumn{2}{c}{DR-AS vs Archak--Ipeirotis} & \multicolumn{2}{c}{DR-AS vs DR} & \multicolumn{2}{c}{G\(\times\)DR-AS vs plain G} \\
\cmidrule(lr){2-3}\cmidrule(lr){4-5}\cmidrule(lr){6-7}
Category & \(\Delta\) EQ & \(\Delta\) QLIKE & \(\Delta\) EQ & \(\Delta\) QLIKE & \(\Delta\) EQ & \(\Delta\) QLIKE \\
\midrule
Sports                 & \(0.065^{*}\)  & \(3.47^{*}\)  & \(0.019\)      & \(0.44^{*}\)  & \(0.143^{*}\) & \(2.00^{*}\) \\
Politics               & \(0.017\)      & \(11.13^{*}\) & \(0.026^{*}\)  & \(5.57^{*}\)  & \(0.081^{*}\) & \(0.59^{*}\) \\
Elections              & \(0.032^{*}\)  & \(15.80^{*}\) & \(0.050^{*}\)  & \(7.89^{*}\)  & \(0.214^{*}\) & \(2.00^{*}\) \\
Economics              & \(0.076^{*}\)  & \(4.25^{*}\)  & \(0.067^{*}\)  & \(2.71^{*}\)  & \(0.166^{*}\) & \(1.84^{*}\) \\
Entertainment          & \(0.051^{*}\)  & \(10.17^{*}\) & \(0.046^{*}\)  & \(2.57^{*}\)  & \(0.216^{*}\) & \(1.66^{*}\) \\
Crypto                 & \(0.038^{*}\)  & \(4.18^{*}\)  & \(0.021^{*}\)  & \(2.53^{*}\)  & \(0.272^{*}\) & \(2.18^{*}\) \\
Mentions               & \(-0.034^{*}\) & \(38.25^{*}\) & \(-0.032^{*}\) & \(1.31^{*}\)  & \(0.134^{*}\) & \(1.05^{*}\) \\
Science and Technology & \(0.022^{*}\)  & \(12.30^{*}\) & \(0.034^{*}\)  & \(4.14^{*}\)  & \(0.270^{*}\) & \(1.64^{*}\) \\
Companies              & \(0.104^{*}\)  & \(14.52^{*}\) & \(0.104^{*}\)  & \(10.58^{*}\) & \(0.241^{*}\) & \(1.47^{*}\) \\
Climate and Weather    & \(-0.033\)     & \(8.35^{*}\)  & \(0.001\)      & \(3.00^{*}\)  & \(0.171^{*}\) & \(1.22^{*}\) \\
Commodities            & \(0.058^{*}\)  & \(6.81^{*}\)  & \(0.045^{*}\)  & \(3.40^{*}\)  & \(0.252^{*}\) & \(1.01^{*}\) \\
\bottomrule
\end{tabular}
\end{table}

%% file: fc_moves/5_2_race_text.tex
Point-estimate winners are a coarse summary, because several specifications
are close within most categories. Table~\ref{tab:category-race} therefore
reports paired comparisons category by category, using the bootstrap of
\ifconf Appendix~\ref{app:pairwise}\else Section~\ref{sec:horserace}\fi{} restricted to one category's contracts, and
Figure~\ref{fig:category-4specs} shows the within-category VW-IS-EQ of four
representative specifications. The structural core transfers: DR-AS improves
significantly on the Archak--Ipeirotis benchmark and on deadline resolution
alone on VW-QLIKE in every major category. On VW-IS-EQ it improves
significantly on each benchmark in eight of eleven categories, and the point
estimate favors DR-AS in nine and ten categories respectively, with Mentions
the one significant reversal. The dynamic layer transfers:
GARCH \(\times\) DR-AS improves significantly on plain GARCH on both scores in
every major category. Its lead over the other multiplicative specifications is less
uniform; GARCH \(\times\) deadline resolution is as good or better in several
mid-sized categories, so that advantage comes from the largest categories,
Sports, Politics, Elections, and Economics.

%% file: fc_moves/5_3_catfit.tex
\subsection{Parameter portability}
\label{sec:category-catfit}

Model portability asks which specification works across categories. Parameter
portability asks whether a given specification needs separate category-level
estimates. For each of the two promoted specifications, the \emph{global fit}
estimates parameters on the pooled training panel and applies them to every
category in the test month; the \emph{per-category fit} re-estimates the same
specification within each category. The expanding-window splits and metrics
are otherwise identical, and both fits are scored with interval quantiles
calibrated on the category. Because this exercise is restricted to the eleven major
categories, score levels are not directly comparable to the pooled panel in
Table~\ref{tab:main-horserace}. We measure the effect of category-specific
refitting by
\[
    \Delta\mathrm{VW\text{-}IS\text{-}EQ}
    =
    \frac{\mathrm{VW\text{-}IS\text{-}EQ}^{\mathrm{catfit}}-\mathrm{VW\text{-}IS\text{-}EQ}^{\mathrm{global}}}
         {\mathrm{VW\text{-}IS\text{-}EQ}^{\mathrm{global}}},
\]
so $\Delta\mathrm{VW\text{-}IS\text{-}EQ}$ is a relative change and  positive values mean that per-category refitting worsens
the interval score.

\begin{table}[htbp]
\centering
\caption{Global fit versus per-category fit over the eleven major categories,
volume-weighted. \(\Delta\mathrm{VW\text{-}IS\text{-}EQ}\) is the relative
change in VW-IS-EQ (percent).}
\label{tab:catfit-global-vs-category}
\small
\setlength{\tabcolsep}{4pt}
\begin{tabular}{lccccc}
\toprule
Specification
& \makecell{VW-IS-EQ\\global}
& \makecell{VW-Cov-EQ\\global}
& \makecell{VW-IS-EQ\\per-cat}
& \makecell{VW-Cov-EQ\\per-cat}
& \(\Delta\mathrm{VW\text{-}IS\text{-}EQ}\) \\
\midrule
DR-AS\,(\(\sqrt V\))                 & 0.645 & 0.940 & 0.657 & 0.928 & \(+1.8\%\) \\
GARCH\(\times\)DR-AS\,(\(\sqrt V\))  & 0.577 & 0.957 & 0.594 & 0.937 & \(+3.0\%\) \\
\bottomrule
\end{tabular}
\end{table}

Table~\ref{tab:catfit-global-vs-category} is a refitting exercise, not a new
pooled horse race: it asks whether holding a specification fixed and estimating
it separately by category improves out-of-sample performance and results showed it does not.
Re-estimating the parameters raises VW-IS-EQ for both specifications.
Volume-weighted coverage falls below the nominal 95\% for both models, with
the largest drop in Politics.
At
the category sample sizes available here, pooling therefore appears to be the
more robust way to estimate the structural family.

\begin{figure}[!htbp]
\centering
\begin{subfigure}[t]{0.49\textwidth}
\centering
\includegraphics[width=\linewidth]{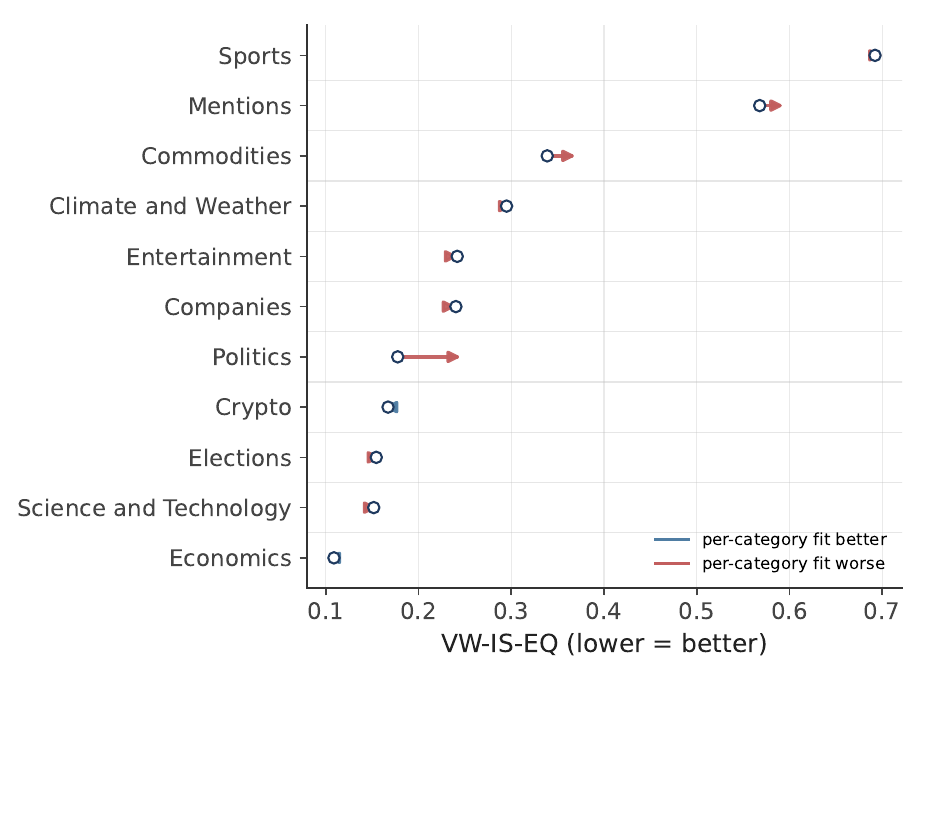}
\caption{Global fit versus per-category fit, by category.}
\label{fig:catfit-paired}
\end{subfigure}
\hfill
\begin{subfigure}[t]{0.49\textwidth}
\centering
\includegraphics[width=\linewidth]{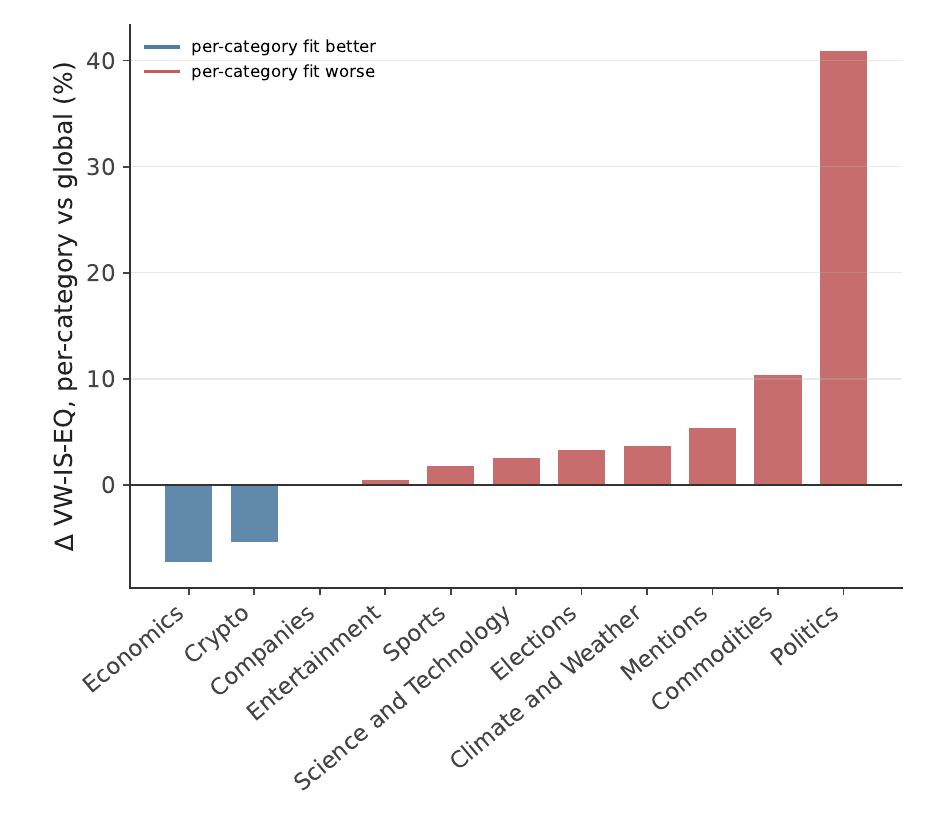}
\caption{Relative change in VW-IS-EQ from global to per-category fit.}
\label{fig:catfit-delta}
\end{subfigure}
\caption{Per-category refitting of GARCH \(\times\) DR-AS\,(\(\sqrt V\)) across
the 11 major categories. Panel (a): the open circle is the category-level
VW-IS-EQ under the global fit; the arrow points to the per-category-fit value.
Panel (b): the relative change \(\Delta\mathrm{VW\text{-}IS\text{-}EQ}\);
negative values (blue) indicate that per-category refitting improves the
score, positive values (red) indicate that it worsens the score.}
\label{fig:catfit-figs}
\end{figure}

Figure~\ref{fig:catfit-figs} repeats the comparison at the category level for
GARCH \(\times\) DR-AS. Refitting improves the score in only two
of the eleven major categories, Economics and Crypto, and worsens it
in the other nine, with the largest deterioration in Politics. The two
largest categories by volume, Sports and Politics, both fail to improve under
refitting, so the volume-weighted comparison in
Table~\ref{tab:catfit-global-vs-category} reflects this deterioration.
Category-specific parameter estimates do not reliably improve on the global
fit.

%% file: fc_moves/5_4_regimes.tex
Portability does not mean equal difficulty. The best global-fit VW-IS-EQ ranges from \(0.12\) in Economics to \(0.71\) in Sports, a roughly sixfold difference. Figures~\ref{fig:category-4specs} and~\ref{fig:category-regimes} suggest that this spread is tied to the category's information environment, not to a need for unrelated model architectures.

Economics is the smooth-resolution endpoint. In Figure~\ref{fig:category-4specs}, the three specifications that add the order-flow term, GARCH dynamics, or both have nearly identical VW-IS-EQ in Economics, between \(0.124\) and \(0.128\), while deadline resolution alone is clearly worse at \(0.195\). Figure~\ref{fig:category-regimes} shows the corresponding large-move pattern. Economics is the only major category whose volume-weighted share of large-move hours, \(3.0\%\), is below its observation share, \(4.9\%\). Trading volume is therefore not concentrated in the large-move tail. This fits the category's economic structure: Economics contracts often resolve around scheduled macroeconomic releases such as CPI, employment, and GDP, where the resolution date is fixed and much of the price movement is the gradual incorporation of public information before the release.

Sports is the event-concentrated endpoint. In Figure~\ref{fig:category-4specs}, the four specifications are again close to each other, but at a much higher level, between \(0.714\) and \(0.762\). The order-flow term and GARCH dynamics add only modest improvement over deadline resolution relative to the overall difficulty of the category. Figure~\ref{fig:category-regimes} shows why the volume-weighted score is so hard to reduce. Only \(4.1\%\) of Sports hours fall in the large-move tail, but those hours carry \(47.8\%\) of Sports volume. Thus nearly half of the volume-weighted Sports evaluation is concentrated in jump-like hours. Sports contracts resolve around discrete game states and events such as goals, injuries, period breaks, and substitutions, whose next-hour price impact is hard to anticipate from smooth forecast-origin state variables. Section~\ref{sec:sports-single-game} separates the single-game markets that account for most of this concentration.

The other major categories lie between these endpoints. Category heterogeneity therefore does not overturn the structural-family result, but it shows where the smooth conditional-variance approximation is weakest. The model is closest to its intended setting in smooth-resolution categories and most incomplete in event-concentrated markets, where an event clock or jump component for public information would be a natural extension.

%% file: fc_moves/5_5_sports_figure.tex
\begin{figure}[!htbp]
\centering
\includegraphics[width=0.85\linewidth]{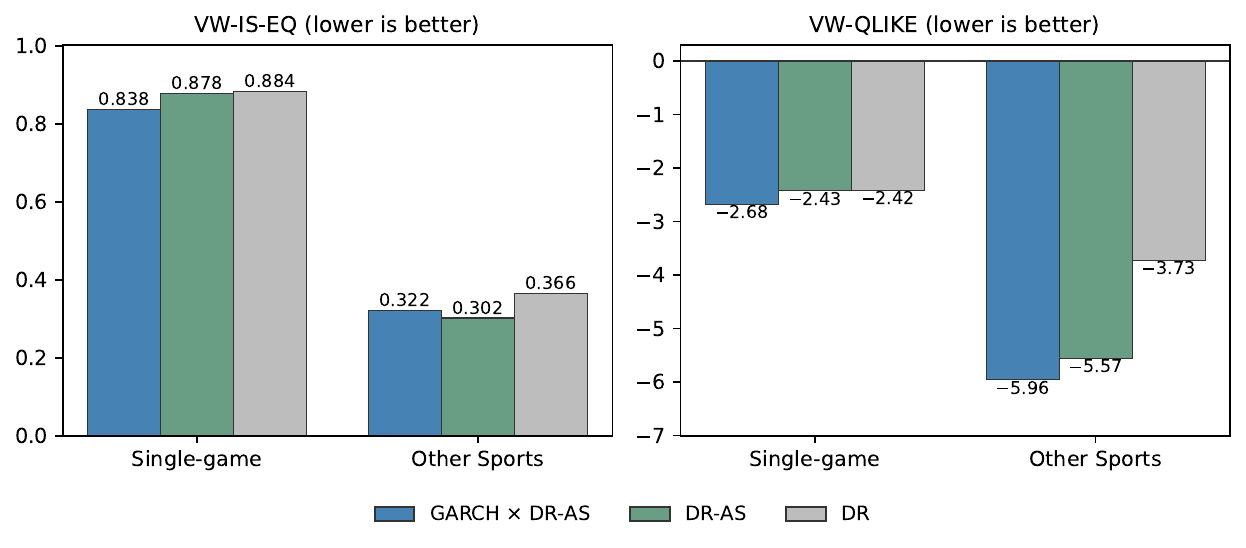}
\caption{VW-IS-EQ (left) and VW-QLIKE (right) under the global fit for
GARCH \(\times\) DR-AS, DR-AS, and deadline resolution on single-game and
other Sports markets. Lower is better for both metrics.}
\label{fig:sports-single-game}
\end{figure}

%% file: 6_conclusion.tex
\section{Conclusion}
\label{sec:conclusion}

This paper builds the case that prediction-market volatility is shaped by features that ordinary asset-price volatility models do not encode: binary payoffs, fixed resolution dates, bounded probability prices, and information-sensitive order flow. In the Kalshi panel, forecasts built from these state variables substantially outperform plain ARCH/GARCH and HAR-RV benchmarks, while a multiplicative GARCH layer provides a further but smaller improvement and reduces the clustering of interval violations that the structural forecast leaves behind to a small residual. The main empirical lesson is therefore that time-series persistence works best after conditioning on the structure of the contract. 

Several extensions follow naturally. First, the arrival rate of informed order flow depends only on trading volume; future work could let it respond to public information events as well, capturing the fact that public information events and informed trading often arrive together. Second, public information moves the price smoothly in the model, and only order flow moves it in jumps; event-heavy markets such as single-game Sports markets suggest the value of adding a jump or event-clock component for public information as well, especially around scheduled announcements, game states, or other discrete information releases. These extensions would move the model from a parsimonious structural forecasting rule toward a richer description of how beliefs evolve in real time.

%% file: 7_appendix.tex
\expandafter\ifx\csname ifconf\endcsname\relax\expandafter\newif\csname ifconf\endcsname\fi
\section{Wright-Fisher Structural Learning Model}
\label{app:wf_structural}

This appendix gives the structural learning model behind the Wright-Fisher deadline-resolution channel of Section~\ref{subsec:public_channel}. Section~\ref{app:public_signal_setup} sets up a primitive Bayesian learning chain in which the market observes a stream of weak binary public signals about the eventual settlement state. Section~\ref{app:weak_convergence} shows that this chain converges weakly to the neutral Wright-Fisher diffusion as the signal strength vanishes. Section~\ref{app:variance_budget} establishes the calendar-time variance-budget identity used to justify the deadline clock.

\subsection{Primitive Public-Signal Experiment and Bayesian Learning}
\label{app:public_signal_setup}

We micro-found the Wright-Fisher channel in information time \(u\). The object of learning is the single terminal settlement state \(\Theta\), not a sequence of new binary outcomes. For \(t<T\), the market does not observe \(\Theta\) directly. Instead, it observes pre-settlement public signals whose distributions depend on that same terminal state. Thus
\[
    \widetilde p_u=\mathbb P(\Theta=1\mid\mathcal F_u^P)
\]
is the market's posterior probability that the final settlement will be YES after observing public information up to information time \(u\). When both channels are present, the same local calculation is applied at the current full posterior \(p_t=\mathbb P(\Theta=1\mid\mathcal F_t)\).

Over a small information-time interval \([u,u+\Delta u]\), the market observes a binary public micro-signal
\[
    X_{u+\Delta u}\in\{0,1\}.
\]
Here \(X=1\) means that the newly arrived public information leans toward a YES settlement, and \(X=0\) means that it leans toward a NO settlement. The variable \(X_{u+\Delta u}\) is not the settlement outcome itself; it is an intermediate observation about the eventual settlement state. Public signal innovations are assumed conditionally independent across disjoint small intervals given the current information and the true state. Set
\[
    \delta=\sqrt{\Delta u}.
\]
The primitive experiment is a mixture of a weak truth-aligned component and a consensus-baseline component:
\begin{equation*}
    X_{u+\Delta u}
    =
    \begin{cases}
        \Theta, & \text{with probability } \delta,\\[2pt]
        Z,      & \text{with probability } 1-\delta,
    \end{cases}
    \qquad
    Z\mid\mathcal F_u^{P}\sim\operatorname{Bernoulli}(\widetilde p_u),
    \qquad
    Z\perp\Theta\mid\mathcal F_u^{P}.
\end{equation*}
The first branch is a reduced-form representation of a public signal component aligned with the eventual settlement state, not literal early settlement. The market observes only \(X_{u+\Delta u}\), not the branch from which it came. Equivalently,
\begin{equation*}
    \mathbb P(X_{u+\Delta u}=1\mid\Theta,\mathcal F_u^{P})
    =(1-\delta)\widetilde p_u+\delta\Theta.
\end{equation*}
Ordinary public information is interpreted relative to the current market consensus \(\widetilde p_u\), but each micro-signal has a small tilt toward the true state. If \(\Theta=1\), then
\begin{equation*}
    \mathbb P(X=1\mid \Theta=1,\mathcal F_u^{P})
    =\widetilde p_u+\delta(1-\widetilde p_u),
\end{equation*}
and if \(\Theta=0\), then
\begin{equation*}
    \mathbb P(X=1\mid \Theta=0,\mathcal F_u^{P})
    =\widetilde p_u(1-\delta).
\end{equation*}
Hence
\begin{equation*}
    \mathbb P(X=1\mid\Theta=1,\mathcal F_u^{P})
    -
    \mathbb P(X=1\mid\Theta=0,\mathcal F_u^{P})
    =\delta>0.
\end{equation*}
The public signal is therefore genuinely truth-bearing: a YES-leaning signal is more likely when the true state is YES, and a NO-leaning signal is more likely when the true state is NO. The adaptive baseline is not a claim that the truth itself is drawn from \(\operatorname{Bernoulli}(\widetilde p_u)\); it is a statement about how public news is read relative to the market's current expectation.

\paraheader{Predictive distribution, Bayes update, and local moments.}
Before observing \(X\), the market averages over the two truth states. Using \(\mathbb P(\Theta=1\mid\mathcal F_u^{P})=\widetilde p_u\),
\begin{align*}
    \mathbb P(X=1\mid\mathcal F_u^{P})
    &=\widetilde p_u[\widetilde p_u+\delta(1-\widetilde p_u)]
      +(1-\widetilde p_u)[\widetilde p_u(1-\delta)] \\
    &=\widetilde p_u.
\end{align*}
Thus
\begin{equation}
    X_{u+\Delta u}\mid\mathcal F_u^{P}
    \sim \operatorname{Bernoulli}(\widetilde p_u).
    \label{eq:public_predictive_distribution}
\end{equation}
This is a predictive statement about the next public signal; the terminal state \(\Theta\) is fixed, and \(\widetilde p_u\) is the market's posterior probability given current information.

Bayes' rule gives the posterior after a YES-leaning public signal:
\begin{align*}
    \widetilde p_{u+\Delta u}^{+}
    &=\mathbb P(\Theta=1\mid X=1,\mathcal F_u^{P}) \\
    &=\frac{\widetilde p_u[\widetilde p_u+\delta(1-\widetilde p_u)]}
            {\mathbb P(X=1\mid\mathcal F_u^{P})} \\
    &=\widetilde p_u+\delta(1-\widetilde p_u).
\end{align*}
After a NO-leaning public signal,
\begin{align*}
    \widetilde p_{u+\Delta u}^{-}
    &=\mathbb P(\Theta=1\mid X=0,\mathcal F_u^{P}) \\
    &=\frac{\widetilde p_u(1-\widetilde p_u)(1-\delta)}
            {1-\widetilde p_u} \\
    &=\widetilde p_u(1-\delta).
\end{align*}
Both cases can be written as the forecast-error update
\begin{equation}
    \widetilde p_{u+\Delta u}-\widetilde p_u
    =\delta(X_{u+\Delta u}-\widetilde p_u).
    \label{eq:public_update}
\end{equation}
The learning rate multiplying the forecast error is \(\delta\); the realized jump size still depends on \(\widetilde p_u\) through the surprise \(X-\widetilde p_u\).

Using \eqref{eq:public_predictive_distribution} and \eqref{eq:public_update},
\begin{align*}
    \mathbb E[\widetilde p_{u+\Delta u}-\widetilde p_u\mid\mathcal F_u^{P}]
    &=\delta\{\mathbb E[X_{u+\Delta u}\mid\mathcal F_u^{P}]-\widetilde p_u\}=0,\\[4pt]
    \operatorname{Var}(\widetilde p_{u+\Delta u}-\widetilde p_u\mid\mathcal F_u^{P})
    &=\delta^2\operatorname{Var}(X_{u+\Delta u}\mid\mathcal F_u^{P}) \\
    &=\widetilde p_u(1-\widetilde p_u)\Delta u.
\end{align*}
The public posterior is therefore a zero-drift martingale in information time, and its local variance is proportional to remaining binary uncertainty. The variance \(\widetilde p_u(1-\widetilde p_u)\) is the predictive variance of a truth-bearing public signal, not the result of drawing the terminal state from the current posterior.

\subsection{Diffusion Limit}
\label{app:weak_convergence}

We now sketch the diffusion-limit argument by which the chain of Section~\ref{app:public_signal_setup} converges weakly to the neutral Wright-Fisher diffusion. The arguments are standard.

Set \(h=\Delta u\) and \(\delta=\sqrt h\), and consider the piecewise-constant interpolation of the chain in Section~\ref{app:public_signal_setup}. Conditional on \(\widetilde p_{kh}=p\), the chain has zero conditional drift (so \(\widetilde p\) is a martingale), conditional second moment \(\mathbb E[(\widetilde p_{(k+1)h}-p)^2\mid\widetilde p_{kh}=p]=h\,p(1-p)\), and bounded one-step jumps \(|\widetilde p_{(k+1)h}-\widetilde p_{kh}|\le\sqrt h\). The jump bound implies the Lindeberg condition, ruling out macroscopic jumps in the limit. A Taylor expansion of any \(f\in C^2([0,1])\) then gives uniform convergence of the discrete generator to \(Af(p)=\tfrac12\,p(1-p)f''(p)\), the generator of the neutral Wright-Fisher diffusion of population genetics \citep{karlin1981second,EthierNorman1977}. The diffusion coefficient \(\sigma(p)=\sqrt{p(1-p)}\) vanishes at the boundary, making \(0\) and \(1\) absorbing, and is \(\tfrac12\)-H\"older on \([0,1]\); the Yamada-Watanabe condition \(\int_{0+} du/u=\infty\) then yields pathwise uniqueness, so the martingale problem for \(A\) is well posed. Combined with compactness of the state space \([0,1]\), the martingale-problem convergence theorem \citep[Ch.~4, Cor.~8.7]{EthierKurtz1986} delivers
\[
    \widetilde p^{\,h}
    \;\Rightarrow\;
    \widetilde p
    \qquad
    \text{in }D([0,\infty),[0,1]),
\]
where the limit \(\widetilde p\) is the neutral Wright-Fisher diffusion
\[
    d\widetilde p_u
    =
    \sqrt{\widetilde p_u(1-\widetilde p_u)}\,d\widetilde B_u^{P},
    \qquad \widetilde p_u\in[0,1],
\]
with \(0\) and \(1\) absorbing.

\subsection{Calendar-Time Variance Budget}
\label{app:variance_budget}

\ifconf
The diffusion limit of Section~\ref{app:weak_convergence} absorbs in \(\{0,1\}\) eventually, but not necessarily by any particular calendar time. \input{fc_moves/2_1_deadline_clock}%

We now establish this identity. Recall the calendar-time Wright-Fisher SDE
\else
We now establish the variance-budget identity \eqref{eq:wf_fundamental_variance} used in Section~\ref{subsec:public_channel} to justify the deadline clock. Recall the calendar-time Wright-Fisher SDE
\fi
\[
    dp_u^{P}
    =
    \sqrt{\frac{p_u^{P}(1-p_u^{P})}{T-u}}\,dW_u^{P}.
\]

Fix \(t<T\) and let \(\tau\in[t,T)\). Applying It\^o's formula to \((p_u^{P})^2\) on \([t,\tau]\) and substituting the SDE,
\begin{equation}
\label{app:A1_1}
    (p_\tau^{P})^2-(p_t^{P})^2
    =
    2\int_t^\tau
    p_u^{P}\sqrt{\frac{p_u^{P}(1-p_u^{P})}{T-u}}\,dW_u^{P}
    +
    \int_t^\tau \frac{p_u^{P}(1-p_u^{P})}{T-u}\,du .
\end{equation}

The stochastic integrand
\[
    H_u
    =
    2p_u^{P}\sqrt{\frac{p_u^{P}(1-p_u^{P})}{T-u}},
    \qquad u\in[t,\tau],
\]
is continuous and adapted. Since \(p_u^{P}\in[0,1]\) and \(T-u\ge T-\tau>0\) on \([t,\tau]\),
\[
    \mathbb E\!\left[\int_t^\tau H_u^2\,du\right]
    \leq
    \int_t^\tau \frac{4}{T-\tau}\,du
    =
    \frac{4(\tau-t)}{T-\tau}
    <\infty .
\]
Thus \(\int_t^\cdot H_u\,dW_u^{P}\) is an \(\mathcal L^2\) martingale and has zero conditional mean given \(\mathcal F_t^{P}\). Taking \(\mathbb E[\cdot\mid\mathcal F_t^{P}]\) on both sides of \eqref{app:A1_1},
\[
    \mathbb E\!\left[
    (p_\tau^{P})^2
    \;\middle|\;\mathcal F_t^{P}
    \right]
    -(p_t^{P})^2
    =
    \mathbb E\!\left[
    \int_t^\tau \frac{p_u^{P}(1-p_u^{P})}{T-u}\,du
    \;\middle|\;\mathcal F_t^{P}
    \right] .
\]
By the martingale property \(\mathbb E[p_\tau^{P}\mid\mathcal F_t^{P}]=p_t^{P}\), the left-hand side equals \(\Var[p_\tau^{P}\mid\mathcal F_t^{P}]\), yielding the general variance-budget identity
\begin{equation}
    \Var[p_\tau^{P}\mid\mathcal F_t^{P}]
    =
    \mathbb E\!\left[
    \int_t^\tau \frac{p_u^{P}(1-p_u^{P})}{T-u}\,du
    \;\middle|\;\mathcal F_t^{P}
    \right]
    \label{app:wf_general_identity}
\end{equation}
for every \(\tau\in[t,T)\), which is \eqref{eq:wf_fundamental_variance}.

It remains to pass \(\tau\uparrow T\). The process \(p_\tau^{P}\) is bounded in \([0,1]\) and absorbs at the boundary, so \(p_\tau^{P}\to p_T^{P}=\Theta\in\{0,1\}\). Bounded convergence gives
\(\mathbb E[(p_\tau^{P})^2\mid\mathcal F_t^{P}]\to\mathbb E[\Theta^2\mid\mathcal F_t^{P}]=\mathbb E[\Theta\mid\mathcal F_t^{P}]=p_t^{P}\), while conditional monotone convergence applies to the right-hand side of \eqref{app:wf_general_identity}. We obtain the terminal case
\[
    \Var[p_T^{P}\mid\mathcal F_t^{P}]
    =
    \mathbb E\!\left[
    \int_t^T \frac{p_u^{P}(1-p_u^{P})}{T-u}\,du
    \;\middle|\;\mathcal F_t^{P}
    \right]
    =
    p_t^{P}(1-p_t^{P}),
\]
which formalizes the statement that the deadline clock spends exactly the remaining binary uncertainty by time \(T\).

The identity \eqref{eq:wf_fundamental_variance} can be integrated in closed form. Applying
It\^o's formula to \(p_u^{P}(1-p_u^{P})\) on \([t,\tau]\) and substituting the SDE,
\begin{equation}
    p_\tau^{P}(1-p_\tau^{P})-p_t^{P}(1-p_t^{P})
    =
    \int_t^\tau (1-2p_u^{P})\sqrt{\frac{p_u^{P}(1-p_u^{P})}{T-u}}\,dW_u^{P}
    -
    \int_t^\tau \frac{p_u^{P}(1-p_u^{P})}{T-u}\,du .
    \label{app:A3_ito_g}
\end{equation}
Since \(\lvert 1-2p\rvert\leq1\) and \(\sqrt{p(1-p)}\leq 1/2\) on \([0,1]\), the stochastic
integrand in \eqref{app:A3_ito_g} is bounded in absolute value by \((T-\tau)^{-1/2}/2\) on
\([t,\tau]\). The stochastic integral is therefore an \(\mathcal L^2\) martingale and has zero
conditional mean given \(\mathcal F_t^{P}\). Write
\[
    m(\tau)
    =
    \mathbb E\!\left[p_\tau^{P}(1-p_\tau^{P})\mid\mathcal F_t^{P}\right],
    \qquad
    m(t)=p_t^{P}(1-p_t^{P}).
\]
Taking \(\mathbb E[\cdot\mid\mathcal F_t^{P}]\) in \eqref{app:A3_ito_g} and using Fubini on the
second integral,
\[
    m(\tau)
    =
    m(t)
    -
    \int_t^\tau \frac{m(u)}{T-u}\,du .
\]
The right-hand side is differentiable in \(\tau\), so \(m\) solves
\(m'(\tau)=-m(\tau)/(T-\tau)\) on \([t,T)\). Integrating
\((\log m)'(\tau)=-1/(T-\tau)\) from \(t\) to \(\tau\) gives
\(\log\!\left[m(\tau)/m(t)\right]=\log\!\left[(T-\tau)/(T-t)\right]\), that is,
\begin{equation}
    m(\tau)
    =
    p_t^{P}(1-p_t^{P})\,\frac{T-\tau}{T-t},
    \qquad t\leq\tau<T .
    \label{app:A3_uncertainty_decay}
\end{equation}
Remaining binary uncertainty therefore decays linearly in calendar time under the deadline clock.

By \eqref{app:A3_uncertainty_decay} the integrand of the variance budget is constant in \(u\),
\[
    \frac{m(u)}{T-u}
    =
    \frac{p_t^{P}(1-p_t^{P})}{T-t},
    \qquad t\leq u<T,
\]
so substituting into \eqref{eq:wf_fundamental_variance} with \(\tau=t+\Delta\),
\[
    \Var\!\left[p_{t+\Delta}^{P}\mid\mathcal F_t^{P}\right]
    =
    \int_t^{t+\Delta}\frac{m(u)}{T-u}\,du
    =
    \frac{p_t^{P}(1-p_t^{P})}{T-t}\,\Delta
\]
for every \(\Delta\leq T-t\), which is the one-step variance \eqref{eq:wf_discrete_variance} used in
Section~\ref{subsec:public_channel}. Setting \(\Delta=T-t\) recovers the terminal case above.

\ifconf
\section{Derivation of the Order-Flow Variance}
\label{app:gm_derivation}

This appendix derives the one-event variance \eqref{eq:gm_exact_variance} of Section~\ref{sec:model_flow}. Fix time \(t\) and write \(p=\mathbb P(\Theta=1\mid\mathcal F^{Q}_t)\) for the order-flow-channel posterior.
\input{fc_moves/2_2_gm_derivation}Because the updated posterior \(p_{\mathrm{after}}=\mathbb P(\Theta=1\mid\mathcal F^{Q}_t,B/S)\) is a conditional martingale, its one-event conditional variance is
\[
    \mathbb P(B\mid\mathcal F^{Q}_t)(p^{+}-p)^2
    +\mathbb P(S\mid\mathcal F^{Q}_t)(p^{-}-p)^2
    =\frac{4\alpha^2p^2(1-p)^2}{1-\alpha^2(2p-1)^2}
    =\frac{s^2}{4}\left[1-\alpha^2(2p-1)^2\right],
\]
which is \eqref{eq:gm_exact_variance}.
\fi

\section{The Variance Decomposition}
\label{app:variance_decomposition}

\ifconf
Collect the order-flow jumps of Section~\ref{sec:model_flow} into the pure-jump martingale \(J_t^{Q}=\sum_{k=1}^{N_t^{Q}}Y_k\), where \(Y_k\) is the posterior jump at the \(k\)-th event, equal to \(p^{+}-p\) or \(p^{-}-p\) evaluated just before it. Evaluating both channels at the full-market price, the price follows the jump diffusion
\begin{equation}
    dp_t
    =
    \sqrt{\frac{p_{t-}(1-p_{t-})}{T-t}}\,dW_t^{P}
    +
    dJ_t^{Q},
    \label{eq:main-sde}
\end{equation}
with \(W^{P}\) the Brownian motion of \eqref{eq:wf_calendar_sde}.
\fi
Write the price process \eqref{eq:main-sde} as \(p_t=p_0+M_t+J_t^{Q}\), where
\[
    M_t
    =
    \int_0^t\sqrt{\frac{p_{u-}(1-p_{u-})}{T-u}}\,dW_u^{P}
\]
is a continuous local martingale and \(J_t^{Q}=\sum_{k\leq N_t^{Q}}Y_k\) collects the order-flow
jumps. Each jump satisfies \(\mathbb E[Y_k\mid\mathcal F_{T_k-}]=0\) and \(\lvert Y_k\rvert\leq1\), so
\(J^{Q}\) is a local martingale. Fix \(t\) and \(\Delta\) with \(t+\Delta<T\). For \(u\leq t+\Delta\) the integrand of
\(M\) is bounded by \((T-t-\Delta)^{-1/2}/2\), since \(\sqrt{p(1-p)}\leq1/2\), so \(M\) is a
square-integrable martingale on \([0,t+\Delta]\). The price is bounded in \([0,1]\) and is therefore
square-integrable as well, and \(J^{Q}=p-p_0-M\) is a difference of square-integrable martingales.
For a square-integrable martingale,
\begin{equation}
    \Var\!\left(p_{t+\Delta}-p_t\mid\mathcal F_t\right)
    =
    \mathbb E\!\left[\langle p\rangle_{t+\Delta}-\langle p\rangle_t\mid\mathcal F_t\right].
    \label{app:var_is_qv}
\end{equation}

The quadratic covariation of a continuous local martingale with a purely discontinuous one is
identically zero,
\[
    [M,J^{Q}]_t
    =
    \sum_{u\leq t}\Delta M_u\,\Delta J_u^{Q}
    =
    0,
\]
because \(M\) has no jumps. Hence \([p]=[M]+[J^{Q}]\), and taking compensators, \(\langle
p\rangle=\langle M\rangle+\langle J^{Q}\rangle\). The continuous channel contributes
\[
    \langle M\rangle_t
    =
    \int_0^t\frac{p_u(1-p_u)}{T-u}\,du ,
\]
and the jump channel contributes the arrival rate times the per-event variance,
\[
    \langle J^{Q}\rangle_t
    =
    \int_0^t\lambda_u\,v^{Q}(p_{u-},\alpha_u)\,du
    =
    \int_0^t K\nu(V_u)\,\frac{s_u^2}{4}\,du ,
\]
where the second equality substitutes \(v^{Q}=(s^2/4)[1-\alpha^2(2p-1)^2]\) from
\eqref{eq:gm_exact_variance} and the parametrization of Section~\ref{sec:model_flow}. Substituting
into \eqref{app:var_is_qv},
\begin{equation}
    \Var\!\left(p_{t+\Delta}-p_t\mid\mathcal F_t\right)
    =
    \mathbb E\!\left[
        \int_t^{t+\Delta}
        \left\{
            \frac{p_u(1-p_u)}{T-u}
            +
            K\nu(V_u)\,\frac{s_u^2}{4}
        \right\}du
        \,\middle|\,\mathcal F_t
    \right].
    \label{app:exact_step_variance}
\end{equation}
The decomposition is exact and carries no cross term.

Write \(n(u)=\mathbb E[p_u(1-p_u)\mid\mathcal F_t]\). It\^o's formula for the jump diffusion
\eqref{eq:main-sde} gives, for \(t\leq\tau<T\),
\[
    p_\tau(1-p_\tau)-p_t(1-p_t)
    =
    \int_t^\tau (1-2p_{u-})\,dp_u
    -
    \int_t^\tau \frac{p_u(1-p_u)}{T-u}\,du
    -
    \sum_{t<u\leq\tau}(\Delta p_u)^2 ,
\]
where the second integral is the second-order term along the continuous part and the sum collects
the jump terms,
\[
    p_u(1-p_u)-p_{u-}(1-p_{u-})-(1-2p_{u-})\,\Delta p_u
    =
    -(\Delta p_u)^2 .
\]
The first integral is a martingale, and the compensator of the jump sum is
\(\int_t^\tau K\nu(V_u)s_u^2/4\,du\) by the parametrization of Section~\ref{sec:model_flow}. Both
integrands are nonnegative and bounded on \([t,\tau]\), so taking \(\mathbb E[\cdot\mid\mathcal F_t]\)
and exchanging expectation and integration gives
\[
    n(\tau)
    =
    n(t)
    -
    \int_t^\tau
    \left\{
        \frac{n(u)}{T-u}
        +
        \mathbb E\!\left[K\nu(V_u)\frac{s_u^2}{4}\,\middle|\,\mathcal F_t\right]
    \right\}du .
\]
Hence \(n\) is absolutely continuous and, for almost every \(u\),
\[
    n'(u)
    =
    -\frac{n(u)}{T-u}
    -
    \mathbb E\!\left[K\nu(V_u)\frac{s_u^2}{4}\,\middle|\,\mathcal F_t\right],
    \qquad
    n(t)=p_t(1-p_t).
\]
Dividing by \(T-u\) gives
\[
    \left(\frac{n(u)}{T-u}\right)'
    =
    -\frac{1}{T-u}\,
    \mathbb E\!\left[K\nu(V_u)\frac{s_u^2}{4}\,\middle|\,\mathcal F_t\right],
\]
so that
\[
    n(u)
    =
    (T-u)
    \left[
        \frac{p_t(1-p_t)}{T-t}
        -
        \int_t^{u}\frac{1}{T-r}\,
        \mathbb E\!\left[K\nu(V_r)\frac{s_r^2}{4}\,\middle|\,\mathcal F_t\right]dr
    \right].
\]
Because \(p\) is a bounded martingale, \(\Var(p_{t+\Delta}-p_t\mid\mathcal F_t)=n(t)-n(t+\Delta)\).
Substituting and writing \(\tau_t=T-t\),
\begin{equation}
    \Var\!\left(p_{t+\Delta}-p_t\mid\mathcal F_t\right)
    =
    \frac{p_t(1-p_t)}{\tau_t}\,\Delta
    +
    (\tau_t-\Delta)
    \int_t^{t+\Delta}\frac{1}{T-r}\,
    \mathbb E\!\left[K\nu(V_r)\frac{s_r^2}{4}\,\middle|\,\mathcal F_t\right]dr .
    \label{app:exact_closed_form}
\end{equation}
As \(\Delta\uparrow\tau_t\) the factor \(\tau_t-\Delta\) drives the second term to zero whenever
the order-flow variance rate stays bounded, and the variance tends to \(p_t(1-p_t)\), the full
budget a martingale settling in \(\{0,1\}\) must spend. Order flow redistributes that budget
towards the present without changing its total.

The first term of \eqref{app:exact_closed_form} is the deadline term of \eqref{eq:main-model}.
Spreads and activity over the coming hour are not observed. Holding them at their forecast-origin
values makes the second term
\[
    K\nu(V_t)\frac{s_t^2}{4}\,(\tau_t-\Delta)\log\!\frac{\tau_t}{\tau_t-\Delta}
    =
    K\nu(V_t)\frac{s_t^2}{4}\,\Delta
    \left[
        1-\frac{\Delta}{2\tau_t}+O\!\left((\Delta/\tau_t)^2\right)
    \right],
\]
of which \eqref{eq:main-model} keeps the leading term. The two agree to first order when
\(\Delta\) is short relative to \(\tau_t\), and separate in the final hours of a contract, where
\eqref{eq:main-model} overstates the order-flow contribution.

\ifconf
\section{Details of the Empirical Design}
\label{app:model-classes}

This appendix gives the full versions of the forecast-origin state and target
(Section~\ref{sec:dataset}), the model classes
(Section~\ref{sec:specifications}), and the parameter estimation and forecast
evaluation (Section~\ref{sec:estimation}), which the main text summarizes.

\subsection{Forecast-origin state and target}

\input{fc_moves/3_1_state_target}\subsection{Model classes}

\input{fc_moves/3_2_classes_intro}
\input{fc_moves/3_2_classes_details}
\subsection{Parameter estimation}

\input{fc_moves/3_3_estimation}\subsection{Forecast evaluation}

\input{fc_moves/3_3_evaluation}\fi

\section{Estimation and Dynamic Recursion Details}
\label{app:estimation_details}

\ifconf Appendix~\ref{app:model-classes}\else The main paper\fi{} defines the forecasting target and the forecast evaluation. This appendix records the additional implementation details needed to interpret the horse race: how fitted scales are estimated, how the multiplicative ARCH/GARCH component is attached to a structural baseline, how the benchmarks are fitted, and how the dynamic state is propagated through contracts.

\paraheader{Quasi-likelihood and fitted scales.}
For each test month \(m\), fitted quantities are estimated using only observations in the expanding training window preceding that month. For a specification with parameter vector \(\theta\), the training criterion is the volume-weighted Gaussian quasi-likelihood
\begin{equation}
    \widehat\theta_m
    \in
    \arg\min_{\theta}
    \sum_{i\in\mathcal T_m^{\mathrm{train}}}
    V_i
    \left\{
        \log h_i^2(\theta)
        +
        \frac{\varepsilon_i^2}{h_i^2(\theta)}
    \right\}.
    \label{eq:app-qmle}
\end{equation}
\(\varepsilon_i=p_{i+1}-p_i\) is the one-hour price innovation. The Gaussian criterion is used only to fit conditional second moments; the interval-score evaluation does not require Gaussian standardized innovations, and the calibrated quantiles of Section~\ref{sec:estimation} absorb any departure from normality. The same forecast-origin volume weights are used in estimation and in the reported VW-IS-EQ and VW-QLIKE.

The deadline resolution and Archak--Ipeirotis benchmark specifications are parameter-free: no fitted scale is applied to either benchmark. DR-AS estimates only the nonnegative order-flow scale \(K\), so \eqref{eq:app-qmle} is a one-dimensional problem. Activity-scaled deadline specifications estimate the positive activity normalization \(\bar V\) in the same way. These scalar quantities are re-estimated separately in each expanding-window training sample.

\paraheader{Multiplicative recursions.}
The multiplicative specifications scale a structural baseline \(B_i\) by a unit-mean dynamic component \(g_i\):
\begin{align}
    h_i^2 &= c\,B_i\,g_i,
    \qquad
    z_i^2 = \frac{\varepsilon_{i-1}^2}{c\,B_{i-1}},
    \qquad
    g_1 = 1,
    \nonumber\\
    \text{ARCH(1)}\times B:\quad
    g_i &= (1-\alpha) + \alpha z_i^2,
    \label{eq:app-mult-arch}\\
    \text{GARCH(1,1)}\times B:\quad
    g_i &= (1-\alpha-\beta) + \alpha z_i^2 + \beta g_{i-1}.
    \label{eq:app-mult-garch}
\end{align}
The baseline \(B_i\) is the structural predictor named in the specification: \(p_i(1-p_i)/\tau_i\) for deadline resolution, \(\varphi^2(\Phi^{-1}(p_i))/\tau_i\) for the Archak--Ipeirotis benchmark, and \(h_i^2(K)\) from \eqref{eq:main-model} for DR-AS. The specifications are fitted in two stages. First, the scale inside \(B_i\) (\(K\) for DR-AS, \(\bar V\) for activity-scaled deadline resolution) is set to its estimate from the structural fit on the same training window and held fixed. Second, \((c,\alpha,\beta)\) are estimated jointly by \eqref{eq:app-qmle} with \(h_i^2=c\,B_i\,g_i\), subject to \(c\ge0\), \(\alpha,\beta\ge0\), and \(\alpha+\beta<1\). The recursion is initialized at the first observation of each contract with \(g_1=1\), equivalently \(z_1^2=1\), so that the first dynamic forecast equals \(c\,B_1\). Because \(g_i\) has unconditional mean one, \(c\,B_i\) is the state-dependent unconditional variance and \(h_i^2\) stays proportional to the structural baseline throughout: when the structural term is small, the dynamic adjustment is small in absolute terms, and there is no additive floor that could dominate the forecast near the price boundaries.

\paraheader{Benchmark recursions.}
The plain dynamic benchmarks use lagged squared innovations but no prediction-market state variables,
\begin{equation}
    \text{ARCH(1)}:\ h_i^2=\omega+\alpha\varepsilon_{i-1}^2,
    \qquad
    \text{GARCH(1,1)}:\ h_i^2=\omega+\alpha\varepsilon_{i-1}^2+\beta h_{i-1}^2,
    \label{eq:app-plain}
\end{equation}
estimated by \eqref{eq:app-qmle} with \(\omega>0\), \(\alpha,\beta\ge0\), and \(\alpha+\beta<1\); the recursion is initialized at the volume-weighted mean squared innovation of the training window. The HAR-RV benchmark regresses the variance on realized variance over the previous hour, day, and week,
\begin{equation}
    h_i^2 = b_0 + b_h\,\mathrm{RV}^{(h)}_i + b_d\,\mathrm{RV}^{(d)}_i + b_w\,\mathrm{RV}^{(w)}_i,
    \qquad
    \mathrm{RV}^{(d)}_i=\frac{1}{24}\sum_{k=0}^{23}\mathrm{RV}^{(h)}_{i-k},
    \qquad
    \mathrm{RV}^{(w)}_i=\frac{1}{168}\sum_{k=0}^{167}\mathrm{RV}^{(h)}_{i-k},
    \label{eq:app-har}
\end{equation}
where \(\mathrm{RV}^{(h)}_i\) is the sum of squared five-minute mid-quote changes over the hour \((t_{i-1},t_i]\), computed from one-minute quotes strictly before the forecast origin, and the daily and weekly averages run over the hours with realized-variance coverage. The coefficients \((b_0,b_h,b_d,b_w)\) are estimated by \eqref{eq:app-qmle} with the slopes constrained to be nonnegative. When a shorter window has no one-minute coverage, its regressor is replaced by the next longer window; hours with no realized-variance history receive the volume-weighted mean squared innovation of the training window.

\paraheader{Contract-level recursion.}
Every recursion is run separately within each contract. Lagged
innovations are taken only from the same contract, and the recursion resets at
contract boundaries. The first dynamic observation in a contract is omitted
because no valid lagged innovation is available, which is why the dynamic
evaluation panel is slightly smaller than the closed-form panel.

For contracts that continue from the training window into a test month, the
variance state is carried forward sequentially using only information available
before each forecast origin. Parameters are fixed at the beginning of the test
month, but the variance state updates as earlier within-contract innovations
become observable. Future test observations are never used to estimate that
month's parameters.

\paraheader{Boundary conventions.}
All model families use common numerical conventions near binary boundaries and
near expiration. These conventions are applied uniformly across specifications,
so relative performance is not driven by singularities in a particular formula.

\ifconf
\section{Stylized Facts: Univariate Binning}
\label{app:stylized}

This appendix gives the full univariate binning analysis summarized in
Section~\ref{sec:stylized}.

\input{fc_moves/3_4_stylized}\fi

\section{Additional Model-Free Binning Diagnostics}
\label{app:binning_figures}

This appendix reports the full set of model-free binning diagnostics summarized in Section~\ref{sec:stylized}. These diagnostics are descriptive. They do not estimate DR-AS, do not control for other state variables, and are not used to rank forecasting models. Their purpose is to show how realized price variation moves univariately with the main observable states. They are computed on the pooled out-of-sample evaluation sample of Section~\ref{sec:estimation}, so the binned hours are exactly the hours scored in the forecast comparison. For each driver \(X\), observations are sorted into pre-specified bins, and the plotted statistic is the volume-weighted root-mean-square next-hour innovation,

\begin{equation}
  \mathrm{VW\text{-}RMS}_b
  =
  \sqrt{
  \frac{\sum_{i\in b} V_{t_i}\varepsilon_i^2}
       {\sum_{i\in b} V_{t_i}}
  }.
  \label{eq:vwrms}
\end{equation}
The driver-strength ratio is the largest bin VW-RMS divided by the smallest. In the figures, boxes are volume-weighted quartiles of \(|\varepsilon_i|\) within each bin, and the bands around VW-RMS are 95\% contract-cluster bootstrap intervals.

\begin{table}[htbp]
\centering
\caption{Complete model-free driver-strength ratios. Each ratio is the maximum
bin VW-RMS divided by the minimum bin VW-RMS.}
\label{tab:driver-summary-full}
\small
\begin{tabular}{clcl}
\toprule
Rank & Driver & Ratio & Interpretation \\
\midrule
1 & Time to resolution \(\tau\) & \(22.6\times\) & Deadline effect \\
2 & \(1/\sqrt{\tau}\) & \(22.6\times\) & Re-expression of deadline scale \\
3 & Time \% to resolution & \(9.0\times\) & Normalized time-to-resolution proxy \\
4 & Category & \(8.0\times\) & Proxy for duration / market type \\
5 & Volume \(V\) & \(7.5\times\) & Activity / information-flow proxy \\
6 & Open interest & \(3.8\times\) & Participation proxy \\
7 & Price \(p\) & \(3.5\times\) & Boundary shape in \(p(1-p)\) \\
8 & Spread \(s\) & \(2.2\times\) & Adverse-selection / liquidity proxy \\
\bottomrule
\end{tabular}
\end{table}

\begin{figure}[htbp]
\centering
\begin{subfigure}{0.48\textwidth}
\includegraphics[width=\linewidth]{figures/Uni_01_price_bins.pdf}
\caption{Price \(p\).}
\end{subfigure}
\hfill
\begin{subfigure}{0.48\textwidth}
\includegraphics[width=\linewidth]{figures/Uni_02_tte_bins.pdf}
\caption{Time to resolution in hours.}
\end{subfigure}

\begin{subfigure}{0.48\textwidth}
\includegraphics[width=\linewidth]{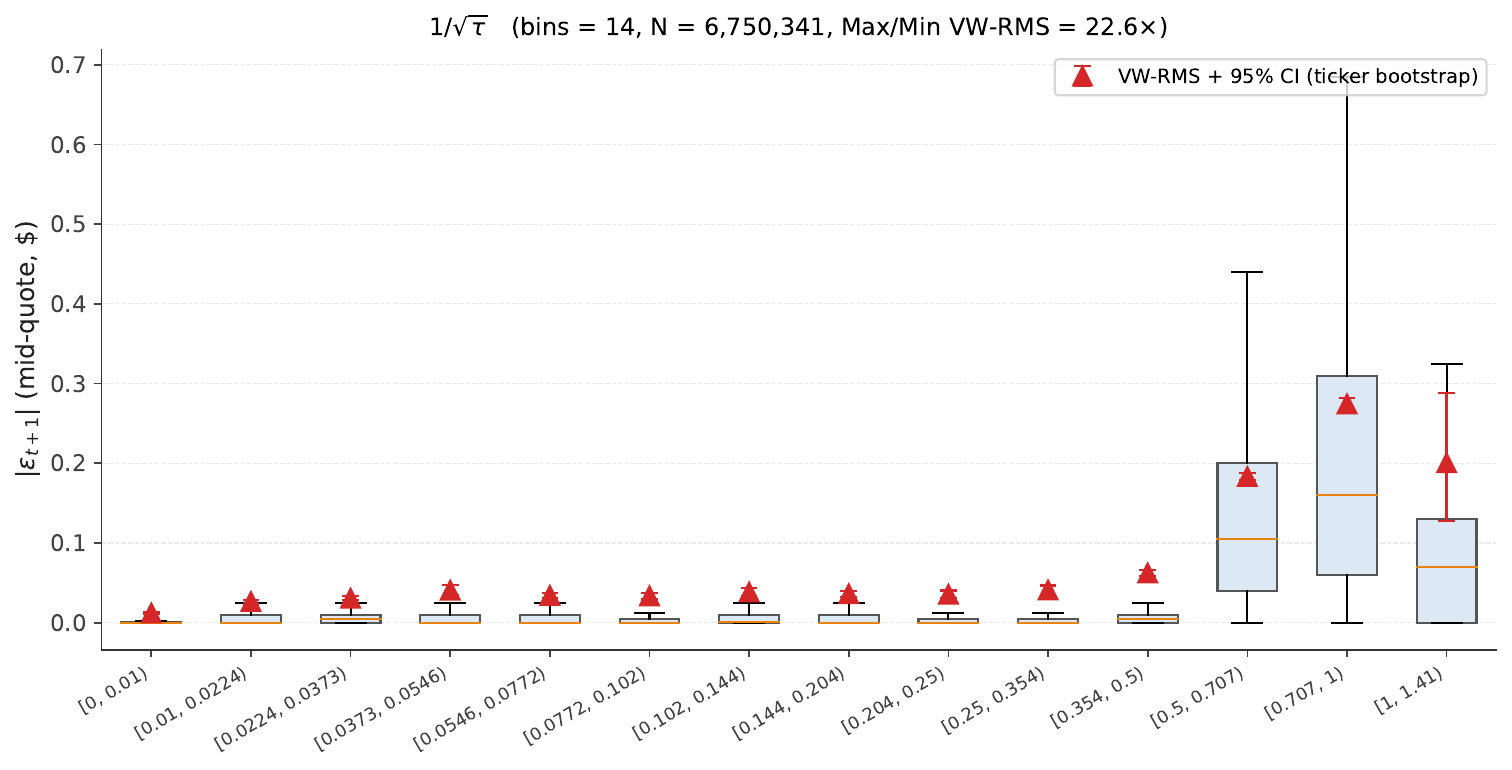}
\caption{\(1/\sqrt{\tau}\).}
\end{subfigure}
\hfill
\begin{subfigure}{0.48\textwidth}
\includegraphics[width=\linewidth]{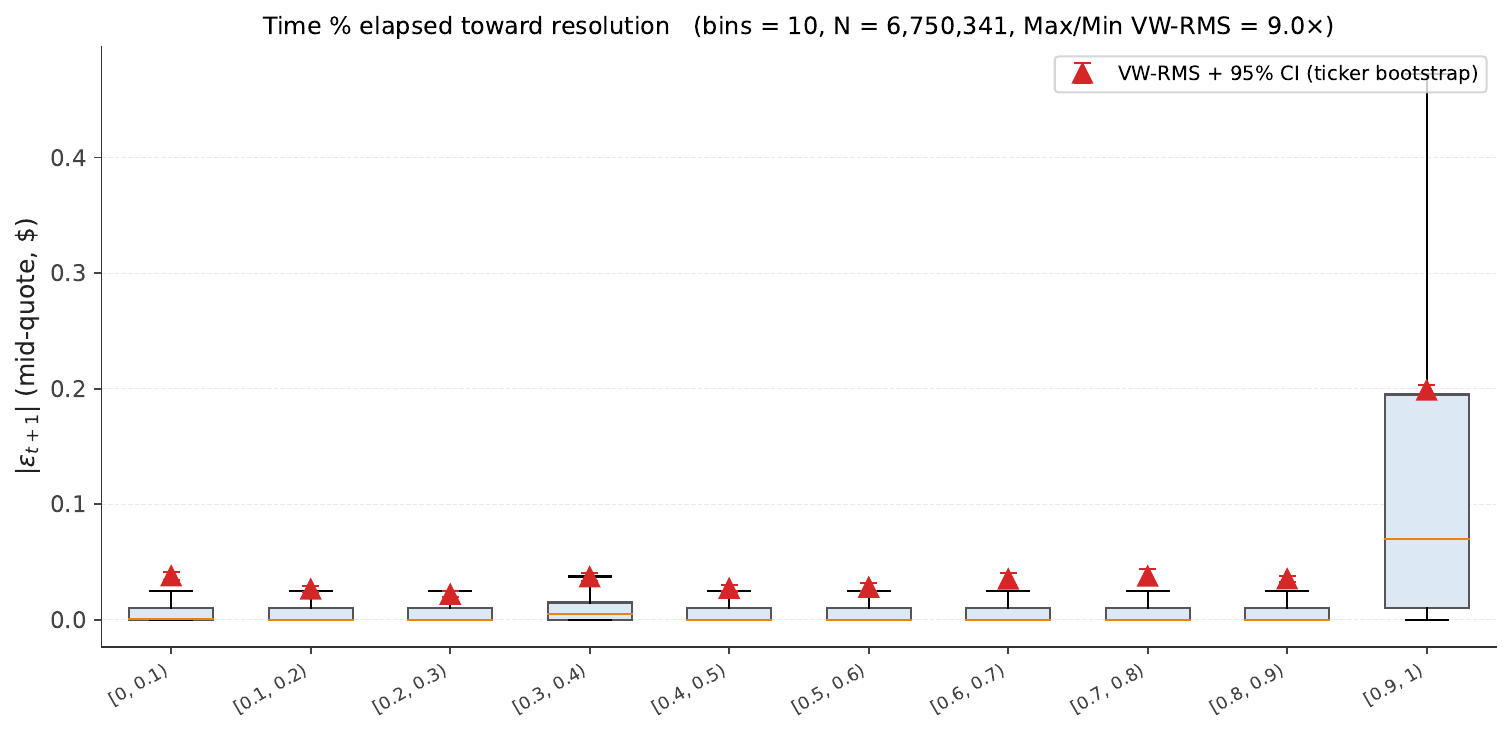}
\caption{Time percent to resolution.}
\end{subfigure}

\caption{Deadline and time-to-resolution diagnostics. Price and time to resolution are the primitive inputs of the deadline-resolution component. The inverse-square-root and time-percent panels are included as alternative time-to-resolution scalings.}
\label{fig:appendix-deadline-bins}
\end{figure}

The first three rows are time to resolution and two monotone re-expressions of it, \(1/\sqrt{\tau}\) and the elapsed share of the contract's life; they should not be read as three separate drivers. For price, the other primitive input of the deadline-resolution component, realized volatility peaks near 50 cents and falls toward the binary boundaries, matching the inverted-U of \(p(1-p)\); it ranks lower in the table because its boundary bins do not collapse to zero, as the mid-quote is still revised through the thin side of the book even when little trading occurs. Category is diagnostic rather than structural: it summarizes differences in market type, contract duration, and trading environment.

\begin{figure}[htbp]
\centering
\begin{subfigure}{0.48\textwidth}
\includegraphics[width=\linewidth]{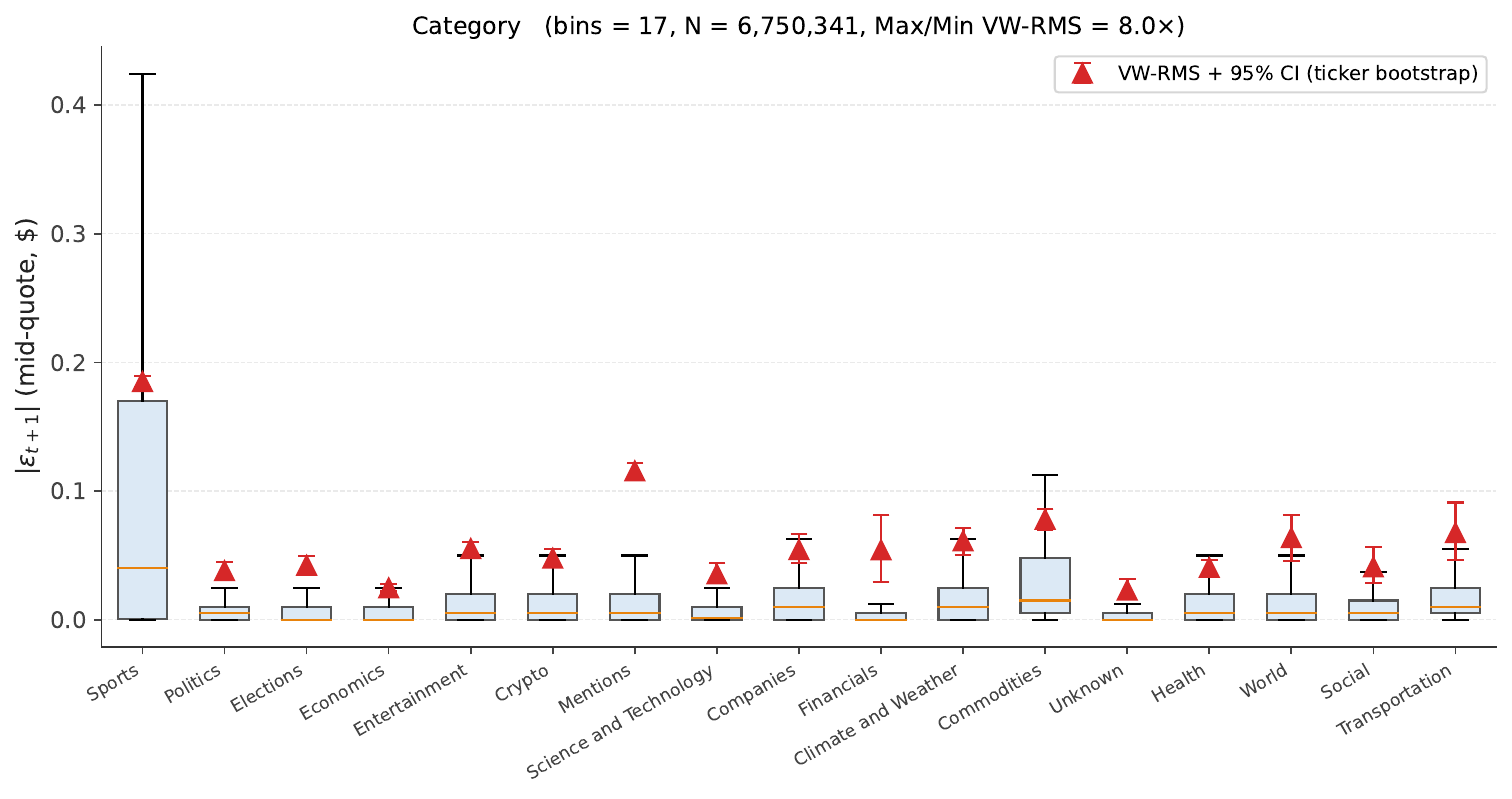}
\caption{Category.}
\end{subfigure}
\hfill
\begin{subfigure}{0.48\textwidth}
\includegraphics[width=\linewidth]{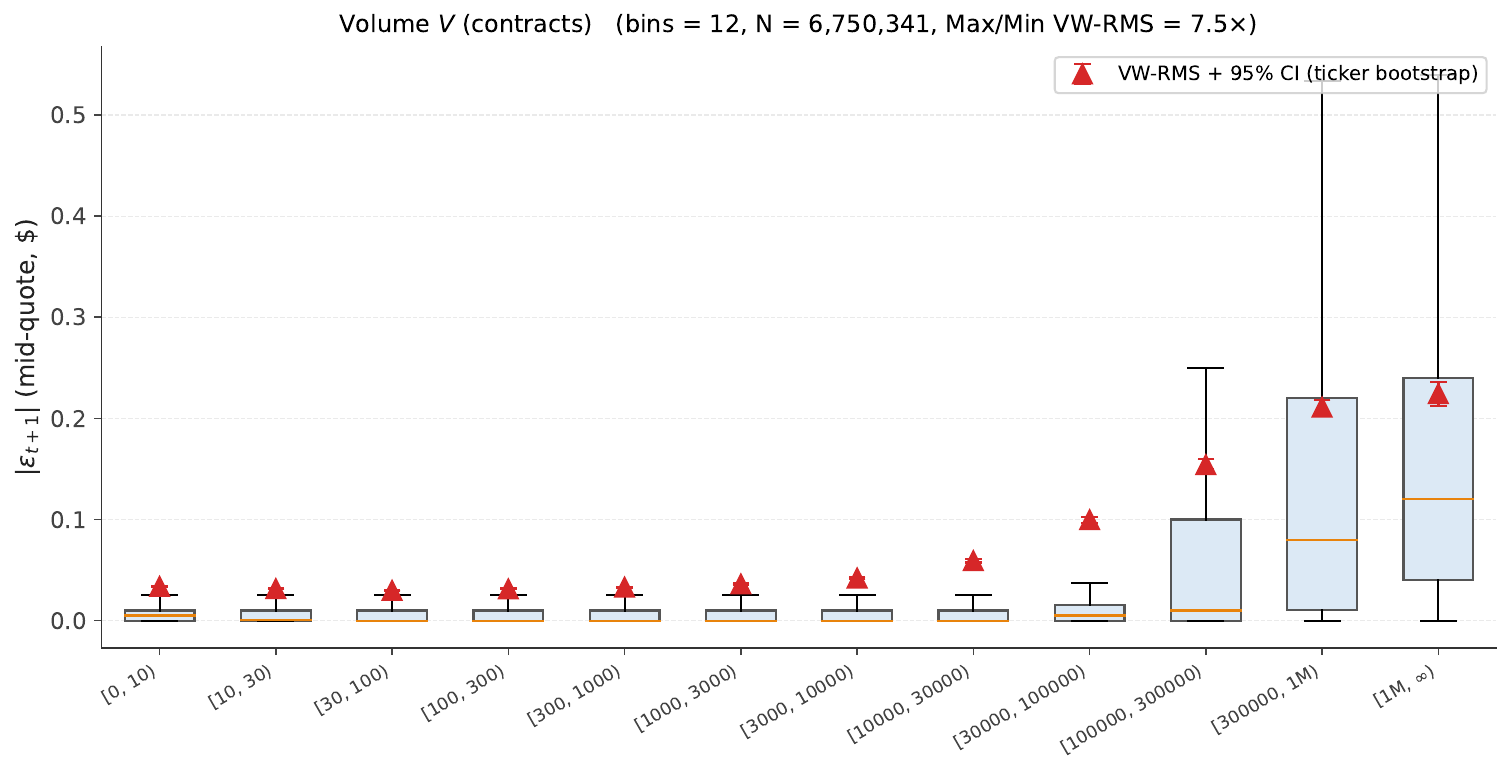}
\caption{Volume \(V\).}
\end{subfigure}

\vspace{0.5em}

\begin{subfigure}{0.48\textwidth}
\includegraphics[width=\linewidth]{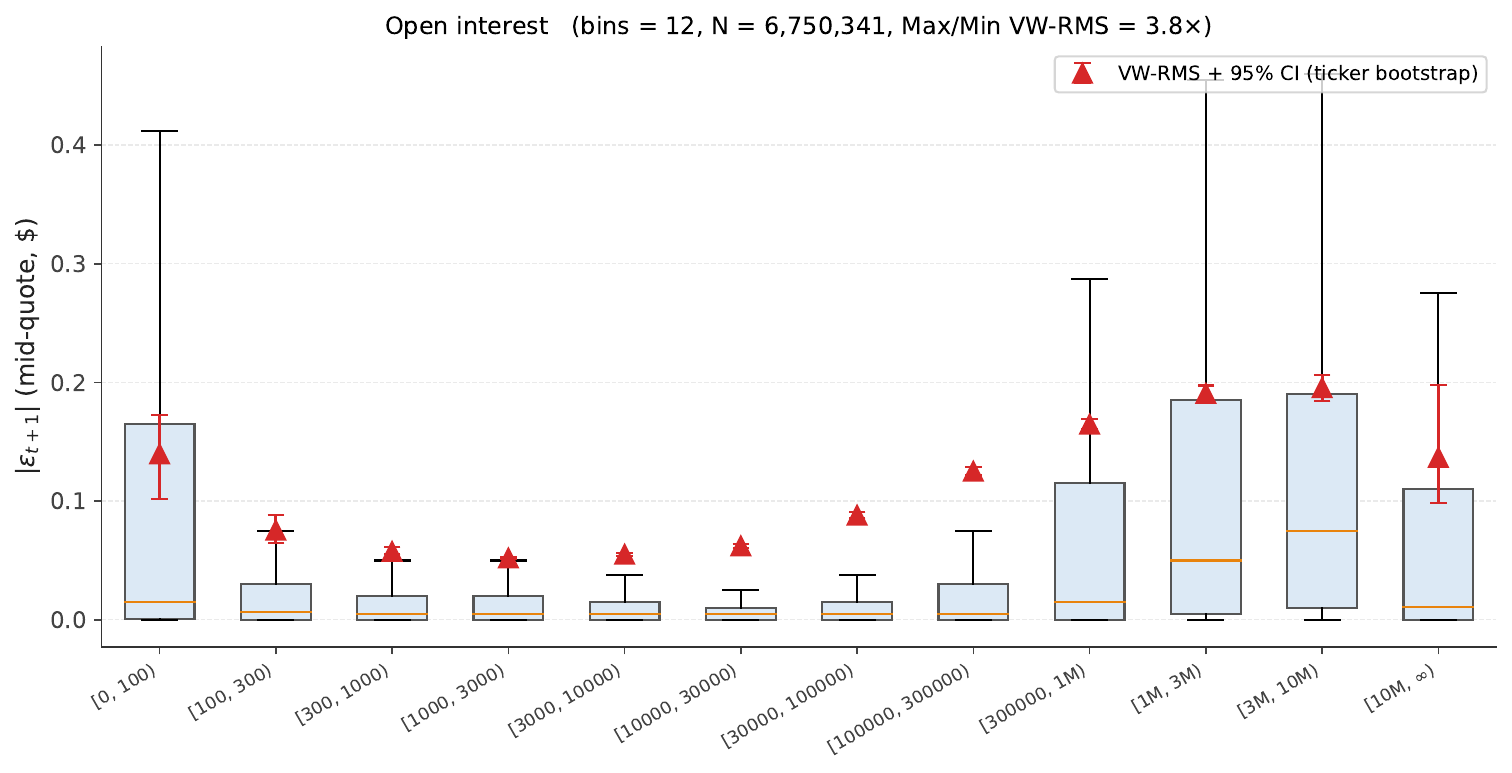}
\caption{Open interest.}
\end{subfigure}
\hfill
\begin{subfigure}{0.48\textwidth}
\includegraphics[width=\linewidth]{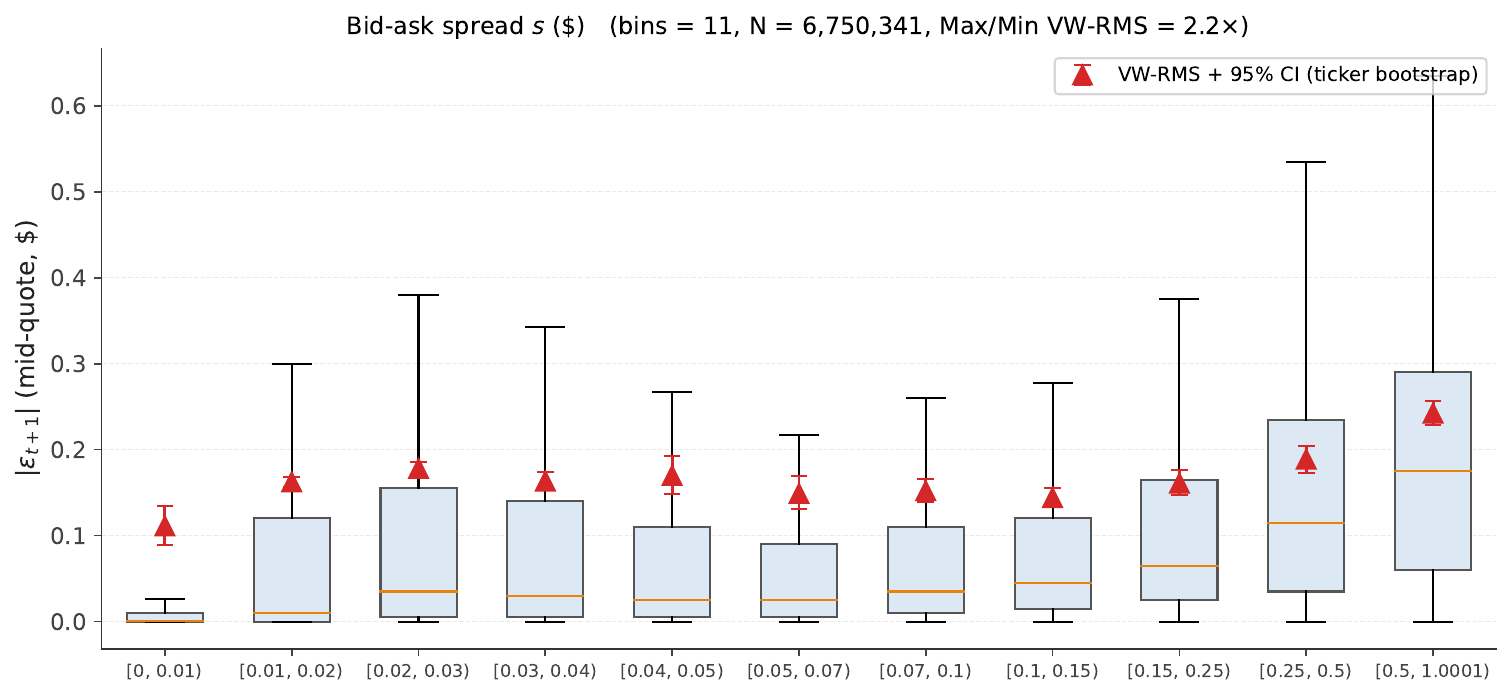}
\caption{Spread \(s\).}
\end{subfigure}

\caption{Market-type, activity, participation, and spread diagnostics. Category and open interest are descriptive controls rather than primitive inputs to DR-AS. Volume and spread are the two observable inputs to the order-flow term, where they enter jointly through \(\nu(V)s^2/4\).}
\label{fig:appendix-market-bins}
\end{figure}

\FloatBarrier

\ifconf
\section{Statistical Tests for the Forecast Ranking}
\label{app:pairwise}

This appendix reports the paired bootstrap tests behind the ranking claims in
Section~\ref{sec:horserace}.

\input{fc_moves/4_0_pairwise}\fi

\ifconf
\section{Activity Proxies within DR-AS}
\label{app:activity}

This appendix gives the full comparison of the activity proxies summarized in
Section~\ref{sec:three-comparisons}.

\input{fc_moves/4_2_activity}\fi

\section{Independence of Interval Violations}
\label{app:conditional_coverage}

This appendix gives the statistic used in Section~\ref{sec:conditional-coverage} and compares it with the pooled likelihood-ratio test of \citet{christoffersen1998evaluating}. Let \(I_{it}\in\{0,1\}\) be the violation indicator for contract \(i\) at hour \(t\), equal to one when the realized price falls outside the interval.

\ifconf

\input{fc_moves/4_3_crate_statistic}\fi
\paraheader{A conditional statistic.}
For contract \(i\), let \(N_i\) be the number of observed hours, \(K_i=\sum_t I_{it}\) the number of violations, \(\mathcal E_i\) the set of valid one-hour transitions (pairs of consecutive observed hours), \(L_i=|\mathcal E_i|\), and
\[
n_{11,i}=\sum_{(u,v)\in\mathcal E_i} I_{iu}I_{iv}
\]
the number of consecutive violation pairs. We condition on \((N_i,K_i,\mathcal E_i)\) and take as the null hypothesis that the \(K_i\) violations are placed uniformly at random over the \(N_i\) observed hours. This conditional null does not involve the contract's violation rate, so heterogeneity in violation rates across contracts cannot induce spurious dependence. Writing \((x)_r=x(x-1)\cdots(x-r+1)\) and \(p_{r,i}=(K_i)_r/(N_i)_r\), the probability that both ends of a valid transition are violations is \(p_{2,i}\), and
\begin{equation}
\mu_i=\E_0[n_{11,i}\mid N_i,K_i,\mathcal E_i]=L_i\,p_{2,i}=L_i\frac{K_i(K_i-1)}{N_i(N_i-1)}.
\label{eq:app-mu}
\end{equation}
The statistic aggregates the excess over contracts,
\begin{equation}
C_{\mathrm{rate}}=\frac{\sum_i (n_{11,i}-\mu_i)}{\sum_i L_i},
\label{eq:app-crate}
\end{equation}
the number of excess consecutive violation pairs per valid transition. It is the exact-conditional, heterogeneity-robust analogue of the alternative in Christoffersen's independence test; in the language of spatial statistics it is a panel join-count statistic under a within-contract permutation null.

The conditional variance is also available in closed form. Two valid transitions that share an hour are dependent. Let \(M_i\) be the number of unordered pairs of distinct valid transitions that share one hour and \(D_i=\binom{L_i}{2}-M_i\) the number of disjoint pairs. Then
\begin{equation}
v_i=\Var_0(n_{11,i}\mid N_i,K_i,\mathcal E_i)
=
L_i\,(p_{2,i}-p_{2,i}^2)+2M_i\,(p_{3,i}-p_{2,i}^2)+2D_i\,(p_{4,i}-p_{2,i}^2),
\label{eq:app-v}
\end{equation}
and if contracts were independent, \(Z_C=\sum_i(n_{11,i}-\mu_i)/\sqrt{\sum_i v_i}\) would be asymptotically standard normal by a cross-sectional central limit theorem. A permutation test that re-draws the violation locations within each contract imposes the same conditional null and gives a finite-sample benchmark for the tail probability; we report its \(z\)-statistic alongside the analytical one.

\paraheader{Inference unit.}
Contracts on the same underlying event may share shocks, so the independence across \(i\) assumed by \(\sum_i v_i\) understates the sampling variation. The confidence intervals in Section~\ref{sec:conditional-coverage} therefore come from an event-cluster bootstrap that resamples events, with all their contracts, with replacement. All statistics in that section are computed hour by hour without volume weights; a volume-weighted version replaces \(n_{11,i}-\mu_i\) and \(L_i\) by \(w_i(n_{11,i}-\mu_i)\) and \(w_iL_i\) for fixed contract weights \(w_i\) and leads to the same conclusions.

\begin{table}[htbp]
\centering
\caption{Independence of interval violations on the calibrated 95\% intervals: heterogeneity-adjusted statistic and pooled Christoffersen benchmark. Coverage is the unweighted share of hours covered. \(C_{\mathrm{rate}}\) is reported \(\times 1000\) with its 95\% event-cluster bootstrap interval and the within-contract permutation \(z\)-statistic. \(\widehat\pi_{01}\) and \(\widehat\pi_{11}\) are the pooled probabilities of a violation after a non-violation and after a violation; \(\mathrm{LR}_{\mathrm{IND}}\) is the pooled likelihood-ratio statistic for their equality (\(\chi^2(1)\), 5\% critical value 3.84).}
\label{tab:app-christoffersen}
\footnotesize
\setlength{\tabcolsep}{4pt}
\begin{tabular}{lccrccr}
\toprule
Specification & Coverage & \(C_{\mathrm{rate}}\times 1000\) [95\% CI] & perm.\ \(z\) & \(\widehat\pi_{01}\) & \(\widehat\pi_{11}\) & \(\mathrm{LR}_{\mathrm{IND}}\) \\
\midrule
DR-AS\,(\(\sqrt V\))                     & 0.919 & \(+4.24\) [\(+3.95,\;+4.50\)] & \(+96.9\)  & 0.050 & 0.221 & 66{,}653 \\
ARCH \(\times\) DR-AS\,(\(\sqrt V\))     & 0.917 & \(-3.99\) [\(-4.31,\;-3.67\)] & \(-92.5\)  & 0.064 & 0.065 & 3.5 \\
GARCH \(\times\) DR-AS\,(\(\sqrt V\))    & 0.955 & \(-0.76\) [\(-0.85,\;-0.67\)] & \(-32.5\)  & 0.035 & 0.042 & 127 \\
\addlinespace
Deadline resolution                      & 0.933 & \(+7.41\) [\(+6.83,\;+7.94\)] & \(+184.2\) & 0.032 & 0.353 & 168{,}297 \\
ARCH \(\times\) deadline resolution      & 0.949 & \(+2.80\) [\(+2.49,\;+3.14\)] & \(+93.3\)  & 0.027 & 0.223 & 61{,}052 \\
GARCH \(\times\) deadline resolution     & 0.957 & \(+1.48\) [\(+1.37,\;+1.61\)] & \(+64.9\)  & 0.030 & 0.115 & 15{,}496 \\
\addlinespace
Archak--Ipeirotis                        & 0.943 & \(+5.64\) [\(+5.16,\;+6.13\)] & \(+161.4\) & 0.026 & 0.329 & 134{,}943 \\
ARCH \(\times\) Archak--Ipeirotis        & 0.938 & \(+4.64\) [\(+4.19,\;+5.11\)] & \(+127.3\) & 0.031 & 0.285 & 104{,}175 \\
GARCH \(\times\) Archak--Ipeirotis       & 0.959 & \(+2.51\) [\(+2.30,\;+2.74\)] & \(+102.7\) & 0.024 & 0.184 & 41{,}744 \\
\addlinespace
ARCH(1) plain                            & 0.986 & \(-0.11\) [\(-0.19,\;-0.02\)] & \(-6.0\)   & 0.011 & 0.122 & 14{,}556 \\
GARCH(1,1) plain                         & 0.991 & \(+0.19\) [\(+0.16,\;+0.24\)] & \(+21.8\)  & 0.007 & 0.074 & 5{,}270 \\
\bottomrule
\end{tabular}
\end{table}

\paraheader{Pooled benchmark.}
Table~\ref{tab:app-christoffersen} reports, for the specifications in Figure~\ref{fig:christoffersen}, the hour-level coverage, \(C_{\mathrm{rate}}\) with its event-cluster interval and permutation \(z\)-statistic, and the pooled Christoffersen quantities: the transition probabilities \(\widehat\pi_{01}\) and \(\widehat\pi_{11}\) estimated on the pooled transitions of all contracts, and the likelihood-ratio statistic \(\mathrm{LR}_{\mathrm{IND}}\) for their equality, referred to a \(\chi^2(1)\) distribution as in \citet{christoffersen1998evaluating}. The pooled statistic rejects independence for every specification but one, with values in the thousands against a 5\% critical value of 3.84, because with millions of transitions even the spurious dependence created by heterogeneous violation rates, described in Section~\ref{sec:conditional-coverage}, is overwhelmingly significant. Read as a ranking rather than as a test, the pooled gap \(\widehat\pi_{11}-\widehat\pi_{01}\) tells the same story as \(C_{\mathrm{rate}}\). The two DR-AS combinations are the only specifications with a pooled gap below \(0.01\), and within each ladder the gap falls from the structural forecast to the ARCH and GARCH multipliers. The pooled statistic misses two things that \(C_{\mathrm{rate}}\) detects. It reads the anti-clustering of ARCH \(\times\) DR-AS as independence, the one case it does not reject, and it reads the plain dynamic benchmarks as clustered, although their pooled gaps of \(0.07\) and \(0.11\) are entirely heterogeneity.

\ifconf
\section{Category-Level Results}
\label{app:category}

This appendix gives the sample composition by category, the per-category
comparisons of the forecast models, the per-category refitting, and the full
discussion of category difficulty and of the Sports subsets that
Section~\ref{sec:universality} summarizes.

\input{fc_moves/5_1_2_composition_portability}
\input{fc_moves/5_2_race_text}
\input{fc_moves/5_3_catfit}\subsection{Smooth resolution and event-concentrated risk}

\input{fc_moves/5_4_regimes}\subsection{Single-game and other Sports markets}
Figure~\ref{fig:sports-single-game} plots the interval and likelihood scores of
Table~\ref{tab:sports-single-game} for deadline resolution, DR-AS, and
GARCH \(\times\) DR-AS.

\input{fc_moves/5_5_sports_figure}\fi
